\documentclass[amsmath,amssymb,floatfix, 11pt]{revtex4}
\usepackage{amsmath}
\usepackage{graphicx}% Include figure files
\usepackage{bm}% bold math
\usepackage{float}
\ifx\pdftexversion\undefined
\usepackage[dvips]{hyperref}
\else
\usepackage{hyperref}
\fi
\hypersetup{
  colorlinks = true, linkcolor = magenta
}

\begin{document}

\def \tr{{\mbox{tr~}}}
\def \ra{{\rightarrow}}
\def \ua{{\uparrow}}
\def \da{{\downarrow}}
\def \be{\begin{equation}}
\def \ee{\end{equation}}
\def \ba{\begin{array}}
\def \ea{\end{array}}
\def \bea{\begin{eqnarray}}
\def \eea{\end{eqnarray}}
\def \nn{\nonumber}
\def \l{\left}
\def \r{\right}
\def \half{{1\over 2}}
\def \etal{{\it {et al}}}
\def \cK{{\mathcal{K}}}
\def \cH{{\cal{H}}}
\def \cM{{\cal{M}}}
\def \cN{{\cal{N}}}
\def \cQ{{\cal Q}}
\def \cI{{\cal I}}
\def \cV{{\cal V}}
\def \cG{{\cal G}}
\def \cF{{\cal F}}
\def \cZ{{\cal Z}}
\def \bS{{\bf S}}
\def \bI{{\bf I}}
\def \bL{{\bf L}}
\def \bG{{\bf G}}
\def \bQ{{\bf Q}}
\def \bR{{\bf R}}
\def \br{{\bf r}}
\def \bu{{\bf u}}
\def \bq{{\bf q}}
\def \bk{{\bf k}}
\def \bK{{\bf K}}
\def \bz{{\bf z}}
\def \bx{{\bf x}}
\def \bpsi{{\bar{\psi}}}
\def \tJ{{\tilde{J}}}
\def \tk{{\tilde{k}}}
\def \tx{{\tilde{x}}}
\def \W{{\Omega}}
\def \e{{\epsilon}}
\def \lam{{\lambda}}
\def \L{{\Lambda}}
\def \a{{\alpha}}
\def \t{{\theta}}
\def \b{{\beta}}
\def \g{{\gamma}}
\def \D{{\Delta}}
\def \d{{\delta}}
\def \w{{\omega}}
\def \s{{\sigma}}
\def \f{{\varphi}}
\def \x{{\chi}}
\def \e{{\epsilon}}
\def \h{{\eta}}
\def \G{{\Gamma}}
\def \z{{\zeta}}
\def \hatt{{\hat{\t}}}
\def \hn{{\bar{n}}}
\def \vk{{\bf{k}}}
\def \vq{{\bf{q}}}
\def \gk{{\g_{\vk}}}
\def \nd{{^{\vphantom{\dagger}}}}
\def \yd{^\dagger}
\def \av#1{{\langle#1\rangle}}
\def \ket#1{{\,|\,#1\,\rangle\,}}
\def \bra#1{{\,\langle\,#1\,|\,}}
\def \braket#1#2{{\,\langle\,#1\,|\,#2\,\rangle\,}}
\newcommand{\cred}[1]{\color{red}#1\color{black}}

%\pagestyle{myheadings}
%\markright{1594/11 Ehud Altman}
\title{
 Non equilibrium quantum dynamics in ultra-cold quantum gases 
      }
\author{Ehud Altman}
\affiliation{Department of Condensed Matter Physics, Weizmann Institute of Science, Rehovot, 7610001, Israel}

%\begin{flushleft}
%\Large{\bf Non equilibrium quantum dynamics in ultra-cold quantum gases}
%\end{flushleft}
\begin{abstract}
Advances in controlling and measuring systems of ultra-cold atoms provided strong motivation to theoretical investigations of quantum dynamics in closed many-body systems. Fundamental questions on quantum dynamics and statistical mechanics are now within experimental reach: How is thermalization achieved in a closed quantum system?  How does quantum dynamics cross over to effective classical physics? Can such a thermal or classical fate be evaded? In these lectures, given at the  Les Houches Summer School of Physics "Strongly Interacting Quantum Systems Out of Equilibrium", I introduce the students to the novel properties that make ultra-cold atomic systems a unique platform for study of non equilibrium quantum  dynamics. I review a selection of recent experimental and theoretical work in which universal features and emergent phenomena in quantum dynamics are highlighted.
\end{abstract}

\maketitle

\tableofcontents
%Outline:
%
%\begin{enumerate}
%\item Introduction to ultra-cold quantum gases
%\item {\em Dynamics of ultra-cold atom interferometers --} A very natural experiment with applications to precision measurement. Exemplify in a simple way a lot of the new concepts: sudden quench, prethermalization, single versus multi-mode dynamics, integrability etc.
%\item{Dynamics of ultra-cold bossing atoms on optical lattices.} The Bose Hubbard model. The superfluid to Mott insulator transition. Effective field theories. The Higgs mode and its decay near the QCP.  
%\item {\em Fast quench across quantum critical points} 
%\begin{enumerate}
%\item Classical order parameter dynamics.
%\item Many-mode quantum evolution: Quantum seeding, creation of topological defects, emergence of classical physics at long times.
%\end{enumerate}
%
%\item {\em slow quench across a quantum critical point}
%\item Generic non-thermal states: dynamics in many-body localized states.
%
%\end{enumerate}

\section{Introduction to ultra-cold quantum gases}

A lot of the recent work on ultra cold atomic systems is focused on using them as simulators to investigate fundamental problems in condensed matter systems.
However it is important to note that many-body systems of ultracold atoms
are not one-to-one analogues of known solid state systems. These are independent
physical systems that have their own special features, and present new types of challenges, and opportunities.
One of the main differences between ensembles of ultracold atoms and solid state systems is in the  experimental
tools available for characterizing many-body states. While many traditional techniques
of condensed matter physics are not easily available (e.g. transport measurements), there
are several tools that are unique to ultracold atoms. This includes time-of-flight
expansion technique, interference experiments, molecular association spectroscopy, and many more.
In some cases, these tools have been used to obtain unique information about
quantum systems. One of the most striking examples is single site resolution
in optical lattices \cite{Bakr2009,Sherson2010}, which allowed exploration of the superfluid to
Mott transition at the unprecedented level. Another example was measurements
of the full distribution functions of the contrast of interference fringes in low dimensional
condensates, which provided information about high order correlation
functions. \cite{Polkovnikov2006Fringe,Gritsev2006,Hadzibabic2006,Gring2012}

An area in which ultracold atoms can make  unique contributions
is understanding non-equilibrium dynamics of quantum many-body systems.
Several factors make ultracold atoms ideally suited for the study of
dynamical phenomena.
\begin{trivlist}
\item {\it Convenient timescales}. Characteristic frequencies of many-body systems of ultracold
atoms correspond to kilohertz. System parameters can be modified and properties resulting from this dynamics
can be measured at these timescales (or faster). It is useful to contrast this
to dynamics of solid state systems, which usually takes place at Giga and Terahertz frequencies, which are
extremely difficult for experimental analysis.
\item{\it Isolation from the environment}. Ultra-cold atomic gases are unique in being essentially closed, almost completely decoupled from an external environment. More traditional condensed matter systems by contrast are always strongly coupled to some kind of thermal bath, usually made of the phonons in the crystal. This distinction does not matter much if the systems are at thermal equilibrium because of the equivalence of ensembles. However, it leads to crucial differences in the non equilibrium dynamics of the two systems.
While the dynamics in standard materials is generically overdamped and classical in nature, atomic systems can follow coherent quantum dynamics over long time scales. A corollary of this is that it is also much harder for systems of ultra-cold atoms to attain thermal equilibrium. Hence
understanding dynamics is important for interpreting experiments even when ultracold atoms are used to explore equilibrium phases.
\item{\it Rich toolbox}. One of the difficulties in studying nonequilibrium dynamics of many-body systems
is the difficulty of characterizing complicated transient states. Experimental techniques that have been developed
for ultracold atoms recently are  well suited to this challenging task. Useful tools include local resolution
\cite{Bakr2009,Sherson2010}, measurements of quantum noise\cite{Altman2004a,Folling2005} and interference fringe statistics \cite{Polkovnikov2006Fringe,Gritsev2006,Hadzibabic2006,Gring2012}.
\end{trivlist}

Before proceeding let me list several fundamental questions  that can be addressed with ultra cold atomic systems and that we will touch upon.

\noindent {\em{Emergent phenomena in quantum dynamics}--} The main theme in the study of complex systems in the last century has been to identify emergent phenomena and understand the universal behavior they exhibit. Examples of successful theories of universal phenomena in equilibrium phases include the Ginsburg- Landau-Wilson theory of broken symmetry phases and critical phenomena, the theory of Fermi liquids, and more recently, the study of quantum hall states, spin liquids and other topological phases. In addition there is a growing understanding of emergent universal phenomena in classical systems out of equilibrium. This includes for example the phenomena of turbulence, coarsening dynamics near phase transitions, the formation of large scale structures such as sand dunes and Raleigh-Bernard cells,  dynamics of active matter such as flocking of animals and more.
 Yet there is very little understanding of non equilibrium many-body phenomena that are inherently quantum in nature.
It is an interesting open question whether the dynamics of complex systems can exhibit emergent universal phenomena in which quantum interference and many-body entanglement play an important role.

\noindent{\em{Thermalization and prethermalization in closed systems}--} Intimately related to establishment of emergent phenomena is the question of thermalization and the approach to equilibrium in closed quantum systems. Cold atomic system are ideal for testing the common belief that generic many-body systems ultimately approach thermal  equilibrium. There are however exceptions to this rule. Integrable models, for example, fail to come to thermal equilibrium because the dynamics is constrained by an infinite set of integrals of motion. In general integrable models require extreme fine tuning. Nevertheless systems of ultra-cold atoms in one dimensional confining potentials often realize very-nearly integrable models. The reason for this is that the interactions naturally realize almost pure two body contact interactions. This fact has been used to demonstrate breaking of integrability in the Lieb-Lininger model of one-dimensional bosons with contact interactions\cite{Kinoshita2006}. This experiment has motivated a lot of theoretical and numerical investigations to characterize the non thermal steady state that is reached in such cases. The latter is usually characterized using a generalized Gibbs ensemble (GGE) that seeks the equilibrium (maximum entropy) state subject to the infinite number of constraints set by the values of the integrals of motion in the initial state\cite{Rigol2008}. 

The systems which realize nearly integrable models usually do include small perturbations that break integrability. It is believed (though not proved) that these perturbations will eventually lead to true thermalization. Hence the (GGE) is considered to be a long lived prethermalized state. An intersting question that we will touch upon concerns the time scale for relaxational dynamics to take over and lead to true thermal equilibrium. 

\noindent{\em Absence of thermalization and many-body localization --} 
A body of recent work starting with a theoretical suggestion \cite{Basko2006,Gornyi2005} (and much earlier conjecture by Anderson \cite{Anderson1958}) points to quantum many-body systems with quenched disorder as providing a generic alternative to thermalization. That is, in contrast to integrable models, lack of thermalization is supposed to be robust to a large class of local perturbations in such closed systems. This phenomenon is known as many-body localization (MBL) Important questions that are beginning to be addressed with ultra-cold atomic systems \cite{Kondov2013,Schreiber2015} concern the dynamical behavior in the MBL state and the nature of the transition from MBL to conventional ergodic dynamics. 
If there is a critical point controlling the transition from a thermalizing to a non thermalizing state, does it involve singularities in correlation functions similar to those that dominate quantum critical phenomena at equilibrium?
%
%\noindent{\em{Relaxational dynamics --}} Generic systems, even closed ones, of course do thermalize. But the relaxational dynamics associated with attaining equilibrium in closed systems may be very different from condensed matter systems that relax toward equilibrium with an external bath.
%The latter are well described by stochastic classical equations\cite{Hohenberg1977} in which the noise term is determined by the temperature of the external bath and the coupling to it through the fluctuation dissipation theorem. It is believed that in generic systems the stochastic Langevin equations do provide an accurate asymptotic description of the attainment of equilibrium. An interesting question is how this type of dynamics emerges at long times from the quantum dynamics of a closed system. Is it possible to experimentally observe and quantify quantum corrections to the relaxational dynamics?

In these lecture notes I focus on universal phenomena in dynamics that can and have been investigated using ultra cold atomic systems. Of course I cannot cover everything, so I picked some illustrative examples that I find interesting from a fundamental perspective and have also been directly investigated experimentally using ultra-cold atomic systems. 
 In section \ref{sec:interf} I  discuss the dynamics of ultra-cold atom interferometers, primarily in one dimension, as case studies of prethermalization and thermalization dynamics. In section \ref{sec:optlat} I turn to the dynamics of bosons in optical lattices. I review the theoretical understanding of the quantum phase transition from a superfluid to a Mott insulator and then discuss emergent dynamical phenomena that have been studied in the vicinity of this transition. In particular I will discuss the nature of the Higgs amplitude mode that emerges near the critical point as well as the modes for decay of super currents due to the enhanced quantum fluctuations near the Mott transition.  In section \ref{sec:quench} I discuss far from equilibrium dynamics involving a quench of system parameters across the superfluid to Mott insulator phase transition. Section \ref{sec:quench1d} reviews some of the recent work on quenches in one dimensional dimensional optical lattices. Finally, in section \ref{sec:MBL} I turn to review the recent progress in theoretical understanding of many-body localization as well as recent experiments that investigated this phenomenon. I should note that in these notes I include the very significant recent progress on many-body localization that took place after the lectures were delivered. I have also tried to make each section self contained so that they can be read independently of each other. However if the reader is not familiar with the physics of Bosons on optical lattices and the Mott transition it would help to read section \ref{sec:optlat} before section \ref{sec:quench}.

\section{Dynamics of ultra-cold atom interferometers: quantum phase diffusion}\label{sec:interf}

Atomic condensates have been used early on as interferometers, utilizing the matter waves for making precision measurements of accelerations. Such experiments constitute a natural and simple example of a dynamical quench experiment, which beside its practical applications raise fundamental questions in non-equilibrium quantum physics. Hence this example will serve here to illustrate several key concepts, which occur frequently in non-equilibrium studies of cold atoms.

An idealized interferometer experiment consists of a macroscopic condensate separated into two wells and prepared with a well-defined relative phase between the wells. When the wells are disconnected the relative phase is expected to evolve in time under the influence of the  potential difference between the two wells. Following this evolution the phase is measured directly by releasing the atoms from the trap and observing the interference pattern established after a time of flight. 

\begin{figure}[t]
 \includegraphics[width=0.9\textwidth]{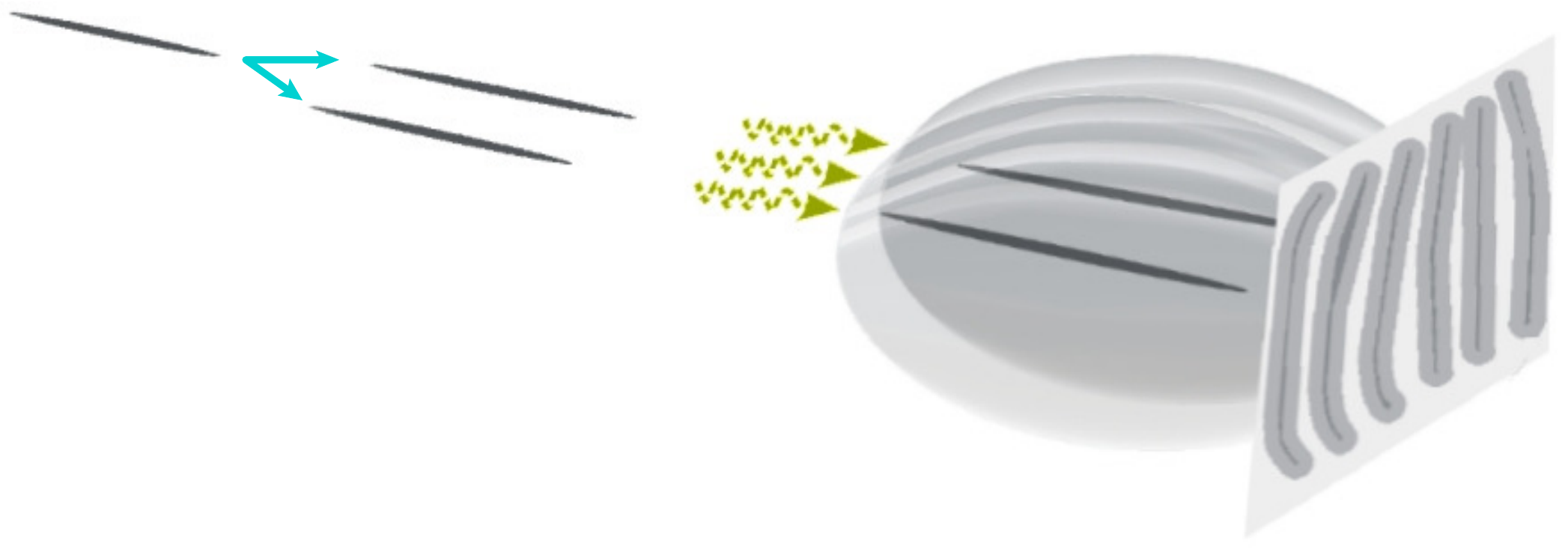}
 \caption{Typical setup for an interferometer measurement. A condensate is split into two decoupled wells such that the relative condensate phase between the wells is initially well defined $\t_-=0$. The two condensates are let to evolve under the influence of the local potentials acting on them for time $t_m$. At time $t_m$ they are released from the trap and their expansion leads to an interference pattern that is observed on a light absorption image.  \label{fig:interf-setup}}
\end{figure}

One method to realize such a preparation protocol was to start with a one condensate in an elongated potential well and raise a radio-frequency induced potential barrier to split the condensate into two one dimensional condensates\cite{Schumm2005}. 
A schematic setup of this nature is illustrated in Fig. \ref{fig:interf-setup}. 
Alternatively, instead of using two separate wells as the two arms of the interferometer one can use two internal states of the atoms, as done in Ref. \cite{Widera2008}. The system is prepared with a condensate in a single internal state, say $\ket{\ua}$. A two photon transition is used to a induce a $\pi/2$ rotation to the state $(\ket{\ua}+\ket{\da})/\sqrt{2}$, this is the analog of the coherent splitting in the double well realization. Thereafter the system is  let to evolve freely under the influence of the Hamiltonian with no coupling between the two internal state. Finally the coherence between the two states remaining after a time $t$ is determined through a Ramsey type measurement. That is, measuring how much of the $\ket{\ua}$ state population is restored following an $-\pi/2$ rotation of the internal state.

\begin{figure}[t]
 \includegraphics[width=0.9\textwidth]{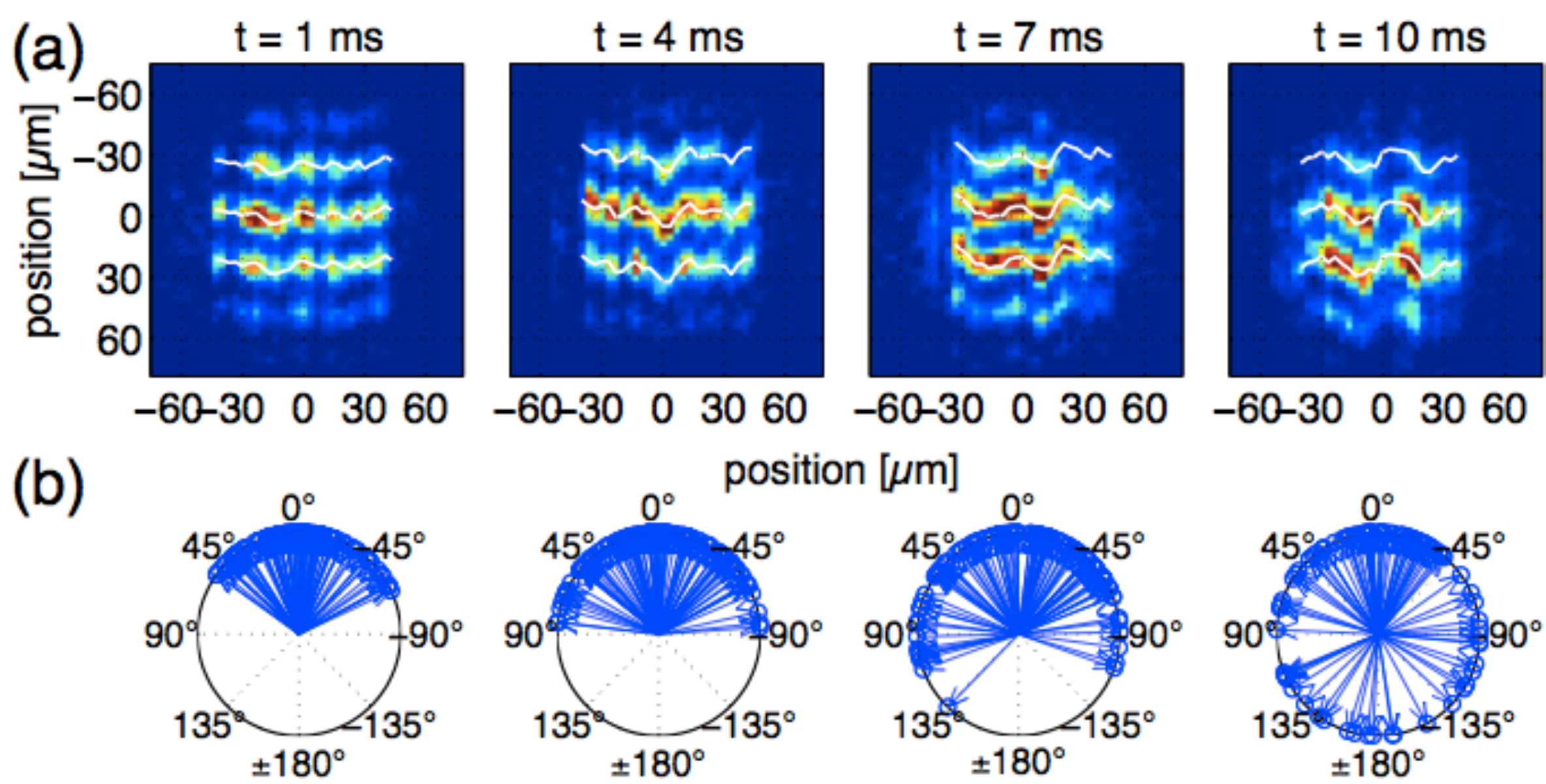}
 \caption{(a) Example of fringes from interfering one-dimensional condensates taken at varying times after the condensate had been coherently split  \cite{Hofferberth2007}. (b) The distribution of relative phase extracted from the interference fringes grows with time. (Reproduced from Ref. \cite{Hofferberth2007}) \label{fig:phase}}
\end{figure}

In reality the relative phase is not determined solely by the external potential difference between the two wells (or two internal spin states). The phase field of an interacting condensate should be viewed as a quantum operator with it's own dynamics. The intrinsic quantum evolution leads to uncertainty in the relative phase, which grows in time and eventually limits the accuracy of the interferometric measurement. 
Thus even if the condensates are subject to exactly the same potential, the relative phase measured by the interference pattern between them will have a growing random component. The rate and functional form with which this uncertainty grows is a hallmark of the many-body dynamics.
Fig. \ref{fig:phase} is an example of an experiment performed by the Vienna group \cite{Hofferberth2007} that shows the distribution of relative phase determined in repeated measurements after different times from the initial preparation. 

It is interesting to point out the essential difference between the problem
we consider here and dephasing of single-particle interference effects as seen, for example, in mesoscopic electron systems. In the latter case the dephasing of a single electron wave-function is a result of the interaction of the single electron with a thermal bath, which consists of the other electrons in the Fermi liquid or of phonons.  Because the bath is thermal and the energy of the injected electron is low, the dephasing problem is ultimately recast in terms of linear response theory \cite{Stern1990}.  

By contrast, dynamic splitting of the
condensate in the ultracold atom interferometer takes the system far from
equilibrium, and the question of phase coherence is then essentially one of
quantum dynamics.  The system is prepared in an initial state determined
by the splitting scheme, which
then evolves under the influence of a completely different Hamiltonian, that
of the split system. It is the first example we encounter of a quantum quench.
Dephasing, from this point of view, is the process that
takes the system to a new steady (or quasi-steady) state. In this respect, the
ultracold atom interferometer is a useful tool for the study of non equilibrium
quantum dynamics. One of our goals is to classify this dynamics into different universality classes.

\subsection{Phase diffusion in the single-mode approximation}\label{sec:single-mode}

Let us start the theoretical discussion from the simplest case of perfect single mode condensates initially coupled by a large Josephson coupling that locks their phases. The Hamiltonian which describes the dynamics of the relative phase $\f$ is given in the number-phase representation by
\be
H= \half U \, n^2 -J\cos\f
\label{H1}
\ee
Here $U$ is the charging energy of the condensate $U=\mu/N$, where $\mu$ is the chemical potential and $N=\av{N_1}=\av{N_2}$ the average particle number in the condensate. Alternatively we can express the charging energy using the contact interaction $u$ and the condensate volume $L^d$ as $U=u/L^d$. The relative particle number $n=(N_1-N_2)/2$ and relative phase $\f$ are conjugate operators obeying the commutation relation $[\f,n]=i$.

At time $t=0$ the Josephson coupling $J$ is shut off and the time evolution begins subject to the interaction hamiltonian alone. The system is probed by releasing the atoms from the trap after a time evolution over a time $t$ subject to the Hamiltonian of decoupled wells $J=0$. The aim is to infer the coherence between the two condensates from the emerging oscillating pattern in the density profile.  

Here it is important to note that such information can only be inferred from averaging the density profile (interference fringes) over many repetitions of the experiment. A single time-of-flight image, originating from a pair of macroscopic Bose condensates, exhibits density modulations $\rho(x)\sim A_0\cos(q_0 x +\f)$ of  undiminishing intensity $A_0\propto N$ but with possibly a completely random phase\cite{Andrews1997,Castin1997,Polkovnikov2006Fringe}. If the condensates are completely in phase with each other, then $\f$ is fixed and the average interference pattern taken over many experimental runs would be the same that seen in a single shot. On the other hand, if the relative phase $\f$ is completely undetermined, the average interference pattern would be vanishing. Therefore the correct measure of coherence is the average interference amplitude remaining after averaging over many shots, which is directly related to the quantum expectation value ${\mathcal A}(t)\sim \av{\cos\f}$. 

The time evolution of the relative phase subject to the Hamiltonian (\ref{H1}) with $J=0$ follows the obvious analogy to spreading of a wave-packet of a particle with mass $m=1/U=L^d/u$, the relative phase fluctuation follows a ballistic evolution
\be
\av{\f(t)^2}=\av{\f(0)^2}+{\av{\d n(0)^2} u^2\over L^{2d}} t^2
\ee
This implies a Gaussian decay of the Fringe amplitude with time
\be
{\mathcal A}(t)\approx A_0 e^{-t^2/t_D^2}
\ee
with the dephasing time scale $\tau_D= (L^d/u)/\sqrt{\av{\d n(0)^2}}= (N/\mu))/\sqrt{\av{\d n(0)^2}}$. Several important insights may already be gleaned from this simple model.
\begin{enumerate}
\item Quantum phase diffusion in a true condensate is a finite size effect. Indeed in the proper thermodynamic limit the broken symmetry state with well defined phase is infinitely long lived.
\item Quantum phase diffusion is driven by interaction. Setting $u=0$ eliminates it.
\item The rate of phase diffusion depends on the initial state. A narrower initial particle number distribution entails longer dephasing time.
\end{enumerate}

The last point implies that the dephasing time $t_D$ depends crucially on the preparation protocol of the interferometer.  In one extreme the initial state is prepared with maximal coherence. That is, a state with all particles in a symmetric superposition between the two wells. In the limit of large particle number N, we can take this state to have a gaussian relative number distribution between the wells with a variance $\av{\d n(0)^2}=N/2$. Accordingly the phase diffusion time in this case is $\tau_D=\sqrt{2N}/\mu$. The 

As we'll see below, such a state can be prepared by a rapid ramp-down of $J$ on a scale faster than the characteristic scale set by the chemical potential $\mu$.  We can increase the coherence time by preparing an initial state with a narrower number distribution. Due to the number phase uncertainty relation, we must pay by having a larger initial phase distribution and hence achieve lower accuracy in phase determination at short times. In quantum optics terminology this is called a squeezed state.

The simplest way to achieve squeezing is to split the condensates more slowly. In the slow extreme of  adiabatic splitting the system ends up in the ground state with completely undetermined relative phase (and vanishing number fluctuations). This state is infinitely long lived, but of course useless for phase determination. More generally we can split the condensates on a time scale $\tau_s$. The frequency scale we should compare to in order to asses whether this drive is slow or fast is the instantaneous gap in the system given roughly by the Josephson frequency $\w_J(t)=\sqrt{U J(t)}$. As long as $\w(t)>1/\tau_s$ the system to a good approximation remains in the instantaneous ground state and the dynamics is essentially adiabatic. However from some point during the split when $\w(t)< 1/\tau_s$, the dynamics should rather be viewed as a sudden split.
The effective initial state is the ground state of the system at the break-point $\w(t^*)=1/\tau_s$.

Using a Harmonic approximation of the junction, $H_{osc}={\mu\over 2N} n^2+{J\over 2}\f^2$, the number fluctuation in the initial state is easily found to be $\sqrt{\av{\d n^2}}\sim \sqrt{\w(t^*)N/\mu} = \sqrt{N/(\mu\tau_s)}$. We then obtain the phase-diffusion time of the squeezed state $\tau_D = \sqrt{N \tau_s/\mu}$. Because of the squeezing, the minimal phase uncertainty is now $\sqrt{\f(0)^2}\sim \sqrt{\mu\tau_s/N}$, larger than the uncertainty $1/\sqrt{N}$ in the case of the (fast) maximally coherent split. 

The single mode phase diffusion considered  in this section was first discussed in the context of Bose condensates by Leggett and Sols\cite{Leggett1991}. It is however a special case of the more general  problem of spontaneous symmetry breaking as discussed originally by Anderson in Ref. \cite{Anderson1952}. A single mode Hamiltonian emerges as a description of the dynamics of the uniform broken symmetry order parameter. In this case it is the dynamics of the  $U(1)$ order parameter, or condensate phase. The dynamics is governed by an effective mass which grows heavier with the system volume. Hence the phase becomes static due to the infinite mass in the thermodynamic limit allowing spontaneous symmetry breaking. In the next section we will  see under what conditions the condensate zero mode can indeed be safely separated  from the linearly dispersing Goldstone modes. In particular we'll show how the multi mode dynamics can dominate the dephasing in interferometers composed of low dimensional condensates.

\subsection{Many modes: hydrodynamic theory of phase diffusion in low dimensional systems}

The single mode approximation described above is expected to be a reasonable description of bulk three dimensional condensates. In this case internal phase fluctuations, quantum or thermal, are innocuous since they do not destroy the broken symmetry. The mode responsible for restoring the symmetry in a finite system is the uniform mode and therefore only its dynamics has to be considered. However, many interference experiments have been performed with highly elongated, essentially one dimensional condensates\cite{Schumm2005,Jo2007,Hofferberth2007,Widera2008}. In this case zero point fluctuations due to long-wavelength phonons destroy the broken symmetry even in the thermodynamic limit. It is  expected that these phonons will have a crucial impact on the loss of phase coherence.

The model Hamiltonian to consider is a direct generalization of the single mode Hamiltonian considered in the previous section
\be
H=H_1+H_2-J(t)\int d^dx \cos(\f_1-\f_2).
\ee
Here $H_{1,2}$ are the Hamiltonians of the individual one (or higher) dimensional condensates and the Josephson coupling $J(t)$ now operates along the entire length (or area) of the condensates. As in the single mode case the system is prepared with strong Josephson coupling, which is thereafter rapidly shut off. In the way this scheme is commonly implemented \cite{Schumm2005,Jo2007}  the system actually starts as a one dimensional condensate in a single tube, which is then split into a double tube over a timescale $\tau_s$ slow compared to transverse energy levels in the tube. At the same time $\tau_s$  can be fast compared to the chemical potential of the condensate, leading to preparation of a coherent state of the relative phase\cite{Schumm2005}. Alternatively $\tau_s$ can be larger than the inverse chemical potential, leading to preparation of a squeezed phase state\cite{Jo2007}. 

We want to track the ensuing dynamics of the relative phase $\f_1-\f_2$  following this preparation stage. Again, this is done by releasing the atoms from the trap at fixed times $t$ after the split and inspecting the resulting "matter wave" interference fringes in the density profile. As in  the single mode case interference fringes can be seen even when the two condensates are independent. The correct measure of the coherence between the two condensates is the average of the fringe amplitude over many repeats of the experiment, directly related to the expectation value 
${\mathcal A}(x,t)\sim \av{\cos[\f_1(x,t)-\f_2(x,t))]}$. What can be conveniently measured in an experiment is in fact the integral of this quantity over a certain averaging length:
\be
{\mathcal A}_L(t)= \av{A_L(t)}\equiv \int_{-L/2}^{L/2} dx\av{\cos[\f_1(x,t)-\f_2(x,t))]}
\label{Aint}
\ee
This quantity of course vanishes for independent condensates. It is important to distinguish ${\mathcal A}_L(t)$ from the quantities $\av{|A_L(t)|}$ and $\av{|A_L(t)|^2}$ that would give a non-vanishing value (dependent on $L$) even for independent condensates \cite{Polkovnikov2006Fringe}.
Fig. \ref{fig:phase} shows an example of interference fringes and the distribution of phases of the quantity $A_L(t)$ measured in repeated experiments.  

Here, in computing the time evolution of the aversge fringe amplitude ${\mathcal A}_L(t)$ we will closely follow the analysis of Ref. \cite{Bistritzer2007}. There, a detailed theoretical description of the dynamics is obtained by approximating the exact Hamiltonians $H_{1,2}$ by the low energy Harmonic-fluid description of the phase fluctuations (phonons) in the superfluid\cite{Popov1972}
\be
H_\a=\int dx \left[\rho_s(\nabla\t_\a)^2
+\kappa^{-1}\,\Pi_\a^2\right].
\label{LL}
\ee
Here $\t_\a$ is  the phase field of the condensate $\a=1$ or $2$ and $\Pi_\a$ is the canonical conjugate smooth density field. $\rho_s$ and $\kappa$ are the macroscopic phase stiffness and compressibility of the condensate respectively. These should in general be considered as phenomenological parameters extracted from experiment. However, for weakly interacting Bose gases the compressibility is only weakly renormalized from the bare value given by the inverse contact interaction $\kappa=g^{-1}=\rho/\mu$ while the stiffness is directly related to the average density as $\rho_s=\rho_0/m$. The short range cutoff of this hydrodynamic theory is the healing length of the condensate $\xi_h=1/\sqrt{4mg\rho_0}$. It will sometimes be convenient, to express quantities in terms of the Luttinger parameter $K=\pi\sqrt{\rho_s\kappa}$ that determines the power-law decay of correlations in 1d.

We note that the anharmonic terms neglected here are irrelevant for the asymptotic low energy response in the ground state, yet they may influence the decoherence dynamics at long times. This will be considered in the next section.

It is convenient to transform to the relative $\f=\t_1-\t_2$ and "center of mass" $\t=(\t_1+\t_2)/2$ phase variables and their respective conjugate momenta $\Pi=\Pi_1+\Pi_2$ and $n=(\Pi_1-\Pi_2)/2$. Within the harmonic-fluid description the relative and "center-of mass" coordinates are decoupled. This is not changed by the dynamic Josephson term, which does not involve the "center of mass" modes at all.
Since the observable we are interested in (\ref{Aint}) involves only the relative phase,
we can forget about the center of mass modes at this level of approximation.

After the Josephson coupling is turned off the relative phase evolves under the influence of a purely Harmonic Hamiltonian. In Fourier space the different spatial modes of the relative degrees of freedom are conveniently decoupled and we have
\be
H_-= {1\over 2}\sum_q \left( {2\kappa^{-1}} |\Pi_q|^2+ {\half\rho_s q^2} |\f_q|^2 \right).
\ee
Assuming that the initial state is a Gaussian wave-function, the time evolution of the mean fringe amplitude, our proxy for phase coherence, is easily computed from the Harmonic theory
\bea
{\mathcal A}(t)&=&\rho\av{e^{i\f(x)}}=\rho e^{-{1\over 2L^d}\sum_{q}\av{|\f_q(t)|^2}}=\rho e^{-g(t)}\label{At}\\
\av{\f_q(t)^2}&=&\av{\f_q(0)^2}\cos^2 (c q t)+{1\over\av{\f_q(0)^2}}\left({\pi\over q K}\right)^2 \sin^2(c q t)
\label{At2}
\eea

To make further progress we need the form of the initial phase distribution, which depends on the preparation scheme.

\subsubsection{Fast split scheme} 
As in the single mode case we begin with a discussion of a fast split. In this case all particles remain in the symmetric superposition between the two wells. Such an initial state is particularly simple to write in the number phase representation if there are many particles per length unit $\xi_h$ that serves as a cutoff to the hydrodynamic theory. This is equivalent to the weak coupling requirement $K>>1$. In this case the Poisson statistics of the particle number is well approximated with Gaussian statistics with $\av{\Pi_q(0)^2}\approx \rho$. Accordingly the phase wave-function is Gaussian in this limit and the variance $\av{\f_q(0)^2}\approx 1/2\rho$ should be plugged into (\ref{At2}). 

We convert the sum over all ${\bf q\ne 0}$ in (\ref{At}) to an integral, while separating out the uniform $q=0$ component, which leads to a unique contribution.  Following a simple change of variables we then have
\be
g(t)\approx {t^2\over \tau_L^2}+ {S_d\over (2\pi)^d}\int_0^{\mu t} dz z^{d-1}\left( {1\over 4 \rho( ct)^d}\cos^2z+\rho(\pi/ K)^2 (ct)^{2-d}{\sin^2 z\over z^2} \right)
\ee
Where $S_d={2\pi^{d/2}\over \G(n/2)}$ is the surface area of a unit sphere in $d$ dimensions  and $\tau_L=\mu^{-1}{\sqrt{\rho L}}=\mu^{-1}\sqrt{N}$. The first term in the integrand saturates to a constant at long times. The second term depends crucially on the spatial dimension $d$. The contribution of the zero mode together with the integral over the internal phonons then gives:
\begin{eqnarray}
\label{At2}
{\mathcal A}(t) \propto
{\mathcal A}_0 \exp[-t^2/\tau_L^2]
\times \left\{
\begin{array}{cc}
\exp[- t/ \tau_Q], & d = 1, \\
(\mu t)^{- \a},\,\, & d = 2, \\
\text{const} \,\,\, & d=3,
\end{array}
\right.
\end{eqnarray}
where $\tau_Q={2\over \rho c}(K/\pi)^2= (2/ \mu) K/\pi^2$.

Let us pause to discuss this result. The contribution of the zero mode is the same in all dimensions. It gives rise to a Gaussian decay on a time scale that diverges in the thermodynamic limit. This is precisely our result obtained in the previous section from the single mode approximation. Now in addition we have included the internal modes at the Gaussian level. In three dimensions we see from (\ref{At2}) that the phase fluctuations do not contribute to the time dependence at long times.  This is tantamount to the statement that in three dimensions broken symmetry is dynamically stable in the thermodynamic limit. On the other hand, in $d=1,2$ the coherence decays even in the thermodynamic limit due to the internal phonons. This is the dynamical counterpart of the Mermin Wagner theorem, showing how the internal (Gaussian) phase fluctuations dynamically destroy long range order. The decay can be traced back to an infrared divergence of the $q$ integral.

\subsubsection{Prethermalization in the fast split scheme}  The harmonic model we are considering now is the simplest example of an integrable model. The different $q$ modes are decoupled from each other and their occupations are conserved quantities. Hence the model does not in general relax to thermal equilibrium. It is interesting to ask what is the nature of the steady state the system reaches after it has dephased. Approach to such a non-thermal steady state is commonly referred to as prethermalization. The name stems from the expectation that there is a much longer time scale over which the dephased state would finally relax to a true thermal state. Thus the prethermalized state is a long lived quasi-steady state established in the system before the onset of true thermalization. In our case the final relaxation to the thermal state is governed by the enharmonic terms. This analysis is postponed to the next section.

Within the Gaussian approximation, the full information on the steady state is held in the two point correlation function (in the relative phase sector):
\bea
C(x,t)&=&\rho^2 \av{e^{i[\f(x,t)-\f(0,t)]}}=\rho^2 e^{-\av{\f(x,t)^2}+\av{\f(x,t)\f(0,t)}}\nn\\
&=& \rho^2 e^{-{1\over L^d}\sum_q\av{|\f_q(t)|^2}[1-\cos({\bf q}x)]}=e^{-f(x,t)}
\label{Ct}
\eea
This expression is directly analogous to the formula for the fringe amplitude (\ref{At}). Similarly the important contribution to the integral $f(x,t)$ comes from the second term in (\ref{At2}) leading to, for $d=1$, 
\be
f(x,t)\approx  { \pi S_d\over K^2}\rho ct\int_0^{\mu t} dz {\sin^2 z\over z^2} \left[1-\cos\left( z {x\over ct}\right)\right].
\ee
This integral exhibits completely different behavior depending on whether the point at  the position $x$ and time $t$ is inside or outside the "light-cone" emanating from the other point  which we have set to at $x=0$ at the time of the split $t=0$. For $ct< x$ the last cosine term is rapidly oscillating and averages to zero. In this case we have $f(t,x)\approx 2g(t)$ and therefore $C(x,t)\approx {\mathcal A}(t)^2$. This agrees with the expectation that for $t<x/c$ there was no time for information to propagate between the two points and therefore the correlation function can be decoupled into the independent expectation values ${\mathcal A}(t)$ at the two points.

On the other hand at long times $t\gg x/c$ we make a different change of variables using $\a=qx=z x/ct$ as an integration variable to obtain:
\be
f(x,t)\approx{ \pi S_d\over K^2}\rho \, x\int_0^{x/\xi_h} d\a  {\sin^2 (\a/2)\over \a^2} \left[ 1- \cos\left( \a {2ct\over x}\right) \right] 
\ee
Now the last cosine term is rapidly oscillating ($ct/x\gg 1$) and we have a time independent result $f(x,t)\approx (\pi/2K)^2 \rho x$.  This implies an exponential decay of the phase correlations in the steady state with a correlation length $\xi_\f= 4K^2/(\pi^2\rho)$. 

Such steady state behavior mimics the exponential decay of phase correlations in equilibrium at a finite temperature, which is characterized by the correlation length $\xi_\f^{\text{eq}}=cK/(\pi T)=\mu K^2/ (\pi\rho T)$. Thus we can assign an effective temperature $T_{eff}=\pi\mu/4$ to the prethermalized state by comparing the correlation length in it to the correlation length in thermal equilibrium. 

It is important to note that the effective temperature $T_{eff}$ is completely independent and, in fact, has nothing to do with the real temperature of the condensate prior to the split. The original, thermal degrees of freedom of that condensate are projected following the rapid split onto the symmetric density and phase modes of the split condensate. The state of the relative degrees of freedom, accessible to interference experiments, is a pure state fully determined by the splitting process.  
 Within the harmonic fluid theory the symmetric and anti-symmetric sectors are decoupled, and so within this approximation the temperature of the original condensate does not affect the dynamics.
 
The prethemalized state reached following a fast split of a one dimensional condensate has been seen in beautiful experiments by Vienna group using atom chips\cite{Gring2012,Smith2013}. These results are based on an earlier theoretical proposal by Kitagawa et. al. \cite{Kitagawa2011} showing that the prethermalized state can be fully characterized by the distribution of the amplitude of fringes seen in an interference experiment. These experiments confirm the applicability of the harmonic theory in the accessible time-scales.

\subsubsection{\em Finite split rate} 

A dynamic split done on a timescale $\tau_s<\mu^{-1}$ can be treated in complete analogy to our treatment of a single mode. For simplicity we separate the two stages of the dynamics. First we treat the splitting process in which the Josephson coupling is gradually turned off. The outcome of this process provides the initial state for calculating the dynamics at later times under the influence of the hamiltonian of the decoupled tubes. Here we give a brief summary of a detailed calculation along these lines found in \cite{Rafi-Thesis}.

To address the splitting dynamics we expand the Josephson coupling to quadratic order to  obtain a time dependent mass term for the relative phase mode
$
H_J \approx  \sum_q  \D(t) |\f_q|^2
$
We then assess for each mode separately whether the splitting on a time-scale $\tau_s$ is slow or fast compared with the characteristic frequency $\w_q=cq$ of this mode.  For modes at wave vectors $q>1/(c\tau_s)$ the split is effectively adiabatic. Therefore these modes end up in the quantum ground state of the final, fully split Hamiltonian. Modes at lower wave-vectors 
 develop adiabatically only to the point that their frequency is $\w_q\sim 1/\tau$ and they remain frozen for the rest of the splitting process.

\subsection{Beyond the Harmonic fluid description of one-dimensional phase diffusion}

The harmonic fluid  hamiltonian (\ref{LL}) is a fixed point of the renormalization group in a one dimensional system, which describes a stable quantum phase of matter known as a Luttinger liquid \cite{Haldane1981}. 
Hence, as long as the macroscopic (fully renormalized) values of $\kappa$ and $\rho_s$ are used,  the harmonic Hamiltonian provides an asymptotically exact description of the ground state and low energy excitations of the system. The neglected anharmonic terms affect the long distance correlation and low frequency response only through the renormalization of the quadratic coefficients $\kappa$ and $\rho_s$ from their bare values.

But in spite of being irrelevant for the linear response of the system {\em in the ground state}, the non linear terms can influence the long time dynamics of the interferometer, which starts in a state of finite energy density after the split. As discussed above, the splitting scheme lead to excitation of modes at all wave vectors, which then evolve independently within the quadratic theory (\ref{LL}). Only the non-linear terms can break this integrability through the coupling they induce between modes at different wavectors. Moreover, the non linear terms lead to coupling between the relative and total density fluctuations of the two split condensates. Recall that the relative fluctuations are produced in a pure state by the splitting scheme while the total density fluctuations had already existed in the original single condensate and presumed to be thermal. Thus the non-linear terms give rise, effectively, to coupling with a thermal bath.

\subsubsection{Self consistent phonon damping}
The long time limit of the evolution of the relative phase due to the equilibration with the thermal bath was considered by Burkov et. al in Ref. \cite{Burkov2007}. The central assumption in this approach is that the relative mode has nearly reached thermal equilibrium with the center of mass modes at a final temperature $T_f$. We note that the time scale it takes the system to reach this near equilibrium regime is set by different processes\cite{Arzamasovs2013}, on which I will comment later on.

The damped relative phase mode near thermal equilibrium
 can be described by phenomenological Langevin dynamics
\be
{\ddot\f}_k + c^2 k^2 \f_k+2\g_k{\dot\f}_k=2 \ \zeta_k
\label{eq:langevin}
\ee
Because the system is near thermal equilibrium the noise satisfies the fluctuation-dissipation theorem with $\av{\z_k(t)\z_{-k}(t')}=2(\mu/\rho) T_f\g_k\d(t-t')$. Eq. (\ref{eq:langevin}) is simply a model of damped phonons with $\g_k$ being the width of the phonon peaks in the structure factor. We expect that the final temperature $T_f$ will be somewhat higher than the initial temperature $T_{i}$ of the unsplit condensate because of the energy added to the system in the split process. For a sudden split the added energy density is $\sim \half \mu/\xi_h$, i.e. chemical potential per healing length. 

% The damping comes from the non linear terms in the Hydrodynamic description of a one-dimensional fluid. 
The kinematic conditions of linearly dispersing phonons in one dimension lead to divergence of the damping rate $\g_k^0$ calculated in to one loop order. % (diagram (a) in Fig \ref{fig:damping}). 
A finite result is obtained, however, if a damping rate is inserted into the phonon Green's function at the outset and then calculated self consistently.  Such a self consistent calculation, first done by Andreev \cite{Andreev1980} (see also \cite{Burkov2007}), gives a non-analytic dependence of $\g_k$ on $k$:
\be
\g_k\approx {1\over 2\pi} \sqrt{\mu T {\mathcal K}} \left({ c |k|/ \mu}\right)^{3/2}
\label{gamma}
\ee 
This expression is valid for wave-vectors $k$ in the linearly dispersing regime, i.e. $ck\ll \mu$ and for weak interactions so that the Luttinger parameter ${\mathcal K}\gg 1$.

The solution for $\f_k(t)$ can be formally written in terms of the Green's function of the damped Harmonic oscillator
\be
\f_k(t)=\int_0^t dt_1 {\cG}(k,t-t_1)\z_k(t_1)
\ee
Where 
\be
\cG(t)={1\over 2\pi}\int d\w e^{i\w t}{1\over \w^2 -\e_k+i\w\g_k}=e^{-\g_k t/2} {\sin\left(\sqrt{\e_k^2-(\g_k/2)^2}\,\,t\right)\over \sqrt{\e_k^2-(\g_k/2)^2}}
%\approx e^{-\g_k t/2}{\sin (ckt)\over c k}
\ee
%In the last approximation we assumed $\g_k\ll c k$. We'll discuss below when this is valid.
Now by plugging in the $\d$-correlated noise near equilibrium we can compute the relative phase fluctuations
\bea
\av{\f_k(t)^2}&=&{2 T\mu\over \rho}\g_k  \int_0^t dt_1 e^{-\g_k (t-t_1)}{\sin^2\left(\sqrt{\e_k^2-(\g_k/2)^2}\,\, (t-t_1)\right)\over\e_k^2-(\g_k/2)^2}\nn\\
&=& {T\mu\over \rho \,\e_k^2} \left[1-e^{-\g_k t}\left(1+{\g_k\over 2\e_k}\sin(2\e_k t)\right)\right]+O\left[\left(\g_k/\e_k\right)^2\right]
\eea
Using the result $\g_k=\eta \e_k^{3/2}$ (with $\eta=\sqrt{T}/(2\pi\mu)$) from Eq. (\ref{gamma}) we find
\bea
\av{\f(x,t)^2}&=&\int_0^\mu d \e \nu_{1d}(\e)\av{\f_\e(t)^2}\approx {T\mu\over \pi\rho c}\int_0^\mu d\e {1\over\e^2}\left(1-e^{-\eta\e^{3/2} t}\right)\nn\\
&\approx&{T\over \cK}\left[\int_0^{(\eta t)^{-2/3}}{\eta t\over \sqrt{\e}} d\e +\int_{(\eta t)^{-2/3}}^\infty {d\e\over \e^2}\right]= {3T\over \cK}(\eta t)^{2/3},
\eea
where we have used the relation valid for weak interaction $\pi\rho c/\mu=\pi\sqrt{\kappa\rho_s}=\cK$. This result implies decay of phase coherence as
\be
{\mathcal{A}}(t)={\mathcal{A}}_0 e^{-(t/t_T)^{2/3}},\label{stretched}
\ee
with $t_T=(2\pi\mu/T^2) \cK$.  $\cK=(\pi/2)\sqrt{\rho/gm}$ is the Luttinger parameter in the weak coupling limit. The non analytic time dependence of the phase coherence stems directly from the non analyticity of the momentum dependent damping rate. 

It is interesting to note that the decoherence driven by thermal fluctuations (\ref{stretched}) , being a stretched exponential, is slower than the dephasing driven by quantum dynamics, which, we have seen, depends exponentially on time. The physical reason behind this somewhat counter intuitive result is that the thermally excited phonons provide a damping mechanism that slows down the unitary phase evolution.

\subsubsection{Kardar-Parisi-Zhang scaling}
The result (\ref{stretched}) relies on the scaling of the phonon damping rate $\g_k\propto |k|^{3/2}$ derived using a self consistent diagrammatic approach\cite{Andreev1980}. Because the approach is not rigorously controlled it would be good to understand the scaling in a different way. 

An alternative viewpoint was provided in Ref. \cite{Kulkarni2013}. In this paper  Kulkarni and Lamacraft  suggested a possible connection between the one dimensional condensate dynamics at finite temperature and the Kardar-Parisi-Zhang (KPZ) equation commonly used to describe randomly growing interfaces. If indeed the dynamics belonged to the same universality class it would immediately imply  as a consequence the anomalous dynamical scaling obtained in the self consistent calculation.

Following Ref. \cite{Kulkarni2013} we illustrate the relation to KPZ dynamics starting from the hydrodynamic equations of the condensate
\bea
\partial_t\rho+\partial_x(v\rho)=0\nn\\
\partial_t v +v\partial_x v+(g/m)\partial_x\rho
\eea
These are the Euler and continuity equations with $v(x,t)$ the velocity field $\rho(x,t)$ the density and $g$ the contact interaction. Linearizing these equations with respect to the velocity field and the deviation $n(x,t)=\rho(x,t)-\rho_0$ from the average density, leads to a wave-equation describing phonons of the system moving to the right or left at the speed of sound $\pm c_0=\sqrt{\rho_0 g/m}$. Note that beyond the linear approximation, the local speed of sound of the fluid depends on the local density $c_\rho=\sqrt{\rho g/m}= c_0\sqrt{\rho/\rho_0}$. Now define the "chiral velocities" as the fluid velocities measured with respect to the local sound velocity of left and right moving waves: $v_\pm=v\pm c_\rho$. In terms of the chiral velocities the equations of motion can be written as:
\be
v_\pm+v_\pm\partial_x v_\pm = {1\over 3}(\partial_t+v_\pm\partial_x)v_\mp
\ee
Kulkarni and Lamacraft pointed to the similarity between these equations and two copies of the, so called,  noisy Burgers equation. This would be precisely the situation if we could replace the right hand sides of the equations by fields $\zeta_\pm(x,t)$ providing Gaussian white noise to the left hand sides. As it stands however, we can think of the left moving modes $v_-$, randomly occupied with the Bose distribution at temperature $T$,  as  a random force field acting on the right movers and similarly of the right movers as providing the noise to the dynamics of the left moving modes. With this approximation the coupled equations reduce to two effectively independent noisy burgers equations.

The connection between the noisy burgers  equation and the KPZ equation is well known. It is easily demonstrated by representing the chiral velocities as spatial derivatives of chiral phases $\f_\pm=(1/m)\partial_x\f_\pm$, just as the actual velocity is related to the condensate phase $v=(1/m)\partial_x\f$. Plugging these relations into the Burgers equations and integrating over $x$ gives  $\partial_t\f_\pm+(1/m)(\partial_x\f_\pm)^2=\zeta_\pm$. There is also a dissipative term $D\partial_x^2\f_\pm$ generally present in the KPZ equation. It is absent here because we started from non dissipative Hamiltonian dynamics. Dissipation however would be generated upon coarse graining because of the coupling between high and low wave vector modes induced by the non linearity. Hence in the long wavelength limit we expect the dynamics to be governed by the KPZ equation:
\be
\partial_t\f_\pm=D\partial_x^2\f_\pm+{\lambda\over 2}(\partial_x\f_\pm)^2+\zeta_\pm
\ee
The height field  (here the chiral phases $\f_\pm$) is known to obey the dynamic scaling $\av{(\f_\pm(x,t)-\f_\pm(x,0))^2}\sim t^{2/3}$. Hence also the phase field $\f=(\f_+ +\f_-)/2$ obeys the same scaling, which leads to the stretched exponential decay of the coherence (\ref{stretched}).

\subsection{Discussion and experimental situation}

Above we have arrived at two seemingly conflicting results. The harmonic fluid theory of quantum phase diffusion predicts exponential decay (\ref{At}) of phase coherence in a long one dimensional system, while the hydrodynamic theory of thermal relaxation predicts a stretched exponential decay (\ref{stretched}). Which will be seen in experiments?
Direct comparison between the thermal and quantum time scales $t_T$ and $t_Q$ suggests that as long as the temperature of the relevant bath is smaller than the chemical potential then the quantum phase diffusion should dominate. By the time the thermal relaxation sets in the phase system is then essentially dephased.  

Moreover, the calculation of the thermal phase relaxation time $t_T$ summarized in the previous section assumes that the process can be described in terms of an effective hydrodynamic theory of a nearly thermalized system. However the system in question is initially very far from equilibrium and, being a one-dimensional system of bosons with short-range interactions, it is very nearly described by an integrable model (Lieb-Liniger model). Clearly the equilibration rate should be proportional to some measure of integrability breaking, which is absent in the expression for $t_T$.  Breaking of integrability in an elongated Bose gas with a tight transverse confinement  stems from effective three particle interactions mediated by virtual occupation of transverse excited states\cite{Mazets2008}. Therefore the integrability breaking should be controlled by the ratio $\mu/\omega_\perp$ of the interaction energy (chemical potential) to the transverse confinement frequency.

Recently, Arzamasovs et. al. \cite{Arzamasovs2013} computed the thermalization rate of a nearly integrable, weakly interacting one dimensional Bose gas, by expanding around the integrable Lieb-Liniger model. Their result can be expressed as 
 \be
 {1\over\tau_R}\sim \mu \left({\mu\over \w_\perp}\right)^2 \left({T\over \mu}\right)^7 {1\over K^3}
 \ee
 From this it is clear that in the relevant experimental regime of $K\sim 10$ and temperature range $T\lesssim \mu$ the system does not equilibrate. The dynamics should be well described by quantum phase diffusion leading to a prethermalized state.
 
The first experiments, which measured the dephasing dynamics seemed to contradict this conclusion, reporting initially a streched exponential decay of the phase coherence\cite{Hofferberth2007} with the exponent $2/3$ expected from the thermal dephasing process. However, improved experiments of the same group done with a very similar apparatus found excellent agreement with prethermalization driven by the harmonic fluid dynamics\cite{Gring2012}  (See Ref. [\onlinecite{Smith2013}] for a detailed reanalysis explaining the problems in the original experiments, which led to the missnterpretation).  

%Another interesting question is whether the anomalous diffusion obtained in the classical analysis can be realized in  another  experimentally feasible situation.  One possibility for a slightly different setup  is to start with decoupled Bose liquids already at thermal equillibrium in temperature $T>\mu$. Then at time $t=0$ couple the two liquids to each other weakly for a short time and observe the dephasing after they are  decouple again. In this case the  phase evolution would  satisfy the requirement of the classical dynamics, near thermalization at a single high temperature, from the outset. Phrasing the problem in this way also reveals that the anomalous diffusion should appear as a specific linear response function of a pair of decoupled one dimensional liquids at high temperature. 

\section{Dynamics of ultra-cold bosons in optical lattices}
\label{sec:optlat}
The natural regime of ultra-cold Bose gases is that of weak interactions. In this regime, the gas is well described by the dynamics of a non-fluctuating classical field through Gross-Pitaevskii equation. Even in low dimensions, where fluctuation effects are inevitably important at long wavelengths they are  hard to observe in small traps. One  approach to enhance fluctuation effects and potentially reach novel quantum phases and dynamics is to load the atoms into optical lattice potentials generated by standing waves of laser light. 

The lattice has two important effects on the quantum gas. One effect is to increase the effective mass of the atoms and thereby quench the kinetic energy with respect to the interactions. The second important effect is to  break Galilean invariance.  This liberates the quantum gas from strong constraints, thus opening the way to realizing alternative quantum phases and new modes of dynamics.  In this regard the observation of a quantum phase transition of bosons in an optical lattice from a superfluid to a Mott insulator \cite{Greiner2002a} has been an important milestone in the study of ultra cold atomic systems. This experiment has lead to intense theoretical and experimental work to understand the dynamics of strongly correlated lattice bosons, and later also of fermions.  

The superfluid to insulator transition of lattice bosons was first discussed by Giamarchi and Schulz\cite{Giamarchi1988} in the context of one dimensional systems and soon after by Fisher et. al. in Ref. \cite{Fisher1992} for higher dimensional systems. Investigation of this physics with ultra-cold atoms in optical lattices was first proposed by Jaksch et. al. \cite{Jaksch1998}. In most of this section I  review the basic theory of the superfluid to Mott insulator transition and describe recent theoretical and experimental work, which advanced our understanding of the universal dynamical response near the critical point. Discussion of far from equilibrium dynamics in this regime is postponed to sections \ref{sec:dynamical-instab} and \ref{sec:quench}.

\subsection{Mott transition in two and three dimensional lattices}

\begin{figure}[t]
\includegraphics[width=0.7\linewidth]{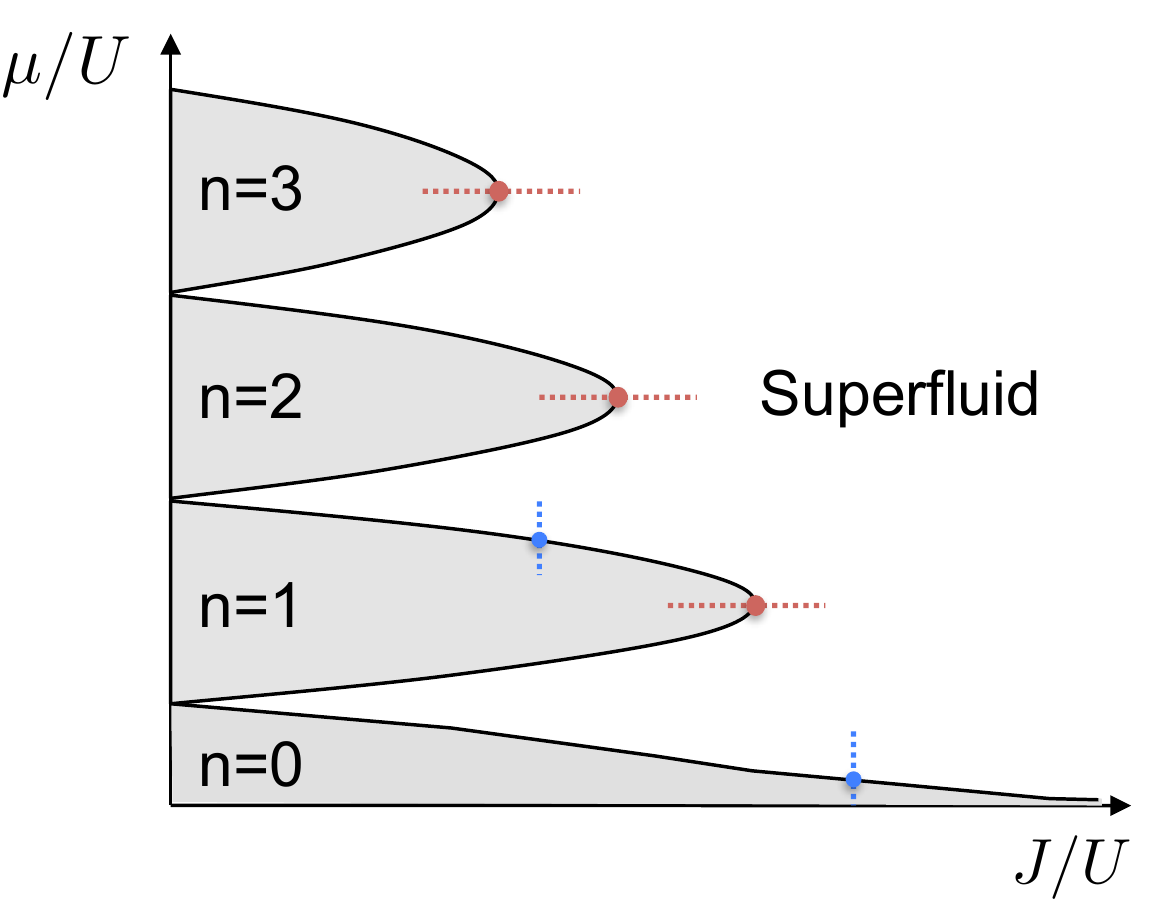}
\caption{Zero temperature phase diagram of the Bose Hubbard model. The grey areas are the incompressible Mott phases with integer filling $n$. The horizontal dashed lines (red) correspond to phase transitions occurring at constant commensurate filling; they are described by the relativistic critical theory (\ref{Seff}). In contrast, the transitions tuned along the vertical dashed lines (blue) are described by the Gross-Pitaefskii action (\ref{SGP}), which also describes the weak interaction regime. The latter can be thought as being near a transition tuned by chemical potential from a vacuum state to the weakly interacting BEC.}
\label{fig:pd}
\end{figure}

Superfluidity in two and three dimensions is perturbatively stable to introducing a weak lattice potential regardless of the interaction strength and independent of the commensurability of the potential with the particle density. To allow the establishment of an insulating phase it is therefore important to work with a strong lattice potential with wide gaps between the lowest and the second Bloch bands. In this case, if both the temperature and the interaction strength are smaller than the band gap, we can use the well known Bose Hubbard model as a microscopic description of bosons in the lowest Bloch band
\be
H=-J\sum_\av{ij} (b\yd_ib\nd_j+H.c) + \half U\sum_i n_i(n_i-1)-\mu\sum_i n_i
\label{BHM}
\ee
The chemical potential $\mu$ is used to control the average lattice filling $n$. For simplicity I will not consider here the effects due to the global trap potential. The values of model parameters expressed in terms of the microscopic couplings are given by:
\bea
J&\approx&\sqrt{16/\pi}\, E_r \left(V_0/ Er\right)^{3/4} e^{-2\sqrt{V_0/E_r}} \\
U&\approx& \sqrt{8/\pi}\, ka\,E_r (V_0/E_r)^{3/4}
\eea
where $k$ is the wave vector of the lattice, $a$ the $s$-wave scattering length corresponding to the two body interaction between atoms and $E_r= \hbar^2 k^2/2m$ is the recoil energy.
For details on the effect of the trap and derivation of the model parameters I refer the reader to Refs. \cite{Jaksch1998,Bloch2008}.
Where the parameter values are

The fact that there must be a phase transition in the ground state of the model as a function of the relative strength of interaction at integer lattice filling can be established by considering the two extreme limits. First consider the limit of weak interactions. For vanishing interaction strength (and $T=0$ all the bosons necessarily condense at the bottom of the tight binding band. Weak interactions compared to the band-width, i.e. $Un<<4zJ$ cannot excite particles far from the band bottom. Therefore in this regime the system can be described by an effective continuum action (Gross-Pitaevskii)
\be
S_{GP}=\int d^d x dt \left\{i\psi^\star \partial_t\psi-{1\over 2m_*}|\nabla \psi|^2+\mu|\psi |^2-u|\psi|^4\right\}
\label{SGP}
\ee
where $m_*$ is the effective mass in the lowest Bloch band, $u=U a^d$ and $a$ is the lattice constant. Hence the physics in this regime is identical to that of a weakly interacting superfluid in the continuum. The effect of the lattice is only to renormalize parameters.

Consider now the opposite regime of strong interactions or small hopping $J$. In particular let us start with the extreme limit of decoupled sites (i.e.  vanishing hopping $J=0$). At integer lattice filling $n$ the system of decoupled has a unique ground state where each site is occupied by exactly $n$ bosons. 
 The elementary excitations above this ground state are gapped (with a gap $U$), and consist of particles (sites with $n+1$ bosons) and holes (sites with $n-1$ bosons). The gap makes this state robust against introducing a small hopping matrix element $J$. The elementary excitations gain a dispersion with band width $W\sim 2 d n J$. However these excitations remain gapped and retain their  $U(1)$ charge quantum number as long as $W<U$. 

The gap to charge excitations implies that the zero temperature Mott phase is incompressible. A chemical potential couples with an opposite sign to particle and to hole excitations, decreasing the gap of one species while increasing it for the other. But as long as both excitations remain gapped, the lattice filling cannot change with chemical potential. Thus the compressibility $\kappa=\partial n/ \partial \mu=0$. 

From the above consideration one can infer that the Mott phases are established as lobes  in the space of chemical potential $\mu$ and hopping $J$, characterized by a constant integer filling (see Fig. \ref{fig:pd}). 
Upon changing the chemical potential the upper (or lower) boundary of the phase is reached when either the particle (or hole) excitations condense. The density begins to vary continuously upon further increase (or decrease) of the chemical potential beyond the boundary with the compressible phase. 
The critical theory, which describes this transition, tuned by the chemical potential, is just the field theory (\ref{SGP}) where the Bose field here describes the low energy particle or hole excitations at the upper or lower phase boundary respectively.  

The transition can also be tuned by varying the tunneling strength (or the interaction) at a fixed integer density. In this case the particle and hole excitations must condense simultaneously, which enforces particle hole symmetry at the critical point. This in turn implies emergent relativistic (i.e. Lorentz invariant) theory in the vicinity of the quantum phase transition, which in the gapped phase indeed must describe particle and anti-particle excitations of equal mass. 

More directly, emergence of Lorentz invariance at the tip of the Mott lobe can be established as follows (see e. g. \cite{Sachdev-Book}). We write down a more general theory than (\ref{SGP}) 
\be
S=\int d^d x dt \left\{iK'\psi^\star \partial_t\psi+K |\partial_t\psi|^2-c^2|\nabla \psi|^2+r(\mu)|\psi |^2-u|\psi|^4\right\}
\label{Sgen}
\ee
In general if both $K'$ and $K$ are non vanishing we can neglect $K$ at the critical point as it is it is irrelevant by simple power-counting compared to the $K'$ term. However we now show that $K'$ must vanish at the tip of the Mott lobe leaving us with a relativistic theory. By requiring invariance under the uniform Gauge transformation (i.e. redefinition of the energy):
\bea
\psi&\to& \psi e^{-i\phi}\nn\\
\mu&\to& \mu+\partial_t\phi
\eea
we find that $K'=\partial r/\partial\mu$. Thus at the tips of the Mott lobes where $r(\mu)$ reaches a minimum as a function of $\mu$, the coefficient $K'$ must vanish.  Although the special transition at the tip may seem highly fine tuned it is actually realized naturally in a uniform system with a fixed (integer) average filling.  

\subsection{The Higgs resonance near the superfluid to Mott insulator transition}

\begin{figure}[t]
\includegraphics[width=0.7\linewidth]{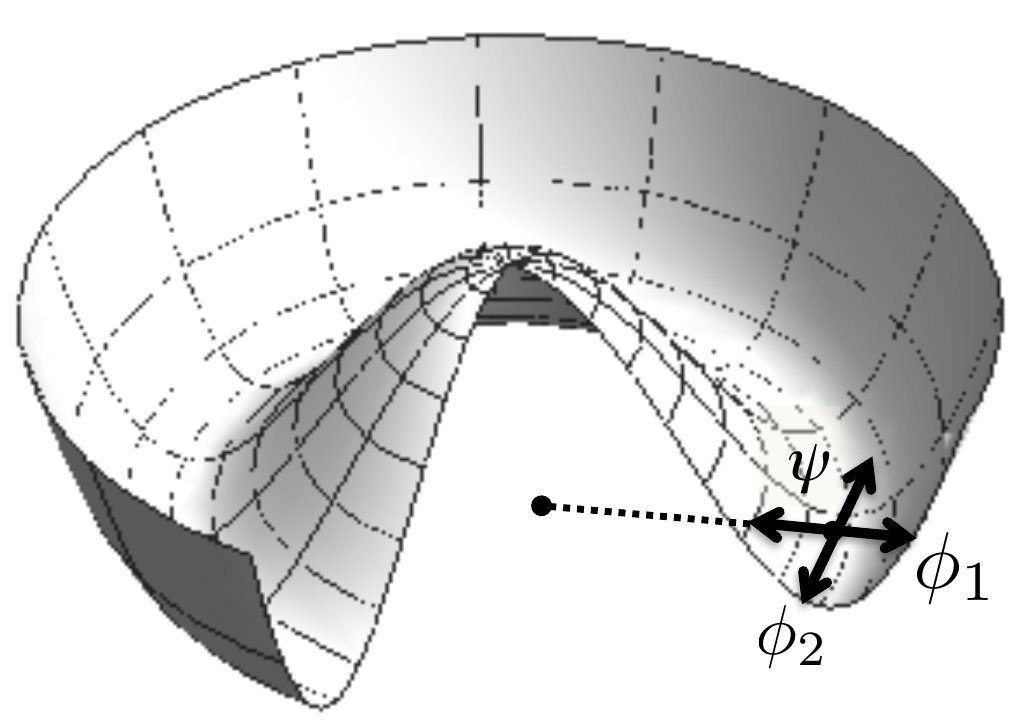}
\caption{Mexican hat potential that describes the dynamics of the order parameter in the superfluid phase. The longitudinal and transverse fluctuations $\phi_1$ and $\phi_2$ are two independent modes in the effective relativistic description of the critical point (\ref{Seff}), the gapless Goldstone mode and the Gapped Higgs excitation. In the Gross-Pitaevskii action (\ref{SGP}) by contrast, the same two fluctuations form a canonically conjugate pair and therefore make up a single mode, the Goldstone mode of the superfluid.}
\label{fig:mex}
\end{figure}

We now discuss how the existence of a quantum phase transition from a superfluid to a Mott insulator impacts the dynamics of bosons in an optical lattice. As discussed above, near the critical point, where the diverging correlation length far exceeds the lattice spacing, the dynamics is described by an effective relativistic field theory
\be
S_{eff}=\int d^d x dt \left\{ |\partial_t\psi|^2-c^2|\nabla \psi|^2-r|\psi |^2-\half u|\psi|^4\right\}
\label{Seff}
\ee
where $c$ is the sound velocity. One of the most interesting consequences of this emergent critical theory is the appearance of a new gapped excitation analogous to the Higgs resonance in particle physics. %Near the critical point the excitation spectrum and decay modes associated with this resonance become universal.

The collective mode structure in the superfluid phase near the Mott transition can be understood by considering the classical oscillations of the order parameter field about its equilibrium broken symmetry state. Note that both the critical action (\ref{Seff}) valid near the superfluid to Mott transition at integer filling and the Gallilean invariant action (\ref{SGP}) of the weakly interacting condensate describe the motion of the order parameter field on a mexican hat potential as illustrated in Fig. \ref{fig:mex}. Naively it looks like in both cases there are two modes, a soft Goldstone mode corresponding to fluctuations of the order parameter along the degenerate minimum of the potential and a massive fluctuation of the order parameter amplitude. This static picture is misleading. It hides a crucial difference in the dynamics which stems from the different kinetic terms. 

In the Gallilean invariant (Gross-Pitaevskii) dynamics the amplitude of the order parameter is proportional to the particle density, i.e. $\psi\sim \sqrt{\rho}e^{i\f}$. Plugging this into the kinetic term of (\ref{SGP}) we get $i\psi^*\dot\psi=\rho\dot\f$. Hence the amplitude and phase mode are in this case a canonical conjugate pair which together make up only one excitation mode, the gapless Goldstone mode (or phonon) of the superfluid. 
On the other hand, in the relativistic theory the phase and amplitude fluctuations are not canonical conjugates. This is easy to see by expanding in small fluctuations around a classical symmetry breaking solution. We take $\psi=\psi_0+\phi_1+i\phi_2$, where $\psi_0=\sqrt{-r/u}$ is a real classical saddle point solution  and the real fields $\phi_1$ and $\phi_2$ represent fluctuations in the massive and soft directions of the potential. A quadratic expansion of the action (\ref{Seff}) in these fluctuation fields gives
\be
S_0= \int d^d x dt\, \left(\dot\phi_1^2-c^2(\nabla\phi_1)^2+2 r \phi_1^2\label{Lr}\right) + \left(\dot\phi_2^2- c^2(\nabla\phi_2)^2\right) .
\ee
Hence in the relativistic theory (\ref{Lr}) $\phi_1$ and $\phi_2$ represent two independent harmonic modes: a massless Goldstone mode $\phi_2$ and a massive amplitude mode $\phi_1$ with a mass that vanishes at the critical point.
The decoupling between phase and amplitude (at least in the quadratic part of the theory) is possible because near the Mott transition {\em the order parameter amplitude is unrelated to the particle density}. Indeed the amplitude vanishes even as we tune the system to the Mott transition at constant particle density. 

The order-parameter amplitude mode is closely analogous to the famous Higgs particle in the standard model of particle physics. In the standard model the order parameter is charged under a local Gauge symmetry, therefore the Goldstone mode is replaced, through the Anderson-Higgs mechanism by a massive Gauge boson. What is left of the order parameter dynamics is then only the amplitude fluctuations, which make up the Higgs Boson. In our case because the condensate is Gauge-neutral there is no "Higgs-mechanism", Goldstone modes remain and coexist with the Higgs amplitude mode.  

The gapless Goldstone modes provide a decay channel, for example through the coupling $u\phi_1\phi_2^2$ obtained upon expanding the original action to cubic order, which can lead to broadening of the Higgs resonance. This observation naturally raises the question if the amplitude mode is observable as a sharp resonance near the critical point. The answer to this question, which turns out to be interesting and subtle was formulated only recently in Refs.\cite{Podolsky2011,Podolsky2012,Gazit2013}. The theory was confirmed when the Higgs was first observed in a system of ultra-cold atoms near the superfluid to Mott insulator transition\cite{Endres2012}. Below I review the theoretical understanding and the measurement scheme used to detect and characterize the Higgs mode.

As a first attempt to address the questions of decay of the Higgs mode and asses if it is visible as a peak in some linear response measurement we may be tempted to 
compute the self energy of the longitudinal mode $\phi_1$. To lowest order in the cubic coupling $u\phi_1\phi_2^2$ the Matsubara self energy is given by
%\be
%\Sigma(q)=u^2\int {d^{d+1}k\over (2\pi)^d} {1\over k^2(k+q)^2\sim\left\{\begin{array}{ll}
%{u^2\over \log q} & d=3\\
%{1\over q} & d=2 \end{array} \right.
%\ee
\be
\Sigma_1(q)=u^2\int {d^{d+1}k\over (2\pi)^d} {1\over k^2(k+q)^2}\sim \begin{cases}
{u^2\over \log q} & d=3\\
{1\over q} & d=2 \end{cases} 
\label{Sigmav}
\ee
In real frequency this implies $\mbox{Im}\Sigma_1(\w,{\bf q}=0)\sim 1/|\w|$ in two dimensions. 
One may worry that the low frequency divergence in the above self energy would wash away or completely mask any peaked response associated with the Higgs mode. But this conclusion is incorrect. As I explain below, the above self energy does not reflect the intrinsic decay of the amplitude mode.

To clarify whether a particular mode would show up as a resonance in a linear response measurement it is important to first specify what 
is the perturbation the system is responding to. A common probe in some systems with broken symmetry is one that couples directly to the order parameter. We can write a perturbation of this type as $H_{vex}=\vec{h}(t)\cdot\vec{\psi}$, where $\vec\psi$ is a vector (e.g. magnetic) order parameter. In magnetic systems, Neutron scattering acts as a vector probe of this type. In the case of a superfluid we can think of the complex order parameter $\psi=\psi_1+i\psi_2$ as a two component vector $\vec\psi=(\psi_1,\psi_2)$. The component of the external vector field acting parallel to the ordering direction $\vec{\psi}_0$ (longitudinal component) couples to the amplitude fluctuation $\s$. Thus the self energy (\ref{Sigmav}), with infra-red divergent spectrum, is related to the linear response of the system to a longitudinal vector probe, i.e. $\mbox{Im}\Sigma_1(\w)\sim \chi_{||}''(\w)$. But the vector perturbation is not a natural one to apply in ultra-cold atomic systems because such a perturbation has to violate charge conservation. 

A simple and direct scheme to measure dynamical response of ultra-cold atoms involves periodic modulation of the optical lattice potential\cite{Schori2004}. For atoms in the lowest Bloch band, described by the Hubbard model (\ref{BHM}), the lattice modulation translates to a modulation of the hopping amplitude $J$. Since the lattice modulation does not break the $U(1)$ symmetry this probe can only couple to scalar terms in the critical action (\ref{Seff}). In particular because the lattice strength is the microscopic parameter used to tune the transition, its modulation translates to a modulation of the tuning parameter $r$, i.e. the perturbation Hamiltonian is $H_s=\lambda(t)|\psi|^2$. 

In order to describe the response to the scalar probe it is convenient to use the polar representation of the order parameter $\psi=(\psi_0+\s)e^{i\t}$. At quadratic order the amplitude and phase fluctuations $\sigma$ and  $\theta$ play the same role as the longitudinal and transverse fluctuations $\phi_1$ and $\phi_2$ respectively. The scalar probe $\lambda$ couples linearly to $\sigma$ in the same way as the longitudinal probe couples to $\phi_1$. However the cubic coupling of the amplitude mode to the phase fluctuations 
is of the form $\sigma\partial_\mu\t\partial^\mu\t$, i.e. the amplitude fluctuations couples to gradients of the phase and not directly to the phase. Indeed this must be the case by the $U(1)$ symmetry. The self energy of the scalar amplitude fluctuation is then given by a similar loop diagram as that of the longitudinal fluctuation (\ref{Sigmav}), but the additional gradients now cancel the denominators to give an infrared convergent result $\Sigma_s(\w)\sim \w^3$. 

It is instructive to reconsider the longitudinal susceptibility in the polar representation as well. As before let us assume, without loss of generality, that the broken symmetry order parameter is real. The longitudinal fluctuation of the order parameter can be written directly in the polar representation
\be
\phi_1=\mbox{Re}\psi-\psi_0= (\psi_0+\s)\cos\t-\psi_0= \s+\half\psi_0 \t^2 + O(\s\t^2)
\ee
From this expression it is clear that the longitudinal fluctuation is not a pure amplitude fluctuation $\sigma$; it is contaminated by pairs of Goldstone modes through the $\t^2$ term. A longitudinal probe therefore directly excites this gapless continuum leading to the infrared divergent response we have obtained above. Compared to the longitudinal probe, the more physical scalar probe also gives a cleaner signature of the Higgs amplitude mode \cite{Podolsky2011}.

Having shown that the response to a scalar probe is infra-red convergent, I now briefly review what is known about the full line-shape and how the Higgs resonance manifests in it. Of particular interest is the question whether a peak associated to the Higgs resonance appears in the scaling limit. We broaden our discussion a bit and consider the response to the scalar probe on both sides of the critical point. In the Mott side there is only a gapped excitation at $q=0$, which therefore cannot decay and corresponds to a real pole at $\D\equiv (g_c-g)^\nu $. We shall now use $\D$ rather than $|g-g_c|$ to parameterize the deviation from the critical point on both sides of the transition. 
\begin{figure}[t]
\includegraphics[width=1\linewidth]{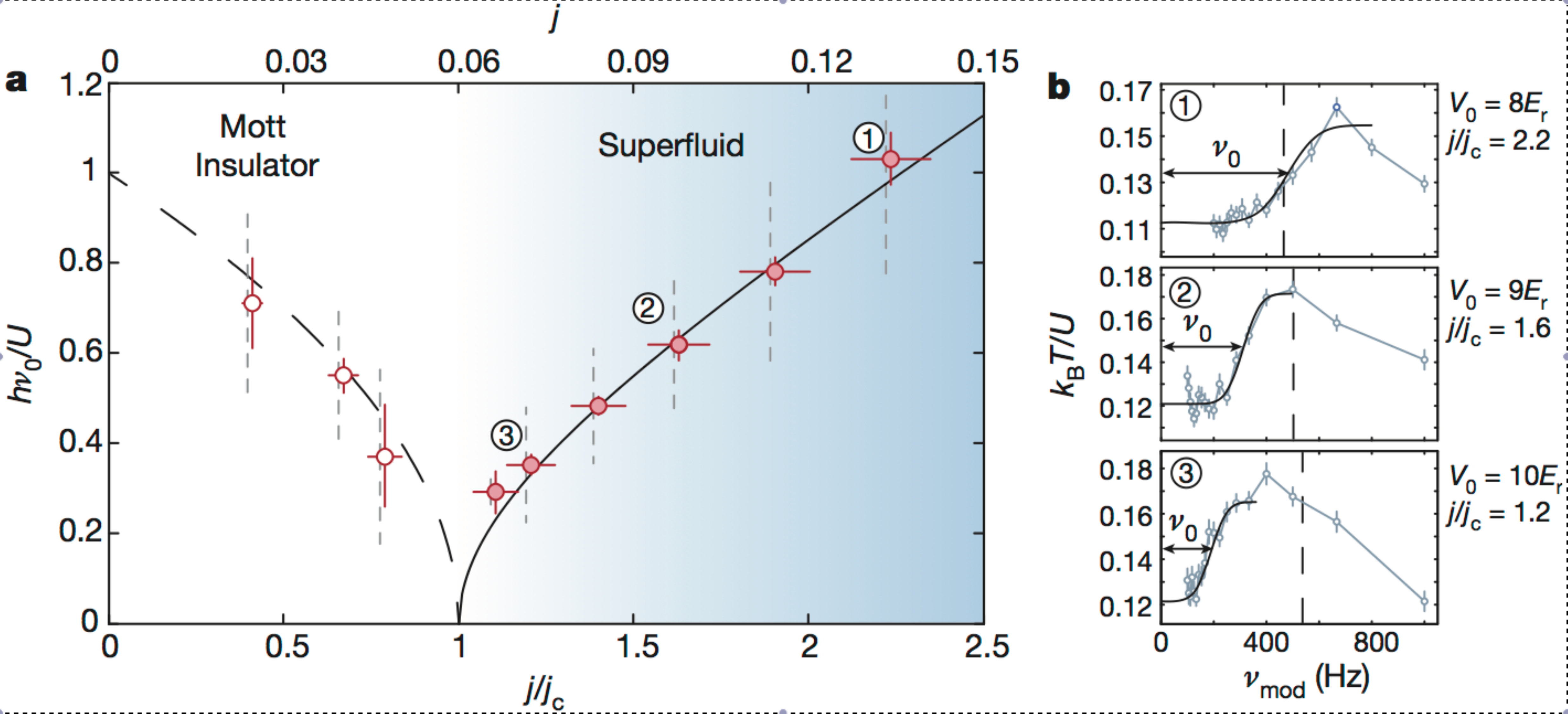}
\caption{(a) Energy of the Higgs excitation in the superfluid phase  ($j/j_c>1$) and of the gapped particle-hole excitation in the Mott insulator  ($j/j_c<1$) measured in Ref. \cite{Endres2012} shows softening of the modes on approaching the critical point. (b) Examples of the measured spectra. The Higgs mode is seen as a sharp leading edge rather than a peak because of in homogenous broadening in the harmonic trap. Figure 
reproduced fro Ref. \cite{Endres2012}.}
\label{fig:HiggsExp}
\end{figure}

The pertinent question concerning the response in the (ordered) superfluid side of the transition is whether a peak appears also on this side in the scaling limit. If it does then the frequency at the peak must also vanish as $|g-g_c|^\nu$ upon approaching the critical point. Equivalently, the scalar susceptibility of the superfluid is expected to follow a scaling form $\chi(\w,\D)\sim \D^{D-2/\nu}\Phi(\w/\D)$ with $\Phi(x)$ a universal scaling function with a peak at $x=x_p\sim 1$ (see \cite{Podolsky2012}). Theoretical analyses using 
Recent numerical results\cite{Pollet2012,Gazit2013,Rancon2014} as well a theoretical analysis\cite{Podolsky2012,Katan2015} indeed find such a a scaling form. It is important to note, however,  that neither method is fully controlled.  In particular,
the quantum Monte-Carlo simulations  performed for this system can give a Matsubara response function, while the analytic continuation to real frequencies requires uncontrolled approximations or otherwise is an exponentially hard problem. Therefore an experimental test of the theoretical and numerical predictions is needed.

An experiment aimed at testing some of the above predictions was reported in Ref. \cite{Endres2012} using bosons in a two dimensional optical lattice. Here, the heating rate 
$\dot E\sim \w\chi_s''(\w)$ due to weak lattice modulations was measured to high accuracy. Typical response spectra and the characteristic mode frequency obtained from them are reproduced here in Fig. \ref{fig:HiggsExp}. 
Note that these spectra exhibit a sharp edge followed by a continuum rather than a peak. This behavior is associated with the inhomogeneity of the trap. The leading edge stems from the response in the middle of the trap, which is closest to the critical point. 
The scaling of the leading edge upon approaching the transition from either side of it agrees quantitatively with the numerical predictions. In particular the ratio of excitation frequency in the superfluid to that in the Mott side is found to be $\w_p/\D\approx 2$, consistent with state of the art numerics \cite{Gazit2013,Rancon2014} and markedly different from the mean field prediction $\w_p/\D=\sqrt{2}$. The conclusion from the theoretical analyses and the experiment is that the Higgs mode in two spatial dimensions is visible and the spectral peak associated with it survives in the scaling limit. 

\subsection{Dynamical instability and decay of super flow in an optical lattice}
\label{sec:dynamical-instab}
Bosons tend to condense and become superlfuids at low temperature. In the last section we have seen that bosons in an optical lattice can lose their superfluid properties at zero temperature through phase quantum transition into the Mott insulating state. I have discussed how this transition manifests in the dynamical linear response of the system at equilibrium. In this section I will discuss the breakdown of superfluidity that occurs due to flow of a super current in the lattice. This is a highly non-equilibrium route for breakdown of superfluidity. In particular I will focis on the interplay between the Mott transition and the critical current in the superfluid phase.

If a superfluid in a Galilean invariant system is carrying a current then this current can be removed by moving to the Galilean frame of the  superfluid. Hence without disorder or a lattice potential to break translational symmetry, a state with any magnitude of current is necessarily stable . Here I will focus on the effect of the lattice. In particular I will ask what is the critical current above which superfluidity breaks down in the different regimes and under what conditions this is indeed the critical current is sharply defined. 

\subsubsection{Mean field critical current}
Let me start with a mean field description of super-currents. Consider a condensate described by the order parameter 
$\psi_i=\sqrt{n_s}e^{i\t_i}$, where $i$ is a lattice site. 
A state with uniform current, say in the $\hat{x}$ direction, is described by a uniform phase twist, $\t_i= p x_i$. 
The super-current density is then 
\be
I={1\over 2 i m_*}\left(\psi^*_{i+{\hat x}}\psi_i-\psi^*_i\psi_{i+\hat{x}}\right)= {n_s\over m_*} \sin(p),
\ee
where $m_*$ is the effective mass of a particle on the lattice, and $p\equiv \t_{i+\hat{x}}-\t_i$ is the phase change across a link in the $\hat{x}$ direction.  

By looking into this expression we can anticipate many of the results that I will discuss in more detail below. In a weakly interacting condensate at $T=0$ the superfluid density is just the total atom density $n_s=n$, essentially independent of the phase twist $p$.
Then the maximal current that the system can carry is $I_c=n/m_*$ and it occurs when the phase twists by $\pi/2$ over a lattice constant ($p_c=\pi/2$).  This result can be understood by considering the single particle dispersion $\e(p)=-m_*^{-1} \cos(p)$, where $p$ is the lattice momentum. A condensate with phase twist $p$ is obtained by condensing the bosons into the state with lattice momentum $p$. When $p$ exceeds $\pi/2$ the local effective mass at that momentum $m(p)^{-1}=\partial^2\epsilon/\partial p^2$ turns negative. The situation near the Mott transition is different. The system there is highly sensitive to increase of the effective mass $m_*$, which leads to vanishing of $n_s$. It would , in the same way, be sensitive to increasing the local mass $m(p)$. Hence in this regime $n_s$ is a decreasing function of $p$ and we reach the critical (maximal) current at a much smaller value of the phase twist $p_c<\pi/2$. 

To see how the instability occurs in the weakly interacting limit we have to consider the relevant equation of motion in this regime, which is the lattice Gross Pitaevskii equation:
\be
i{d\psi_i\over d t}=-J\sum_{\hat{\d}}\psi_{i+\delta}+U|\psi_i|^2\psi_i
\ee
This mean field description provides an accurate description of the dynamics for $U\ll Jn$. The idea now is to linearize the equation in the fluctuations $\d n$ and $\f$ around the uniform current solution $\psi_i=\sqrt{n+\d n}e^{i p x_i +\f}$. For $p=0$ solving these linear equations simply gives the linearly dispersing phonons (Bogoliubov modes ) of the superfluid. However, beyond the critical value $p=\pi/2$ there are eigenmodes that take imaginary values, indicating a dynamical instability\cite{Wu2001}. 

It is interesting to convert the critical current for dynamical instability to a critical flow velocity. In doing so we get $v_c=p_c/m_*={\pi\over 2} J a$. This is much larger than Landau's criterion, which gives the much smaller sound velocity, $v_s=\sqrt{JUn}a$ as the maximal stable flow velocity. In terms of the collective modes, Landau's criterion corresponds to the point where the excitation with negative momenta turn negative because of the doppler shift. Negative frequencies, unlike imaginary ones, do not in by themselves imply an instability. The idea of Landau's criterion is that if we add static impurities to the system, then they will induce scattering that creates pairs of negative and positive frequency modes that will lead to decay of the super current. But in a pure lattice system, the current will decay only when the dynamical instability is reached. The dynamical instability in a weakly interacting lattice condensate has been seen experimentally in Ref. \cite{Fallani2004}.

Let us now turn to the strongly interacting regime, near the Mott transition, where the correlation length $\xi\gg a$. Here the dynamics is described by the effective continuum relativistic action (\ref{Seff}). Treating this theory within the mean field approximation, we obtain the saddle-point equation of motion
\be
\ddot\psi=\nabla^2\psi+\tilde{r}\psi-|\psi|^2\psi
\label{RGP}
\ee
where I have rescaled time and the field: $t\to c t$, $\psi\to c^{-1}\sqrt{u} \psi$ and $\tilde{r}=r/c^2$.
In these units we can write the correlation length as $\xi=1/\sqrt{\tilde{r}}$.
This equation admits the stationary solutions:
\be
\psi(x,{\bf{z}})=\sqrt{r-p^2}e^{ip x}
\label{psip}
\ee
where $\bf{z}$ denote the spatial coordinates transverse to $\hat{x}$. From this it is obvious that the solutions disappear and therefore the current carrying states cannot be stable for phase twist $p\ge\sqrt{r}=1/\xi$.
To find the precise critical twist for a dynamical instability I expand Eq. (\ref{RGP}) following Ref. \cite{Altman2005a} in small fluctuations around the stationary solutions. In the long wavelength limit $\bf{q}\to 0$ this gives two modes with frequencies:
\bea
\w_1^2({\bf{q}})&=&2(r-p^2)+{r+p^2\over r-p^2} q_x^2 +{\bf{q}}_\perp^2\\
\w_2^2({\bf{q}})&=& q_x^2 {r-3 p^2\over r-p^2}+{\bf q}_\perp^2
\eea
The first mode $\w_1$ is the generalization of the amplitude (Higgs) mode to the finite current ($p\ne 0$) situation. It is gapped and has positive frequency unless $p^2>r$. The second mode $\w_2$ is the gapless phase (phonon) mode. This mode becomes unstable for $p>p_c=\sqrt{r/3}=1/(\sqrt{3}\xi)$. We see that the critical momentum vanishes as $1/\xi$ upon approaching the critical current. 

\begin{figure}[t]
\includegraphics[width=0.6\linewidth]{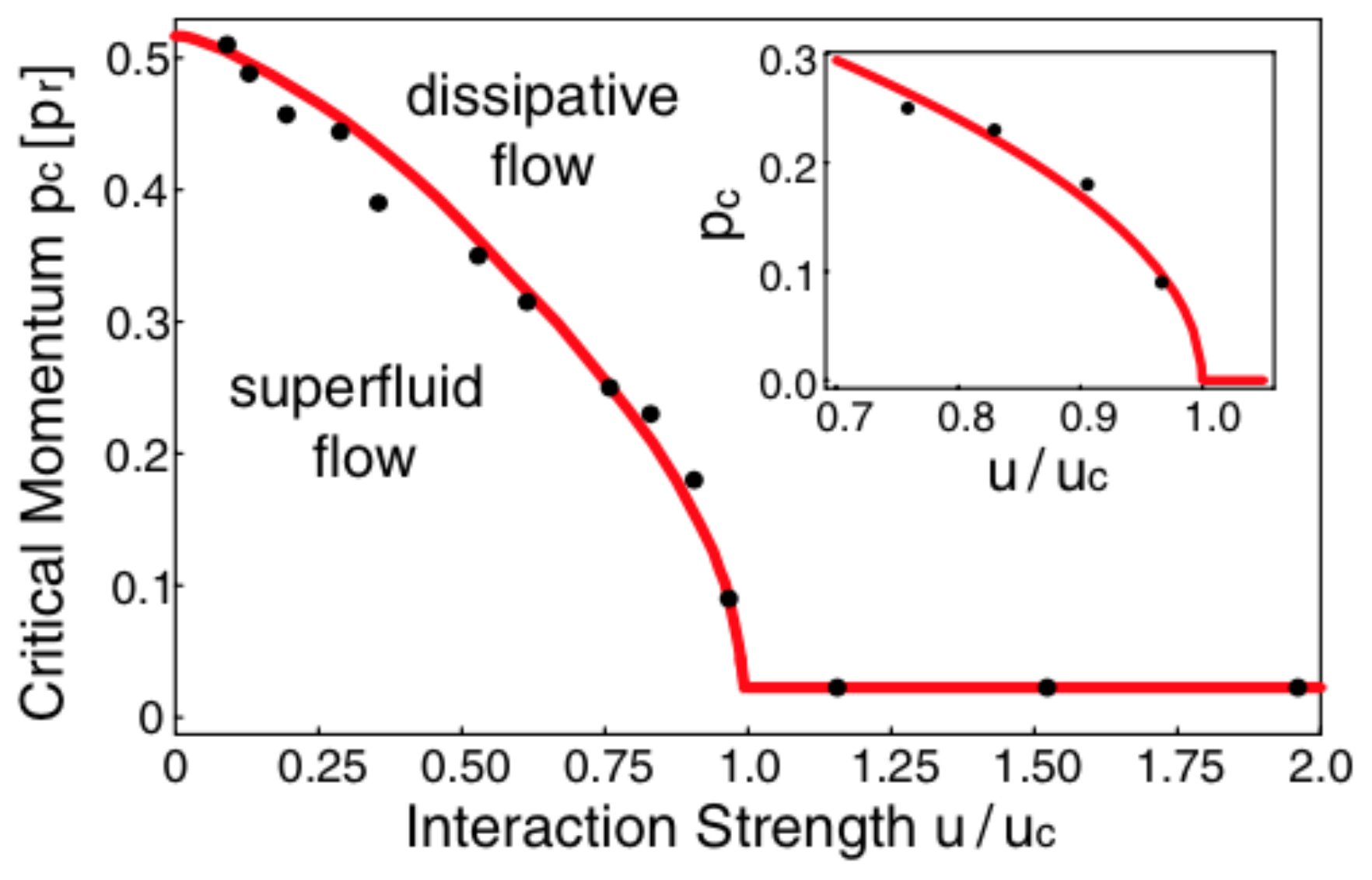}
\caption{Measured critical momentum  (phase twist) as a function of the lattice strength near the superfluid to Mott transition. The red line is the mean field approximation for the critical momentum discussed in this section (Adapted from Ref. \cite{Mun2007}).}
\label{fig:pc}
\end{figure}

The behavior of the critical current near the Mott transition in a three dimensional optical lattice has been investigated in experiment\cite{Mun2007}. A current was induced by moving the optical lattice, whereas the main observable was the existence of a sharp peak in the momentum distribution function as measured in a standard time of flight expansion scheme. The critical momentum was the was determined as the flow momentum at which the condensate peak disappeared from the time-of flight absorption image. Fig. \ref{fig:pc} reproduced from Ref. \cite{Mun2007} shows the measured critical current versus the lattice depth, which is in excellent agreement with the mean field result discussed above.

\subsubsection{Current decay below the critical current}

So far we discussed the breakdown of superfluidity at the critical current within a mean field analysis. This seems to be a good enough approximation to describe experiments in three dimensional lattices \cite{Mun2007}. However it is of fundamental interest to ask whether there is a possibility for current to decay, due to fluctuations, even when the flow is slower than the critical flow.  

The fact that the flow below the critical current is a stable solution of the classical equation of motion implies that in order to decay, the field configuration must {\em tunnel} through a classically forbidden region of phase space. In field theory such tunneling of a macroscopic field configuration out of a meta-stable state was first termed by Coleman as the "Fate of the false vacuum"\cite{Coleman1977} (see also the erratum \cite{Coleman1977erratum} as well as the followup paper discussing quantum corrections to the semiclassical theory  \cite{Callan1977}). 
In our case, the false vacuum is the state with a twist; it must go over an action barrier corresponding to creation of phase slip or soliton in space-time to unwind the twist. A similar approach was taken earlier to describe the current decay due to thermal activation of phase slips in superconducting wires below the critical current\cite{Langer1967,McCumber1970} 

For now, calculation of the tunneling probability of a multi-dimensional field configuration looks like a formidable problem. However, following Ref. \cite{Polkovnikov2005} we will simplify it by restricting ourselves to the asymptotic behavior of the decay rate near the classical critical current, i.e. as $p\to p_c$ from below, where the action barrier is low. I will show that in this regime the pertinent information about the solution can be extracted from a simple scaling ansatz.

To understand how the scaling approach works consider first the toy problem of escape from a metastable state in single particle quantum mechanics (or field theory in $d=0$ dimensions). As usual to facilitate a semiclassical approximation of the tunneling path we rotate the action to imaginary time  
\be
S_{0d}=\int d\tau \left[\dot\phi^2 +V(\phi)\right].
\label{S0d}
\ee
Here $V[\phi]$ is the potential, which for a small barrier can generically be written  as a cubic function $V(\phi)=\epsilon\phi^2-\phi^3$. The limit $\epsilon\to 0$, at which the barrier vanishes and the particle at $\phi=0$ becomes classically unstable, is analogous to reaching the classical critical current in the field theory above. Within the semi-classical theory the decay rate is given by 
$
\G \sim \G_0 \exp\left[-{1\over\hbar} \,S_{sp}(\epsilon)\right]
$
where $S_{sp}(\epsilon)$ is the saddle-point action and $\G_0$ is an "attempt-rate" obtained from integrating over the Gaussian fluctuations around the Saddle-point (see e.g. \cite{AltlandBook} for a complete pedagogical treatment of this problem). 

We now pull the $\epsilon$ dependence out of the action (\ref{S0d}) by applying the following rescaling $\phi\to \phi/\epsilon$ and $\tau\to \sqrt{\epsilon} \,\tau$. This leads to the action
\be
S_{0d}=\epsilon^{5/2} \int d\tau \left[\dot\phi^2 +\phi^2-\phi^3\right]
\ee
Consequently the decay rate in the semi-classical approximation is given by $\G_0 \exp\left[-{1\over\hbar} \e^{5/2}\,\tilde{s}_{sp}\right]$, where $\tilde{s}_{sp}$ is just a number independent of $\epsilon$. In this way we have managed to obtain the parametric dependence on the vanishing barrier hight without solving the saddle-point equations. This will be needed only to obtain the number $\tilde{s}_{sp}$. In the case of the toy model this number can be computed exactly because there is an exact solution to the saddle-point equations, while in more interesting situations it can be approximated   using a variational ansatz.   
%
%\be
%-2\ddot\phi+2\phi-3\phi^2=0
%\ee
%In this toy model, they are exactly solvable, yielding the bounce solution
%\be
%\phi={1\over \cosh (t/\sqrt{2})}
%\ee

Let me now turn to the real situation at hand. The first task is to write down an action, analogous to (\ref{S0d}) that can describe the tunneling of the field configuration out of the current carrying state close to the critical current. I start with effective action of the Mott transition in imaginary time, using units such that $c=1$. Furthermore I rescale all lengths and times with the correlation length, i.e. $x_\a\to x_\a/\xi$. In these units 
\be
S=\int d^{d+1} x\left\{|\nabla\psi|^2+ r|\psi|^2+\half |\psi|^4\right\}
\ee 
Next, following \cite{Polkovnikov2005} we express $\psi$ using the fluctuations around the stable current carrying state with phase twist $k$ (note that I am using $k=p\xi$ which describes the twist, or momentum, in units of $1/\xi$ appropriate for the above action)
\be
\psi(x,{\bf z}= \sqrt{1-k^2}\left[1+\eta(x,{\bf z}\right]e^{i k x+i\phi(x,{\bf z})}.
\ee
Here the coordinate $x$ is the direction of the current, while ${\bf z}$ denotes all other coordinates, including imaginary time, transverse to the current. In order to allow possibility for tunneling across the small barrier we must expand the action to cubic order in the fluctuations $\eta$ and $\phi$.
The two fluctuations are decoupled at the quadratic level after the transformation
\be
\eta\to\eta-{k\over 1-k^2}\partial_x\eta
\ee
Now the gapped amplitude fluctuation can be disregarded and we are left with a cubic action that describes the action barrier in terms of the phase mode
\be
S=\int d^d z dx\left\{\half (\partial_{\bf z}\partial_x\phi)^2+{2\over 3}(\partial_{\bf z}\phi)^2+\half (\partial_x^2\phi)^2+2\sqrt{3}\epsilon (\partial_x\phi)^2-{2\over\sqrt{3}}(\partial_x\phi)^3\right\}
\label{Sphi}
\ee
where we have denoted $\epsilon\equiv k_c-k$. After apply the rescaling
\be
x\to {x\over 2\cdot 3^{1/4}\sqrt{\epsilon}},~ z\to {z\over 6\epsilon},~\phi\to{3^{3/4}\over 2}\sqrt{\epsilon} \phi
\ee
we obtain to leading order in $\epsilon$
\be
S={3^{9/4-d}\over 2^d}( k_c-k)^{2.5-d}\int d^d z dx \left\{(\nabla\phi)^2+(\partial^2_x\phi)^2-(\partial_x\phi)^3\right\}
\ee
From this we conclude that the current decay rate is
\be
\G_d=\G_{0} e^{-C_d (k_c-k)^{2.5-d}}
\ee
where $C_d$ is a number, to be discussed below, that can be obtained with a variational calculation.
The physical picture behind this scaling solution is the following. The rescaling was done after identifying natural scales in the action (\ref{Sphi}) that become singular at the classical critical twist. The natural length in the direction of the current is $x_{||}\sim 1/\sqrt{k_c-k}$ and in the transverse and time direction it is $x_\perp\sim 1/(k_c-k)$. Rescaling with these lengths amounts to the postulate that the critical instanton has these spatial extent in the respective directions of space time. In addition the energy barrier scales as $E_{inst}\sim \epsilon^3$. Putting all this together we see that the instant on action must behave as $S_{inst}\sim E_{inst} \times x_{||} \times x_{\perp}^d\sim (k_c-k)^{2.5-d}$. 

The precise value of the action depends on the detailed shape (functional form) of the instanton, which is encapsulated in the constant $C_d$. These constants have been obtained using a variational calculation in Ref. \cite{Polkovnikov2005}. The results are $C_1=73$ and $C_2=67$. In three dimensions the tunneling action appears to diverge as $k\to k_c$, that is, creating critical instanton solutions incurs a diverging action cost whereas in one and two dimensions the cost of the critical instantons vanishes. Therefore in three dimensions the system would actually take the alternative route of creating finite size (non critical instantons) that incur a finite action cost in the limit $k\to k_c$. A variational calculation in three dimensions gives $\G_3\sim e^{-4.3}$.\cite{Polkovnikov2005}. So, while in lower dimensions the probability for current decay goes to 1 continuously on approaching the critical current in three dimensions it jumps from a small decay probability below the critical current to 1 above it. This explains why the mean field picture worked so well in describing the experiments done in three dimensional optical lattices \cite{Mun2007}.

\subsection{The Mott transition in one dimension}

So far our discussion was aimed at lattices in two or higher spatial dimensions. The Mott transition in a one dimensional system is different in several important aspects. Here I give a very brief review of the essential universal features of the one-dimensional problem. For a more detailed account of the theory and relation to experiments with ultra-cold atoms I refer the reader to recent reviews, such as in Ref. \cite{Cazalilla2011}.

The arguments given above for the existence of an incompressible Mott phase in higher dimensions relied on having a strong lattice potential. In particular, the extreme limit of decoupled sites makes no sense otherwise. More generally it can be shown that a superfluid in two or higher dimensions is perturbatively stable to the presence of a {\em weak} lattice potential, regardless of the commensurability.  In one dimension, by contrast, the Mott transition can occur in presence of an arbitrarily weak lattice potential commensurate with the particle density if the repulsive interactions are strong enough. 

This can be seen by starting from the universal long wave-length description of a Galilean invariant superfluid, the harmonic fluid theory already encountered in section \ref{sec:interf}
\be
H_0=\half \int dx\left[ \kappa^{-1} \left({1\over \pi}\partial_x\phi\right)^2 + \rho_s(\partial_x\t)^2\right].
\ee
Here $\t(x)$ is the condensate phase field $\pi^{-1}\partial\phi$ is the smooth density field conjugate to the phase, $[\t(x),\partial\phi(x')]=i\pi\d(x-x')$. $\rho_s=\rho_0/m$ is the superfluid stiffness of a Galilean invariant superfluid and $\kappa$ is the macroscopic compressibility, which in a weakly interacting gas is simply the inverse interaction constant.  The algebraic decay of long distance correlations is fully controlled by the Luttinger parameter $K=\pi\sqrt{\kappa\rho_s}$.

The full density operator can be written in terms of the operator $\phi(x)$ as\cite{Haldane1981} 
\be
\rho(x)=(\rho_0-{\pi^{-1}}\partial_x\phi)\sum_{m=0}^{\infty} \cos[2m\pi\rho_0-2 m\phi(x)]
\ee
Accordingly, a commensurate lattice potential $V(x)=V_0\cos(2\pi\rho_0 x)$, which couples to the particle density has a non oscillating contribution to the Hamiltonian:
\be
H_{lat}=\int dxV(x)\rho(x) = V_0\int dx \cos(2\phi) 
\label{HSG}
\ee

The low energy Hamiltonian is therefore given by the well known Sine-Gordon model. Scaling analysis shows that the lattice potential embodied in the cosine term becomes a relevant perturbation if the interaction is sufficiently strong so that  the Luttinger parameter $K<2$. In this case the field $\phi(x)$ is locked to a minimum of the cosine potential even for arbitrarily small value of $V_0$. $\phi(x)$ can be viewed as the displacement of atoms from a putative lattice arrangement, where a uniform shift of $\phi\to\phi+\pi$ corresponds to translating all atoms by one lattice constant. Therefore locking of $\phi$ to $0$ or $\pi$ corresponds to the same Mott state.

It is interesting that the mechanism that the mechanism that drives localization in the regime of weak lattice potential is an interference effect, coherent back-scattering enhanced by interactions. This is rather different than the mechanism discussed in the case of a strong lattice, which stems from a local gap apparent in the atomic (single site) limit.
Nonetheless I emphasize that the long wavelength physics effective near the critical point in one dimension is universal and always described by the sine-gordon mode (\ref{HSG}) even when the microscopic physics is modeled by the one band Hubbard model (\ref{BHM}). Of course the coefficients of the sine-gordon model, have in general a very complex connection to the parameters of the microscopic model, and usually should be regarded as phenomenological parameters. But the crucial point is that any other non linear terms allowed by the $U(1)$ and lattice translational symmetries are less relevant than the cosine under rescaling than the cosine term in (\ref{HSG}).

\section{Fast quench dynamics of bosons in optical lattices}\label{sec:quench}

Here I will concentrate on quantum quenches, in which the system parameters are varied in such a way that dynamically drives the system across a quantum phase transition. There are important theoretical motivations to study such quenches. In many cases the quantum states on the two sides of the transition are conventional in the sense that they have a classical mean field description That is, they can be represented well by a non-entangled wave function that can be factorized as a direct product of single site wave functions. For example the Mott phase admits such a classical description in terms of the site occupation numbers, whereas the superfluid phase is classical when represented in terms of the condensate phase. But because of the non-trivial commutation relations between the phase and the occupation number, the two representations are not compatible with each other and neither can provide an effective classical description of the time evolution when the system is driven across the phase transition. This makes the dynamics non trivial and inherently quantum mechanical. Nonetheless thanks to the universality associated with quantum critical points there is hope of gaining theoretical insight into some aspects of the many-body quantum dynamics despite its complexity. In the following sections we will discuss several issues that have been addressed both theoretically, and motivated by experiments with ultra-cold atoms.

\subsection{Collapse and revival of the condensate on quenching to decoupled sites}
An extreme example of a quench from a superfluid to a Mott insulator was demonstrated in an early experiment by Greiner et. al. \cite{Greiner2002b} soon after the observation of the superfluid to Mott quantum phase transition. The experimental protocol was as follows:
(i) The system was prepared deep in the superfluid phase. (ii) Starting from the superfluid state, the optical lattice was suddenly ramped up, essentially shutting off the tunneling between sites completely. (iii) The system was probed by releasing the atoms from the trap (and simultaneously shutting down the optical lattice) at varying times after the quench and observing the resulting time of flight image. 

 The initial state, being a Bose condensate, is expected to exhibit sharp interference peaks at zero momentum and at all other reciprocal lattice vectors. When the system is probed at later times of the evolution the peaks disappear and reappear periodically over a time-period set by the on-site interaction.  

The result is easily understood in the extreme case studied in Ref. \cite{Greiner2002b}.  Because hopping was turned off in the quench, the final Hamiltonian consists only of the onsite interaction term in decoupled sites:
\be
H_f= \half U \sum_i n_i(n_i-1)-\mu n_i
\ee
Furthermore, if the initial state is deep in the superfluid phase then it is well described by a site factorizable state with a coherent state wavefunction $\ket{z}= \exp(-|z|^2- z b\yd)\ket{0}$ at each lattice site. Solving for the time evolution therefore reduces to a single site problem, with the wave function of the site given by
\be
e^{-iH_f t} \ket{\psi(0)}=e^{-|z|^2}\sum_{n=0}^{\infty}{z^n\over\sqrt{n!}}e^{-i\half Un(n-1)t+i\mu nt}\ket{n}
\ee
Here $\ket{n}$ is the state of a site occupied by $n$ bosons. Recall that $|z|^2=\av{n}$. The weight of the sharp peaks in the momentum distribution is proportional to the square-modulus of the condensate order parameter, which is easily computed from the above wave function:
\be
|\av{a_i}|^2= |z|^2 e^{-4|z|^2 \sin^2(U t/2)}.
\label{collapse}
\ee
According to this expression the interference peaks decays (or "collapses") on a time scale $t_C\approx (U |z|)^{-1}=1/ (U\sqrt{\av{n}})$. They then fully revive at integer multiples of the time $t_R=2\pi/U$.  The two time scales $t_C$ and $t_R$ are widely separated when the site occupation is large, $|z|^2\gg 1$. Only in this case is there a nearly total collapse of the interference peak.

Note that the collapse time, and in fact the whole process, is identical to the single mode phase diffusion calculated in section \ref{sec:single-mode}. During the collapse the coherent state $\ket{z}$ with a well defined phase $\f=\arg (z)$ gradually becomes a superposition of a growing number of coherent states with concomitantly growing phase uncertainty. 
However, analysis of the problem in the coherent state basis reveals an interesting effect occurring exactly at half the time between revivals, when the interference peaks are at their lowest. To see the effect consider the "paired" order parameter:
\be
|\av{a_i^2}|= |z|^4 e^{-4|z|^2\sin^2(Ut)}
\ee
We see that at times $t_q=(2q+1)\pi/U$, with $q$ integer ,there are revivals of the pair order parameter $\av{a}$ whereas the condensate $\av{a}$ is fully collapsed. Indeed the wave function at these times is a superposition of precisely two coherent states $\ket{\psi}=(\ket{z}+\ket{-z})/2$. This state therefore breaks the $U(1)$ symmetry as does the initial state, but has a residual $Z_2$ symmetry associated with phase rotations by $\pi$.

The time of flight image at the times $t_q$, when  paired condensate order is established are not expected to exhibit any interference peaks. This special order can be detected however by looking at the noise correlations in the time of flight image\cite{Altman2004a}. Specifically, it would manifest as sharp peaks in the correlation between fluctuations at $k$ and $-k$.

\subsection{Beyond decoupled sites: Final hamiltonian near the transition}

So far I discussed the extreme quench of the superfluid to a lattice of decoupled sites. Now I turn to the more general case in which the lattice strength is suddenly increased but to a point where there is still non negligible tunneling between the sites. 
One way to approach this problem is to limit ourselves to the vicinity of the critical point, considering quenches in which both the initial and the final Hamiltonians are close to the quantum phase transition. In this regime one might expect the dynamics to be described by the effective field theory (\ref{Seff}).

The classical equations of motion implied by this effective theory is the non linear wave equation:
\be
{\ddot \phi}=\rho_s\nabla^2\phi-r\psi-u|\phi|^2\phi
\ee
If we assume, for simplicity that the initial state was a homogenous mean field superfluid then we can drop the spatial derivatives. This is then the equation of motion of a classical particle of mass $m=1$ moving on the mexican hat potential. 
The quench amounts to preparing the particle with some value $\phi_i$ larger in modulus than the equilibrium value. In a small quench the new equilibrium value $\phi_0=\sqrt{-r/u}$ is only slightly smaller than the initial value. The field $\phi$ will undergo nearly harmonic oscillations around the new minimum position. This is precisely the Higgs, or amplitude mode. In the other extreme of a large quench the final Hamiltonian is in the Mott insulating regime so that the new equilibrium field is $\phi_0=0$ ($r>0$). 

Within the simple single mode picture there is a "dynamical transition" between the large quench and small quench regime\cite{Sciolla2010}. For fixed initial state the transition occurs when the energy of the initial (static) field configuration in the new potential  is just enough to climb to the top of the Mexican hat. At this point the time of oscillation diverges. However, this singularity may well be a peculiarity of the classical mean field dynamics. 
Suppose we allow for quantum fluctuations (uncertainty) in the initial field configuration and its conjugate momenta (we'll see below how this can be done), but evolve each instance of the distribution according to the classical equations of motion. This can be shown to be the leading correction in the semiclassical approximation\cite{Polkovnikov2003}. In this case only one special instance of zero measure in the entire distribution follows an evolution with a diverging timescales, while generic instances of the ensemble are always off of the critical trajectory.   
 
\subsection{Sudden quench from the Mott insulator to the superfluid side}
This  model with a single complex field (i.e. O(2) ) describes the the superfluid to Mott insulator transition at commensurate filling. During the quench the tuning parameter of the transition is rapidly changed. If we are in the disordered phase initially then $\phi(x)$ starts with small quantum fluctuations around that average value $\phi=0$. When the sign of the tuning parameter $r$ is changed these fluctuations start to grow and we expect eventually develop into a superfluid order parameter. Our discussion of such quenches across symmetry breaking transitions will follow closely a nice recent review by Lamacraft and Moore \cite{Lamacraft-Moore-Review}.  

Before getting into more details let me make a few general remarks on what we expect from such evolution. Quantum mechanics is certainly crucial in the initial stage of the order parameter growth dynamics when the quench is  from a disordered state (e.g. Mott insulator) to an ordered phase.  However there is quite general agreement that when the order parameter is already locally formed it's subsequent dynamics becomes essentially classical. The degree of freedom that grows and becomes an order parameter is intrinsically quantum in that it is composed of non commuting components. However as its expectation value grows the effect of the commutator becomes negligible compared to the expectation values and therefore is irrelevant in determination of correlation functions. Second the coherent degrees of freedom that we denote as order parameter are supplemented by a continuum of "incoherent degrees of freedom" such as phonons that are also generated in the quench. Due to interaction terms in the action the order parameter rapidly becomes entangled with the phonons then we  should no longer discuss  superpositions of  order parameter configurations. Thus we expect that the order parameter becomes classical both in the single particle sense that commutators are unimportant and in the many-body sense  due to absence of superposition states. Rather we expect that the system reverts to effective classical (Langevin) dynamics in which the "incoherent" degrees of freedom serve as a bath for the order parameter dynamics. The description of the quantum evolution is important however to setup the correct initial state for the later classical coarsening.
An interesting question for future investigation is to what extent the classical correlations that emerge in the later dynamics can retain signatures of non-trivial quantum correlations that were present in the initial state.  

\subsection{Model system: inverted harmonic oscillators}

I will now discuss an effective model, following \cite{Lamacraft2007} that illustrates how the order parameter dynamics becomes classical  at rather short times after a quantum quench from a disordered phase into the broken symmetry phase.
In the initial stages of order parameter growth, while the fluctuations $\phi^2$ are still small it is reasonable to neglect the interaction term. In this case we can decouple the modes in momentum space, where each mode at momentum $q$ is described by an independent quadratic theory. 
\be
H_0=\half\sum_q\left[|\Pi_q|^2+(r+q^2)|\phi_q|^2\right]
\label{eq:har}
\ee
When we quench from a positive to a negative value of $r$, a subset of the modes with sufficiently small $q$ are suddenly subject to an inverted harmonic potential and therefore become unstable. Though the fate of these modes is very simple we can use it to illustrate a lot of important concepts. First and foremost we can show precisely how these modes become effectively classical.

\noindent{\em Single mode case--} Let us first concentrate on a single unstable mode described by
\be
H=\half p^2 + \half\w^2 x^2
\ee
a toy model for the evolution of the uniform $q=0$ mode which happens to be the most unstable. The Heisenberg evolution of the quantum operators is  given by:
\bea
x(t)=x(0)\coth(|\w|t)+{p(0)\over |\w|}\sinh(|\w|t)\to \half e^{|\w|t}\left(x(0)+{p(0)\over |\w|}\right)\nn\\
p(t)=p(0)\coth(|\w|t)+x(0)|\w|\sinh(|\w|t)\to  \half e^{|\w|t}\left(\,p(0)+|\w|x(0)\,\right)=|\w|x(t)
\eea
As a first sign that the dynamics becomes classical we see that in the long time limit the operators become effectively commuting (or more precisely, the non commuting part becomes negligible in the computation of correlations).
Using the above expressions we can easily calculate any correlation function at time $t$ in terms of correlations in the initial state. For example the fluctuation of the position is
\bea
\av{x(t)^2}=\av{x(0)^2}\coth^2(|\w|t)+\av{p(0)^2}\sinh^2(|\w|t)
\to {1\over 4}\left(\av{x(0)^2}+{\av{p(0)^2}\over|\w|^2}\right)e^{2|\w|t}
\eea
Assuming the initial state is the ground state of a stable harmonic oscillator with real frequency $\w_0$ then
\be
\av{x(t)^2}={1\over 2\w_0}\coth^2(|\w|t)+{\w_0\over 2|\w|^2}\sinh^2(|\w|t)\to{1\over 8}\left({1\over w_0}+{\w_0\over |\w|^2}\right)e^{2|\w|t}
\label{eq:x2}
\ee
We can interpret this time dependence as the classical evolution $x_c(t)=x_0e^{|\w|t}$, with the initial position $x_0$ being a Gaussian random variable with variance $x_0^2={1\over 8}\left({1\over w_0}+{\w_0\over |\w|^2}\right)$. The momentum at the same late time will be fully determined by the position. This claim can be substantiated by inspecting the Wigner distribution associated with the wave-function at late times.

The Schr\"odinger wave function in the position representation is given by
\be
\psi(x,t)\to A(t) \exp\left[-i {x^2|\w|\over2}\left(1-2e^{i\t}e^{-2|\w|t}\right)\right]
\ee
where $A(t)$ is a normalization constant and $\t=\tan^{-1}({\w_0/|\w|})$. This can be seen by showing that $\psi(x,t)$ is annihilated by the long time limit of $a(t)=x(t)+i\,p(t)/\w_i$. The Wigner distribution associated with the wave-function at late times is 
\be
f(p,x,t)=\int dx'\psi*(x+x'/2,t)\psi(x-x'/2,t)e^{-ipx'}= B(t) \exp\left(-\a(t)x^2-{(p-\b(t) x)^2\over \a(t)}\right)
\ee
where $\a(t)=2|\w|\sin 2\t e^{-2|\w|t}$, $\b(t)=|\w|(1-2\cos 2\t e^{-2|\w|t})$. Because of the decay of $\b(t)$ the Wigner distribution becomes increasingly squeezed in one direction. Specifically, the $p$ distribution approaches a delta-function $\d(p-|\w|x)$, concentrated on the phase space trajectory of a classical particle.

\noindent{\em Many-mode Harmonic model --} We now proceed to discuss the many mode harmonic model given by Eq. (\ref{eq:har}). In this case each mode is described by a different Harmonic oscillator with a frequency dispersing as $\w_q=\sqrt{r+q^2}$. We see that only a subset of the modes having $q<\sqrt{|r|}$ become unstable. The scale $r^{-1/2}$ gives a natural length scale for the inhomogeneity of the order parameter immediately after the quench. Because the order parameter at two points removed further than $r^{-1/2}$ grow in random uncorrelated directions, this also sets the distance between topological defects in the order parameter field immediately after the quench. From the simple arguments above we would predict a vortex density of $n_v\sim r$ 

\noindent{quantum coarsening} To substantiate this point and derive the subsequent evolution we can compute the time dependence of the field correlation function in the harmonic approximation and infer the evolution of the vortex density.
This analysis was carried out by Austen Lamacraft \cite{Lamacraft2007} in an attempt to explain experiments with spinor condensates rapidly quenched across a quantum phase transition to an easy plane ferromagnetic state \cite{sadler2006}. 

The off-diagonal correlation function 
\be
C(x,t)=\av{\phi^*(x,t)\phi(x,t)}=\int {d^d k\over (2\pi)^d} e^{i k x}\av{\phi_k\phi_{-k}}
\ee
can be evaluated directly from Eq. (\ref{eq:x2}), which holds for each $k$ mode. It is convenient for simplicity to consider the case where we quench from deep in the insulating phase into the ordered phase not too far from the transition. In this case $r_0\gg r$ and we get a simple expression:
\be
\av{\phi^*(x,t)\phi(x',t)}\approx {\sqrt{r_0}\over 2}\int {d^d k\over (2\pi)^d} e^{ik (x-x')}{\sinh^2\left(t\sqrt{|r|-k^2}\right)\over |r|-k^2}
\label{phiphi}
\ee
We see that the value of this correlation function is determined to a large extent by causality. First, at $t=0$ two different points $x$ and $x'$ are uncorrelated due to the deep disordered initial state we have taken. Furthermore for any value of $t$ such that  $2kt>x-x'$, i.e. outside of the forward light-cone, we can close the contour of integration in the upper half plane and since there are no poles in the upper half plane the integral remains precisely zero. The correlations at distances within the light-cone $x-x'<2kt$ grow exponentially reflecting the exponential order parameter growth. 

From the behavior of the corelation function we can infer the density of topological defects. Note that since we have taken the initial state to be uncorrelated there is an initial large density of topological defects regulated only by a high energy cutoff that starts to fall with time. The density of defects (generic zeroes) in A Gaussian wavefunction is given by the Halperin-Liu-Mazenko formula\cite{Liu1992}:
\be
n_v= {1\over 2\pi}C"(0)/C(0)={1\over 2\pi C(0)}\int {d^dk\over (2\pi)^d} k^2 \av{\phi_k\phi_{-k}} 
\ee  
Note that this formula can be understood as the average square wave-vector, weighted by the mode occupation, of the inhomogeneous field. Applying this formula to the correlation function found above we have:
\be
n_v(t)={\sqrt{|r|}\over 4\pi t}
\ee

Of course this type of coarsening does not continue for ever. When $|\phi|^2\sim {|r|/ u}$, the non linear term in the action becomes important leading to two effects. First, it stops the exponential growth of the order parameter and second it affects coupling between the modes that leads to dissipation.
Given the exponential growth of the order parameter (\ref{phiphi}), this occurs at time:
\be
t^*\approx {1\over 2\sqrt{|r|}}\log \left(8 r_f^2\over u\sqrt{r_i}\right)
\ee 
This leads to a formula for the topological defect density at the characterisic
time $t^*$
\be
n_v(t^*) \sim  r\log^{-1} \left(8 r_f^2\over u\sqrt{r_i}\right)
\ee
which agrees with the anticipated result up-to the logarithmic factor.

\noindent{\em Dynamics at later times --} The quadratic approximation has to break down at times longer than $t^*$. First the exponential growth is halted by the non linear terms. Second the coupling between the modes will give rise to effective noise and dissipation that is expected to eventually lead to thermalization. Because the order parameter field is by now essentially classical and moreover coupled to a continuum of phonon modes, the latter stage of the dynamics is expected be described by classical Langevin dynamics. Exactly which model to use from the alphabetical classification of reference \cite{Hohenberg1977} is determined by the conservation laws. 

\subsection{Defect density beyond the Gaussian theory -- Kibble-Zurek scaling}

The fact that the quench takes place in the vicinity of a quantum critical point suggests using the scaling properties of the critical point to make statements valid beyond the Gaussian theory. However, taking the initial state of the sudden quench to be far from the critical point as we did in the previous section is not compatible with such an analysis.  Here we discuss an alternative scenario that will be amenable to scaling analysis. The analysis I present below is due to Polkovnikov \cite{Polkovnikov2005universal}  and is closely related to scaling arguments by Zurek for quench across a classical phase transition\cite{Zurek1985}. 

We consider a system with continuous time dependence described by the Hamiltonian
\be
H=H_0+\lambda(t) V
\ee
Here $H_0$ denotes the Hamiltonian at the quantum critical point and $V$ is a relevant operator that allows tuning across the critical point. 
\be
\lambda(t)=v{t^r\over r!}
\ee
is the time dependent field that drives the transition. The special case $\a=1$ describes a linear ramp across the transition. On the other hand the limit $\a\to\infty$ describes a sudden quench with an amplitude set by the coefficient $v$.

The basic idea behind the Kibble-Zurek scaling analysis is to compare the instantaneous rate of change $R=\dot\lambda/\lambda$ to the instantaneous gap $\D$. When still far from the critical point the time evolution can be considered as adiabatic as long as $R<\D$. As we approach the critical point the gap vanishes as $\D\sim \lambda^{\nu z}$, so eventually the rate of change becomes fast relative to the gap $R>\D$. From here the time evolution can be considered as adiabatic until the gap reemerges on the other side of the transition. According to this reasoning the crossover from adiabatic to sudden occurs at the value of $\lambda$ where the equation $\dot\lambda/\lambda=\D$ or
\be
\a v^{1/\a} \lambda_*^{-1/\a}=\lambda_*^{\nu z}
\ee
So, 
\be
\lambda_*=\a^{\a\over \a\nu z+1} v^{1\over \a\nu z+1}
\ee
The order parameter configuration (correlations) freeze to the structure they had at the point $\lambda(t_*)=\lambda_*$, so the vortex density is set by the correlation length at his point $n_v=\xi_*^{-2}$. Using the scaling $\xi=1/\lambda^\nu$ we obtain the vortex density:
\be
n_{v}(t_*)\sim v^{2\nu\over \a\nu z+1}
\ee 

The Kibble-Zurek arguments involving a crossover from sudden to adiabatic evolution help to illustrate the physics, however we can obtain the same result from more general pure scaling arguments, which will allow further generalization. 

First note that the scaling of $\xi_*$ with $v$ can be found if we know the scaling dimension $[v]$ of $v$ since $v\sim 1/\xi^{[v]}$. Now we use $\lambda=v t^\a$, so $[v]=[\lambda]-\a[ t]=\nu^{-1}+\a z$. Inverting the relation we have the characteristic scale $\xi_*=v^{-1/(\nu^{-1}+\a z)}$, leading to the Kibble-Zurek result for the vortex density. We can generalize this to describe the behavior of any correlation function after the quench. For example a two point equal time correlation function must be proportional to a scaling function 
\be
\av{{\hat{O}}(x){\hat{O}}(0)}= {1\over \xi_*^{2\eta}} {\mathcal F}\left({x\over \xi_*}\right)
\ee
where $\eta$ is the scaling dimension of the operator $O$ and $\xi_*$ was found above. As an example we can take the order parameter corerlations in the one dimensional Ising model $\av{\s^z(x,t)\s^z(0,t)}$. The scaling dimension of $\s^z$ at the Ising critical point is $\eta=1/8$, $z=1$ and $\nu=1$. Moreover since correlations decay exponentially in the disordered phase we expect ${\mathcal{F}(y)}\sim e^{-y}$ at long distances. Using the the sclaing of $\xi_*$ with $v$ we expect the following behavior of correlations at long distances as a function of the ramp rate $v$:
\be
\av{\s^z(x,t)\s^z(0,t)} =v^{{1\over 4}\left({1\over 1+\a}\right) }\exp\left(-x v^{1\over1+\a }\right)
\ee

\section{Quench dynamics and equilibration in closed 1d systems}\label{sec:quench1d}

In the previous section I discussed a dynamic quench that takes the system across a quantum critical point. The discussion of the quantum dynamics was confined to the behavior of the correlations at short times following the quench. I pointed to reasons why the dynamics becomes classical at long times. The unitary quantum dynamics gives way to a simpler coarse grained description in terms of classical Langevin equations as the system approaches thermal equilibrium. 

One dimensional systems could be an exception to this rule for at least the following reasons:
\begin{enumerate}
\item  Quantum fluctuations in one dimension generally inhibit order parameter growth eliminating one reason for classical behavior.
\item There are integrable one dimensional models that cannot thermalize. Even when integrability is broken, routes to relaxation are more restricted in one-dimension than higher dimensions,  therefore one may expect a parametrically long period of time separating the initial simple unitary evolution and eventual thermalization. Such a state is often referred to as a pre-thermalization. \end{enumerate}

For these reasons significant theoretical efforts have been made in recent years to understand the evolution of correlations following a quench in one dimension. But even in one dimension the problem is very hard and in general intractable. Strictly speaking, the Hilbert space needed for calculation grows exponentially  with the system size. The problem is especially acute in one dimension no single dominant mode  grows to macroscopic values and dominates the dynamics. On the other hand, we know of an efficient way to simulate quantum ground states in one dimension using Matrix product states or DMRG (see lectures by Uli Schollow\"ock \cite{Schol2011}).
 This method relies on theorems that guarantee a low entanglement entropy in the ground state. However during quantum dynamics remote parts of the system gradually get more entangled giving rise to growth of entanglement entropy that rapidly inhibits efficient simulation.
\subsection{Growth of entanglement entropy}
It is natural to take the initial state in a quantum quench as a ground state of some local Hamiltonian. Such an initial state is characterized by low entanglement entropy, i.e. at most area law up to logarithmic corrections. When this state starts to evolve under the influence of a different Hamiltonian the entanglement entropy will in general grow. The way it grows is important and interesting for several reasons.
First, the growth of entanglement entropy limits the ability to simulate the time evolution using Matrix product state, time dependent DMRG or related methods. Second, if the entanglement entropy is considered from the information theoretic perspective its time evolution measures how fast information propagates between different regions of the system following the quench. Finally, the growth of entanglement entropy can tell us something about the approach to equilibrium and the nature of the long time equilibrium state itself. For instance, does the system approach equilibrium in a finite time scale, independent of the system size? Does the entanglement entropy of a finite sub-system approach the same value expected if it had reached true thermal equilibrium?

The Lieb-Robinson bounds \cite{Lieb1972} on propagation of correlations also set the limit on the growth of the entanglement entropy to be linear in time \cite{calabrese2005evolution,Bravyi2006}. In clean systems, as far as we know this limit is always saturated. The intuitive picture is that the excitations of the system by the quench leads to uniform parametric excitation of quasi-particle pairs (e.g. particle hole pairs) that travel in opposite direction. Since each particle hole pair is in general entangled, the crossing of the interface by one of the partners adds one unit to the entanglement entropy between the two halves of the system. Let us consider only the fastest of those, which travel at the Lieb-Robinson velocity $v_{m}$. Then the number of "particle-hole" pairs shared between the two halves of the system grows linearly with time, and hence the entropy grows as $S(t)\sim v_m t$. 

\subsection{Evolution of correlations following a quench in one-dimension}

As mentioned above, the rapid growth of entanglement entropy in the evolution following a quench prohibits direct numerical simulation of generic models in extended systems, at least for long times. 
 Some progress can be made by resorting to exactly solvable models and benchmarking against numerical simulations of short to intermediate time scales. 
Because of the relevance to recent experiments with ultra-cold atoms in optical lattices (see e.g. \cite{Cheneau2012}), this kind of approach has produced somewhat of an industry. So far, however I believe that it is hard to draw general conclusions from the results. I will therefore make only a brief review of this topic focussing on examples where exactly solvable models can be used to demonstrate a more general rule. 

\subsubsection{Decay of antiferromagnetic order in a spin-$\half$ chain: gapless final state}

A natural model in which to consider the decay of order parameter correlations in one dimension is a quantum spin chain. For example, we want to look at the spin-$1/2$ xxz chain:
\be
H=J\sum_i\left\{S^x_i S^x_{i+1}+S^y_iS^y_{i+1}+\D S^z_i S^z_{i+1}\right\}
\label{eq:xxz}
\ee
Note that the spin model can be thought of as a hard-core boson model with hopping $J/2$ where $b\yd_i=S^+_i$. Alternatively this model is mapped exactly through a Jordan Wigner transformation to a model of fermions with hopping $J/2$ and nearest neighbor interaction $\D$. The ground state of this model is a $z$-antiferromagnet with gapped excitations for $\D>1$ and a gapless phase with power-law correlations (Luttinger liquid) for $0\le\D<1$. 

Let us consider a scenario in which a system, prepared with antiferomagnetic order, is quenched to the state with $0\le\D<1$.
The question we ask concerns the decay of the antiferromagnetic order parameter
\be
m(t)={1\over N}\sum_j (-1)^{j} \av{S^z_j}
\ee

A closely related problem has beed addressed directly by a recent experiment using bosons in optical lattices\cite{Trotzky2012} with the boson density in the lattice playing the role of the $S^z$ component of the spin. 

Of course the experiment is not described by exactly the model presented above, however both systems have the same long wavelength limit. It is therefore tempting to take a universal approach to this class of problems using the long-wavelength continuum limit, in this case a Luttinger liquid, to describe the evolution following the quench. The general framework for doing this using conformal field theory has been set up by Calabrese and Cardy\cite{Calabrese2006} (see also Ref. \cite{Cazalilla2006} for the specific case of a Luttinger liquid).
%{\cred{If there is time explain the basic idea behind this approach.}}

In the case under study here we can also make a direct calculation within the Harmonic Luttinger liquid framework. In terms of the bosonized fields the order parameter is given by $\cos(2\phi)$. Initially it is concentrated near one of the two degenerate minima. The time evolution of this operator is now in direct correspondence with the decay of phase coherence $\cos(\t)$ in an interferometer.
The result, by duality to the latter problem, is:
\be
m_s(t)\sim e^{-t/\tau}
\ee    
where $\tau={2\over \pi K \D_{0}}$.

In order to assess the validity of the low energy approach we shall benchmark it against an exactly solvable model \cite{Barmettler2009,Barmettler2010}. Consider for this purpose a quench to a final Hamiltonian with $\D=0$, that is the "$xx$ model". The $xx$ model can be mapped through a Jordan-Wigner transformation to a model of free Fermions in which we can calculate the order parameter evolution exactly.

We shall take the initial state to be a charge density wave (of the Fermion density):
\be
\ket{\Psi_0}=\prod_{k=-\pi/2}^{\pi/2}\left(u_k+v_k c\yd_{k+\pi}c\nd_k\right)\ket{FS}=\prod_k\left(u_kc\yd_k+v_kc\yd_{k+\pi}\right)\ket{0}
\ee
Note that the case $u_k=v_k=1/\sqrt{2}$ corresponds to a perfect Neel state. A more general state arising from a self-consistent mean field theory of the gapped phase has, in complete analogy with a BCS state:
\be
\av{c\yd_{k+\pi}c\nd_k}_0=u_k v_k= {\D_0\over 2\sqrt{\e_k^2+\D_0^2}}
\ee
where $\e_k=-J\cos k$ is the fermion dispersion. Using the exact Heisenberg evolution of the fermions $c_k(t)=c_k e^{-i t\epsilon_k}$ we can compute the time dependence of the order parameter:
\be
m_s(t)= {1\over N}\sum_k e^{i\,2\e_k t}\bra{\psi_0}c\yd_k c\nd_{k+\pi}\ket{\psi_0}={1\over\pi}\int_{-J}^0 d\e{\cos(2t\e)\over\sqrt{J^2-\e^2}}{\D_0\over\sqrt{\e^2+\D_0^2}}
\ee
Let us first consider the case of a perfect Neel initial state. In this case the factor $u_k v_k$ is a constant $1/2$ and we can do the integral exactly with the result 
\be
m_s(t)=\half J_0(2Jt)\xrightarrow{Jt\gg1} {1\over \sqrt{4\pi Jt}}\cos (2Jt-\pi/4)
\ee 
This result has no trace of the exponential decay implied by the analysis of the CFT in the long wavelength limit.  
In fact since the Neel state is obtained as an initial state with $\D_0\to \infty$, the CFT would predict a decay on a vanishing time scale.
The source of the discrepancy is that the initial state occupies many high energy modes at the band edges and is therefore far from the range where the low energy theory is expected to hold. The above integral and hence the decay of the staggered moment is dominated by the van-hove singularity at the band edge.
We note that an analysis using a linearized spectrum yields $(\L t)^{-1}\sin(\L t)$, where $\L$ is the high frequency cutoff.

For a better comparison to the CFT result we now consider an initial state that is a weak charge density wave characterized by $\D_0\ll J$. This state is much closer to the gapless Luttinger liquid. In this case there are two main contributions to the integral. The first coming from $\e\approx 0$ near the Fermi points, and second from the square-root singularities at $\e=\pm J$. Adding up these two contributions we have:
\be
m_s(t)\approx {\D_s\over \pi J} K_0(2\D_0 t)+{\D_0\over 2J} J_0(2Jt)\,\xrightarrow{\D_0 t\gg 1}\,{1\over \sqrt{4\pi Jt}}\left\{\sqrt{{\D_0\over J}}e^{-2\D_0 t}+{\D_0\over J}\cos(2Jt-\pi/4)\right\}
\ee
In addition to the oscillatory decay identical to the one found in the case of a Neel state, we find here a non oscillatory exponential decay, with a time scale $\tau=(2\D_0)^{-1}$ in agreement with the CFT result. Comparing the two terms we see however that at times $t>\D_0^{-1}\ln(J/\D_0)$ the oscillatory decay dominates the order parameter dynamics. Nevertheless, in contrast to the case of a Neel initial state the low energy modes contribute to the dynamics over a parametrically long period of time.

There are a few more comments I wish to make regarding the applicability of low energy theories. First it is interesting to note a symmetry of the dynamics, which allows us to map the dynamics subject to the Hamiltonian with $\D>0$ exactly onto the dynamics subject to the Hamiltonian with $\D<0$. In particular the time evolution of an initial Neel state under the antiferromagnetic Hamiltonian ($\D=1$) is identical to its evolution under the ferromagnetic hamiltonian ($\D=-1$). This is in spite of the fact that the low energy description associated with the two hamiltonians are completely different. In the antiferromagnetic interaction the low energy modes are linear in $k$, whereas in the ferromagnet they are quadratic.  This result is a corollary of a more general theorem to be proven below cencerning the dynamics under the influence of a hamiltonian $H$ versus the dynamics under $-H$.

\section{Quantum dynamics in random systems: many-body localization}\label{sec:MBL}

\subsection{Introduction}
The quantum dynamics of  isolated random systems, such as systems of ultra-cold atoms in a random potential, has  attracted a great deal of attention recently as a possible alternative paradigm to thermalization. The conventional wisdom is that 
closed many-body systems, following unitary time evolution ultimately reach thermal equilibrium. 
In a quantum system, the process of thermalization involves the entanglement of local degrees of freedom with many distant ones and thereby the loss of any quantum information that may have been stored locally in the initial state. In absence of accessible quantum correlations the long time dynamics of such thermalizing systems is well described by classical hydrodynamics (i.e. slow fluctuations of conserved quantities and order parameter fields).\cite{Lux2014,Hohenberg1977} It is of fundamental interest to find systems that defy the conventional paradigm and in which quantum correlations do persist and affect the long time behavior. 

Integrable quantum systems, mentioned in the previous section, evade thermalization because their time evolution is constrained by an infinite set of integrals of motion. These are, however, highly fine tuned models. Any small perturbation from the exact integrable Hamiltonian will lead to eventual thermalization. Indeed, the main theme of the previous section as well as of section \ref{sec:interf} was prethermalization. Namely, on short to intermediate time scale the system reaches a quasi steady state that can be understood based on the integrable models, while the slower processes which lead to thermalization dominate only at parametrically longer time scales. 
In contrast, it is expected that systems with quenched randomness provide a generic alternative to thermalization. That is, the non thermal behavior would be robust to moderate changes to system coupling constants (disorder, interactions, etc).

The idea that isolated many-body systems subject to sufficiently strong quenched randomness would fail to thermalize was first conjectured by Anderson in his original paper on localization although in calculation he could only address a one particle problem\cite{Anderson1958}. This idea was revitalized in the last decade when convincing theoretical arguments were given for the stability of anderson localization to moderate interaction between the particles even in highly excited many-body states (i.e. states with non vanishing energy density)\cite{Basko2006,Gornyi2005,Imbrie2014}.

Theoretical studies in the last few years have shown that the many-body localized state can be viewed as a distinct dynamical phase of matter, which exhibits clear universal signatures in dynamics  \cite{Znidaric2008,Pal2010,Bardarson2012,Bauer2013,Vosk2013,Serbyn2013,Serbyn2013a,Nandkishore2014,Andraschko2014,Vasseur2014,Serbyn2014a,BarLev2014,Agarwal2014}. In particular these studies show that, similarly to integrable systems, such localized Hamiltonians are characterized by an infinite set of integrals of motion. In this case they are local only up to possible exponentially decaying tails. However, in contrast to integrable systems, the conservation laws can be slightly modified, but not destroyed by moderate changes in the interaction parameters. The integrals of motion are destroyed, or become fully delocalized, only in a phase transition that marks a singularity in the system dynamics. This localization transition defines a sharp boundary between a macroscopic system dominated by quantum physics even at long times to one that is governed by emergent classical dynamics. Some understanding of the transition is beginning to emerge based on general arguments\cite{Grover2014} and novel renormalization group approaches \cite{Vosk2014,Potter2015}.

The physics of many-body localization is beginning to be addressed in experiments\cite{Ovadia2014,Kondov2013}. In an interesting recent development the breaking of ergodicity due to many-body localization was directly observed for the first time and studied in detail using ultra-cold atoms in an optical lattice\cite{Schreiber2015}. The main focus of this section will be to explain the current experimental effort. I will briefly review the essential theoretical notions needed to understand the experimental developments on many-body localizations. I refer the reader to recent reviews\cite{Altman2014,Nandkishore2014} for a more detailed introduction of the theory.

\subsection{Stability of Anderson localization to interactions and the many-body mobility edge}

%Anderson's original paper on localization was motivated by the question of spin diffusion due to dipolar interactions between randomly placed spins (e.g. Phosphorus dopants in Silicon).  I will come back and concentrate on spin systems and particularly on one dimensional spin chains below. But to start I want to mention that the more recent surge on many-body localization began with Refs.\cite{Basko2006,Gornyi2005}, which considered weakly interacting fermions in a random continuous potential. 
%The focus in these works was on obtaining a stability criterion for Anderson localization in the presence of interactions at elevated temperature or energy density. This is an important point to keep in mind, the term many-body localization refers to the dynamical behavior of states at high (i.e. extensive) energies, not ground states.

Let me for a start  consider the system of weakly interacting fermions discussed by Basko et. al. \cite{Basko2006}. 
As done in that work, I will assume the particles are subject to sufficiently strong disorder so that all of the single particle eigenstates are localized. For example the localizing single particle part of the Hamiltonian could be a tight binding model supplemented by strong on-site disorder.
\be
H=-t\sum_\av{ij} c\yd_i c\nd_j +\mbox{H.c.}+\sum_i V_i n_i +\sum_\av{ij}U n_i n_j
\ee
 Without the interaction $U$ such a system is definitely non ergodic if all the single particle states are localized. In particular, a density fluctuation present in the initial state will not be able to relax to the equilibrium distribution. This is true regardless of the energy of the initial state.  The question, as formulated for example in Refs. \cite{Basko2006,Gornyi2005}, is whether this non ergodic state is robust to adding weak interactions between the particles. 

To elucidate the effect of interactions it is convenient to work with the Fock basis $\ket{\{n_\a \}}$ describing occupations of the single particle localized eigenstates $\ket{\a}$ localized around real space positions $\vec{x}_\a$. 
\be
H=\sum_\a \e_\a n_\a + \sum_{\a\b\gamma\delta} \tilde{U}_{\a\b\g\d}c\yd_\a c\yd_\b c\nd_\g c\nd_\d
\label{Hfer}
\ee  
The non interacting part of the Hamiltonian is represented here by the diagonal first term, where $\e_\a$ are the random energies of the single particle localized states.  The short range density-density interaction $Un_in_j$, which is diagonal in the original lattice basis, has off diagonal terms in the single particle eigenstate basis which can hop particles between localized states centered at different locations. Note that $\a,\b,\g\,\d$ should be centered within a localization distance $\xi$ away from each other to have a significant matrix element. 

Although it can mediate hopping, the interaction does not necessarily delocalize the particles. At strong disorder the interaction induced hopping often takes a particle to a state that is highly off resonant with the initial state. If the system is at a finite temperature (or energy density) the missing energy for a hop may be obtained by de-exciting, at the same time, a particle hole excitation with the right energy. Hence the particle hole excitations can potentially provide an effective intrinsic bath which can enable incoherent hopping processes just as phonons do when the system is not isolated. The difference from a phonon bath, however, is that the particle hole excitations are themselves localized and therefore have a discrete local spectrum. A crude way to assess whether the particle hole excitations can indeed give rise to incoherent hopping is to compute the putative hopping rate using the Fermi-Golden rule and then verify that the obtained rate is larger than the level spacing of the putative bath, in which case the particle-holes may be considered as an effective continuum bath. This leads to the following criterion for delocalization valid at low temperatures $T\tilde{U}/\Delta_\xi^2>1$ where $T$ is the temperature $\Delta_\xi$ is the level spacing within a localization volume. In other words there is a critical temperature $T_c\approx \D_\xi^2/\tilde{U}$ below which the system is strictly localized according to this argument.

The above argument is perturbative and one might worry that a higher order process involving a large number of particle-hole excitations may ultimately delocalize the particles. To address this issue, Basko et. al. \cite{Basko2006} invoked an approximate mapping to a single particle problem taking the localized Fock states $\ket{\{n_\a\}}$ to be sites in a multi-dimensional random graph containing $2^N$ sites where $N$ is the number of real space sites. The random local energy associated with each site is the energy of the Fock state computed within the non-interacting Hamiltonian (i.e. $E[\{n_\a\}]=\sum_\a \e_\a n_\a$). The interaction matrix elements  $U_{\a\b\g\d}$ can be viewed as an effective hopping that connects between the sites  of the multidimensional lattice. A system in a given (extensive) energy  explores only the states of the Fock space lattice that are compatible with that energy. Hence the effective connectivity of the Fock space lattice grows sharply with the energy density. Finally the approximate localization criterion $T_c\approx \D_\xi^2/\tilde{U}$ is obtained (for low temperatures) from applying Anderson's criterion for delocalization \cite{Anderson1958} to the Fock space lattice with a connectivity set by the temperature (or energy density). 
 
It is important to note the crucial difference between the many-body mobility edge discussed above and the conventional notion of a mobility edge in the single particle spectrum. In a (weakly interacting) system with a single particle mobility edge there will always be a finite density of particles thermally excited above the mobility edge, which will thus contribute to thermally activated transport. This is not the case with a many-body mobility edge. The critical energy corresponding to the many-body mobility edge is generically extensive like almost all the many-body spectrum. Now suppose the system is prepared in a thermal state (say by coupling to an external bath and then decoupling it) with average energy below the many-body mobility edge. Because the energy fluctuations in the thermal ensemble are sub extensive, the probability to find the system with energy above the mobility edge vanishes in the thermodynamic limit. Hence below the critical energy density or critical temperature the response of the system must be strictly insulating.  

Before closing this section, we note that a Hamiltonian similar to (\ref{Hfer}) can also be written for spins
\be
H_{spin} = \sum_i h_i S^z_i +\sum_{ij} J_{ij}^{\a\b}S^\a_i S^\b_j 
\ee
Here the first term describes non interacting spins subject to a local random magnetic field. The interaction term can potentially give rise to energy transport. But the same arguments as given for the fermion system imply that the localized state should be stable toward moderate interactions between the spins. Note that the spin Hamiltonian does not have to conserve spin; dipolar interactions for example are not spin symmetric. But localization is still well defined via the absence of energy transport. 

\subsection{Numerical studies of the localized phase}

Models with bounded spectrum, such as spin chains and lattice fermions, are convenient for doing numerical simulations. It was first pointed out in ref. \cite{Oganesyan2007} that in such models the entire many-body spectrum can be localized for sufficiently strong disorder when the mobility edges coming from the bottom and top of the many-body spectrum join in the middle of it. The situation in which the entire spectrum is localized can be viewed as localization at infinite temperature, namely the ensemble in which ever energy eigenstate is equally likely. By tuning the disorder or interaction strength we may drive a many-body localization transition at infinite temperature. 

The case of fully localized spectrum is also highly relevant to studies of time evolution following a quench from a simple product state. Most simple initial state are composed mostly of eigenstates in the middle of the spectrum. Hence the change in the long time dynamics will change sharply when the many-body mobility edge closes in the middle of the spectrum.

I start by describing DMRG calculations of the time evolution following a quench in an XXZ model with random field\cite{Znidaric2008,Bardarson2012}:
\be
H=-J\sum_i \left[(S^+_i S^-_{i+1} +\mbox{H.c.}) +J_z S^z_i S^z_{i+1} \right]+\sum_i h_i S^z_i.
\label{xxz}
\ee
Note, that this model is equivalent, through a Jordan-Wigner transformation to a model of interacting spineless fermions in a random lattice potential with $J_z$ the interaction strength.
The time evolution dictated by this Hamiltonian was calculated starting  from a product state of the spins randomly oriented oriented up or down along $S^z$.  

In clean systems, encountered in previous sections such dynamics could only be computed for very short times. The problem was the rapid growth of entanglement entropy that soon prohibited efficient description in terms of matrix product states. In a localized system, however, one might expect that no correlations are formed across the middle partition over distances larger than some (localization) length scale $\xi$ (Here $\xi$ is given in units of the lattice constant). If this was true, then the entanglement entropy  would be bounded from above by a number of order $\xi$, thus enabling DMRG calculations to long times.

The numerical calculations, however, found a surprising result\cite{Znidaric2008,Bardarson2012}. 
The expected saturation of the entanglement entropy occurred only for $J_z=0$, which corresponds to the special case of Anderson-localization of non-interacting fermions. For any other value of $J_z$ the entanglement entropy grew as $S(t)\sim \log t$ at long times. While the entanglement  growth is unbounded it is nonetheless rather slow and therefore allows DMRG calculations to be taken to long times.
 For small value of the interaction $J_z$, the onset of the logarithmic growth of $S(t)$ was delayed to a time of order $\hbar/J_z$. On the other hand the pre factor of the logarithmic growth was roughly independent of interactions. Now the logarithmic growth of the entanglement entropy is understood to be a ubiquitous feature of many-body localization.  Below I discuss how it can be understood within a simple theory of this state.

The fact that the entanglement entropy grows in an unbounded way, even if very slowly, immediately raises the question if the system ultimately thermalize. This question was addressed using numerical simulations in Ref. \cite{Bardarson2012}, where it was shown that the entanglement entropy of a finite subsystem with length $L$ saturates to a constant $S(L)= s_\infty L$. The saturation time is exponentially large in $L$.
Crucially however the saturation value, although it follows a volume law, was found to be significantly lower than the value expected at thermal equilibrium given the system energy.  

\subsection{Effective theory of the many-body localized phase}\label{sec:MBLeff}

The simple universal behavior, such as the ubiquitous slow logarithmic growth of the entanglement entropy, seen in numerical simulations of the many-body localized localized state call for a simple effective description of the long time dynamics in this state. In the more familiar context of low energy equilibrium physics the renormalization group provides a  conceptual framework to describe the universal features of quantum phases and phase transitions. The idea is that  if we are interested only in low energies and in measurements done with low resolution then systems that may be microscopically very different can be described by the same effective theory provided they are in the same phase. In principle these effective theories are obtained as fixed points of the renormalization group: a successive elimination of short wavelength (or high energy) fluctuations and rescaling. The fixed point theory is universal and in most cases simpler than the original microscopic theory. A case in point is Fermi liquid theory which is essentially an integrable theory. Once found to be the correct RG fixed point\cite{Shankar1994}, or simply postulated by Landau, it I can be used as a powerful phenomenological description of the phase. 

It is natural to ask if there is an analogous framework to describe the long time dynamics in the localized phase.
The problem is that the quench dynamics involves all energy scales, so one cannot integrate out high energy modes and focus on the low energies. Indeed, as remarked above, the dynamics is dominated by very high energy eigenstates in the middle of the many-body spectrum. Nonetheless a renormalization group framework can be applied as shown in Refs. \cite{Vosk2013,Pekker2014,Vosk2014}. The basic philosophy idea is to integrate out high frequency modes in order to gain a simpler description of the evolution at long times, this is similar to RG schemes applied to classical dynamical systems (see for example \cite{Mathey2010}). In the quantum context, targeting low frequencies corresponds to focusing on small energy differences between eigenstates rather than on low absolute energies as in the equilibrium scheme. Application of the RG leads to a simple fixed point Hamiltonian that can be used as an effective theory of the MBL state. In generic cases the form of the effective theory can also be postulated based on general arguments and used as a phenomenological description of the state\cite{Serbyn2013a,Huse2014}.  %My focus in these notes will not be on how the RG scheme works or the derivation of the effective theory using it, but rather on the utility of the effective hamiltonian as a phenomenological description of the MBL state. 

%But first, I give only a flavor of how the RG scheme is implemented. 
Before focusing on the effective theory of the localized phase I give a brief qualitative description of the RG scheme. For more details of how exactly it works I refer the reader to Refs. \cite{Vosk2013,Pekker2014,Vosk2014,Altman2014}
The idea is similar to the real-space strong disorder RG (SDRG) scheme developed by Dasgupta and Ma \cite{Dasgupta1980}.
Given a Hamiltonian such as (\ref{xxz}) we pick out at every step the fastest spins in the chain, namely, those subject to the largest couplings. These couplings give rise to a local gap or frequency scale $\Omega$ which is our running high frequency cutoff. If the disorder is strong then neighboring spins would typically be affected by much weaker couplings and therefore are essentially frozen on the time scale $1/\Omega$. The effective Hamiltonian for the dynamics, coarse grained to longer times, is obtained through a unitary transformation of the Hamiltonian (moving to the rotating frame of the fast spins), which freezes out the fast spins and generates effective weaker couplings between the remaining slow ones nearby. The transformation that decouples the fast spins is found perturbatively in $J_{typ}/\Omega$ where $J_{typ}$ is the value of typical couplings on the chain.  Because the RG is implemented as a series of unitary transformations, degrees of freedom are not eliminated from the Hilbert space. Instead at every step another local operator is transformed (frozen) into an integral of motion. The fixed point Hamiltonian is therefore finally obtained as a Hamiltonian written entirely in terms of conserved quantities that are perturbatively related to the bare local operators, such as $S^z_i $ in (\ref{xxz}).  

Thus, lack of thermalization in the MBL state is seen to be the result of infinite number of (quasi) local conservation rules that are revealed in the course of the RG flow. As in the case of integrable models, expectation values of integrals of motion that are constrained by the initial conditions prevent relaxation to a true equilibrium. The important difference of the MBL state from conventional integrable models however is that the effective theory is a stable RG fixed point and it is therefore robust to moderate changes in the microscopic coupling constants (disorder strength, interaction strength, etc). Hence, in contrast to integrable models, the many-body localized state proves to be a generic alternative to thermalization dynamics.

An effective theory written entirely using integrals of motion was also postulated, based on general arguments, as a phenomenological description of the MBL phase \cite{Serbyn2013a,Huse2014}. It is simple to understand how such a theory emerges by considering the weakly interacting  fermion model (\ref{Hfer}). In this basis it is clear that the occupations of the single particle localized states $n_\a =\sum_{ij}\phi^\star_{\a}(x_i)\phi_\a(x_j) c\yd_i c\nd_j$ are conserved quantities of the non interacting model. Note that these are not strictly local because the single particle wave functions $\phi_\a(x)$ have exponential tails away from their center at $x_\a$. 

Starting from the non-interacting system one can derive an effective Hamiltonian perturbatively in $U$. The off diagonal terms of the interaction $U$ in the Fock basis are off resonant and hence do not contribute, at lowest order, to the effective Hamiltonian. To first order in $U$ the diagonal matrix elements of the interaction give rise to an effective interaction $\sum_{\a\b}V_{\a\b} n_\a n_\b$, where $V_{\a\b}\sim U e^{-|{\bf x}_\a-{\bf x}_\b|/\xi}$. Physically, an interaction between electrons occupying distant localized states is generated due to the exponentially small spatial overlap of the two localized wave functions.  
Higher orders in perturbation theory renormalize the integrals of motion $n_\a\to \tilde{n}_\a$ where $\tilde n_\a$ may now contain terms higher than quadratic terms in the fermion operators. High orders in perturbation theory also produce interaction terms involving multiple integrals of motion, such as $V_{\a\b\g} \tilde{n}_\a \tilde{n}_\b \tilde{n}_\g$. 

Similar considerations apply to the spin model (\ref{xxz}) and suggest an effective model of the form\cite{Serbyn2013a,Huse2014}
\be
H_{eff}= \sum_i \tilde{h}_i \tau^z_i+\sum_{i,j} \sum_n\tilde{J}_{ij}\,\tau^z_i \,\hat{B}_{ij}\,\tau^z_j
\label{Heff}
\ee
Here the integrals of motion $\tau^z_i$ are Pauli operators that are related by a quasi-local unitary transformation to the original Pauli operators $\s^z_i$ of the original spin. Following Ref. \cite{Huse2014} we will term the quasi-local $\tau$ operators l-bits or "logical bits", while the strictly local $\sigma$ operators will be termed p-bits or "physical bits".
 The $\hat{B}_{ij}$ are general operators including contributions from all possible strings of $\tau^z_k$ and $\tau^0_k$ (unit matrices) extending between sites $i$ and $j$. The non-local interactions $\tilde{J}_{ij}$ fall off exponentially with distance as $J^z e^{-|x_i-x_j|/\xi}$.
 
The effective model (\ref{Heff}) provides a simple explanation of the logarithmic growth of entanglement entropy with time\cite{Serbyn2013,Huse2014} and, furthermore, allows to predict a host of other intriguing properties of the many-body localized state. 
The growth of entanglement entropy can be understood as follows. Consider two l-bits from the two sides of a given partition of the system that are separated by a long distance $r$. In the initial state these l-bits would generally not be in their eigenstates even if the local p-bits are prepared in a well defined state. Therefore the interaction between the l-bits generates entanglement after a characteristic time of the order of $t=J_{ij}^{-1}\approx J_z^{-1}e^{|r|/\xi}$.
By this time all the l-bits within the interval of length $r$ between the two chosen l-bits will have had a chance to entangle with each other as much as the constraints set by the initial state allow. Hence, the entanglement entropy at this time would grow to  $S=s_* r$. That is $s_*$ is the average diagonal entropy of single l-bits in the initial state, $s_*=\rho_{\ua\ua}\log \rho_{\ua\ua}+\rho_{\da\da}\log \rho_{\da\da}$. Now, inverting the relation between $t$ and $r$ gives $S(t)=s_* \xi \log(J_z t)$. The Von-Neumann entanglement entropy of a finite subsystem of length $L$ saturates to a volume law $S(L)=s_* L$ with a coefficient $s_*$ that depends on the initial state and would generally be lower than the thermal value.

We see that the logarithmic growth of entanglement stems from dephasing of oscillations in presence of a diagonal interaction term that is present only in the case of interacting fermions. Hence this logarithmic growth is a distinguishing feature between a non-interacting Anderson localized state and an interacting MBL state. The entanglement entropy is, however, not readily accessible to experimental probes. It is desirable to identify indications of this physics in quantities that are directly measurable. In particular I will now consider the relaxation dynamics of observables in the many-body localized phase following a quench.

For simplicity I start by considering the relaxation dynamics of the l-bits ${\bf\tau}_i$ although they are not directly accessible in experiments. I will then generalize to the accessible physical bits ${\bf\s}_i$. The relaxation dynamics of the integrals of motion $\tau^z_i$ is trivial; being integrals of motion they simply do not decay, i.e. $\av{\tau^z_i(t)}=\av{\tau^z_i(0)}$. What about the non conserved components $\av{\tau^{x,y}_i(t)}$ ? If there are no interactions then $\av{\tau^{x,y}_i}$ simply precess around the $\tau^z$ axis with random independent frequencies set by the local  fields $h_i$. With interactions the time evolution is more complicated because of the generated entanglement. Let us assume the system starts in a product state  
\begin{equation}
    |\Psi_0\rangle = \prod_{j=1}^L (A^j_\ua\ket{\ua}_j+ A^j_\da\ket{\da}_j)=\sum_{\tau_1, \ldots, \tau_L}\prod_{j=1}^{L}A^j_{\tau_i}|\{\tau\}\rangle 
\end{equation}
%with $A_{\uparrow}^j = \cos (\theta_j/2)$ and $A_{\downarrow}^j = \sin (\theta_j/2)$. 
The time dependent state, evolving under the influence of the effective Hamiltonian (\ref{Heff}) is 
\be
\sum_{\{\tau\}}\prod_{j=1}^{L}A^j_{\tau_i}e^{-iE[\{\tau\}]t}  \{\tau\}\rangle 
\ee
where $E[\{\tau\}] = \sum_i h_i\tau_i + \sum_{i,j}V_{ij}\tau_i\tau_j$ and $\{\tau\}$ is short for the $\{\tau_1,\ldots,\tau_L\}$. The expectation value of the  l-bit operator in this state is
\be
\av{\tau^x_i(t)}=\sum_{\{\tau\}_{j\ne i}}e^{i \w_i[\{\tau\}]\,t}(A^i_\uparrow)^* A^i_\da \prod_{j\ne i}|A^j_{\tau_j}|^2 +\mbox{c.c.}
\label{taux}
\ee
where $\w_i[\{\tau\}]\equiv E[\uparrow_i,\{\tau\}]-E[\da_i,\{\tau\}]$ is the energy change due to flipping spin $i$ given the configuration $\{\tau\}$ of all other spins.
The reason this is a sum over all $2^L$ basis states of the chain is that the spin $i$ could in principle be entangled with the entire chain. But we know that after evolution time $t$, the spin can only be significantly entangled with the $2l(t)\sim 2\xi\log(J_z t)$ spins nearest to it. Mathematically this is seen from (\ref{taux}) by noting that the phase $\w_i[\{\tau\}]\, t$ is essentially unchanged by flipping $\tau_j$ if  $|x_j-x_i| \gg l(t)$. Hence the phase factor can be taken out of the sums over such remote  spins $\tau_j$ and these sum then simply give $1$. We remain with a sum over $N_\w(t)=2^{2l(t)}$ terms oscillating at independent random frequencies. Hence $\av{\tau^x_i(t)}$ will appear as random noise with a fluctuation amplitude $\sim 1/\sqrt{N_\w}$. Because $N_\w$ grows with time the noise strength will correspondingly decrease in time as 
\be
\left(\,\overline{\av{\tau^x_i(t)}^2}\,\right)^{1/2}=N_\w(t)^{-1/2}\sim e^{-\xi\ln 2 \,\ln (J_z t)}=\left({1\over J_z t}\right)^{\xi \ln2}
\ee 

In practice the system can be prepared with well defined states of the physical bits $\vec{ \s}_j$ rather than the l-bits $\vec{\tau}_j$. However, these two sets of operators are perturbatively related so that we may write $\s^z_i=Z\tau^z_i + \ldots$. If we prepare the system with non vanishing expectation value $\av{\s^z_i}_0$ then a component of this expectation value cannot decay and we expect $\av{\s^z}(t)$ to relax to a non vanishing stationry value at long times  $\sim Z\av{\tau^z}_0\sim |Z|^2$. The non conserved component of $\s^z_i$ on the other hand, is expected to undergo oscillations at $2^{2 l(t)}$ independent frequencies just as discussed above for the non-conserved operator $\tau^x_i$. 

To summarize, we expect local physical operators, such as the local density in a fermion system, that are perturbatively related to conserved quantities, to attain a non equilibrium stationary value at long times. Moreover, the same dephasing process that gives rise to the logarithmic growth of entanglement entropy leads to a power-law decay of the temporal noise in these observables. The exponent associated with this power-law decay is non universal (depends on initial state and on disorder strength); it is directly related to the prefactor of the logarithmic growth of entanglement entropy.

Before proceeding we briefly mention other features of the many-body localized phase that can be understood within the effective theories. The first intriguing property is the persistence of quantum coherence in the MBL phase. Suppose we want to use the accessible p-bits as q-bits. We have seen that if we prepare such a q-bit in a state with definite $\s^z_i$ then some memory of the initial state remains forever. On the other hand if we prepare the q-bit to point in the xy plane, then it starts to dephase and entangle with many-other spins. One may worry that the quantum coherence that existed in the initial state (the phase, or the direction, on the xy plane) is lost irreversibly. However, using the effective Hamiltonian (\ref{Heff}), it has been shown that the coherence, i.e. the initial state of the q-bit, can be partially retrieved even after infinite time using protocols that address only the accessible p-bit\cite{Serbyn2014,Bahri2013}. 

Another intriguing feature of many-body localization is that there can be more than one many-body localized phase\cite{Huse2013}. These can exhibit, for example, broken symmetry and even topological order and protected edge states \cite{Bahri2013}. Note that these orders occur in high energy states, even where they are not allowed to exist in an equilibrium situation. As an example consider the one dimensional random transverse field Ising model:
\be
H=\sum_i h_i\s^x_i +\sum_i J_i\s^z_i \s^z_{i+1} +V\sum_i \s^x_i\s^x_{i+1}
\ee
Note that with the added coupling $V$ the model cannot be mapped to free fermions. There are two simple limits to this model that lead to two distinct prototype dynamical phases.
The first is the free spin limit where the couplings $J_i$ and $V_i$ are set to zero. In this case $\s^x_i$ are clearly integrals of motion and the eigenstates of this system do not break the $Z_2$ symmetry. This "paramagnetic" state persists upon adding moderate interactions $J_i$ and $V_i$ as the integrals of motion $\s^x_i$ transform to closely related l-bits $\tau^x_i$. 
The other simple limit is where $h_i=V_i=0$. In this case the eigenstates are frozen random spin configurations in the $\s^z$ basis, such as $\ket{\ua\ua,\da\ua\ua\ua\da\ldots\da }$, which clearly break the $Z_2$ symmetry. For each eigenstate we have $\bra{\psi_n}\s^z_i\ket{\psi_n}\ne 0$.
Again, the integrals of motion do not change qualitatively by adding moderate coupling $h_i$ and $V_i$. This state can be viewed as an eigenstate glass because of the frozen spin order in eigenstates. The two distinct localized phase, eigenstate glass and paramagnet are separated by a dynamical quantum phase transition that can be described using the dynamical renormalization group approach mentioned above\cite{Pekker2014,Vosk2014}.

Note that, while there is spin order in eigenstates, i.e. $\bra{\psi_n}\s^z_i\ket{\psi_n}\ne 0$, the order vanishes in a thermal average. Hence from the thermodynamic perspective the glass is just a paramagnet . However it is in principle easy to detect the glass order dynamically. Because $\s^z_i$ have an overlap with true conserved quantities the autocorrelation function $\av{\s^z_i(t)\s^z(0)}$ must approach a non-vanishing constant at long times. This is true for any initial state including if the system is prepared in a thermal state.

\begin{figure}[t]
 \includegraphics[width=0.7\textwidth]{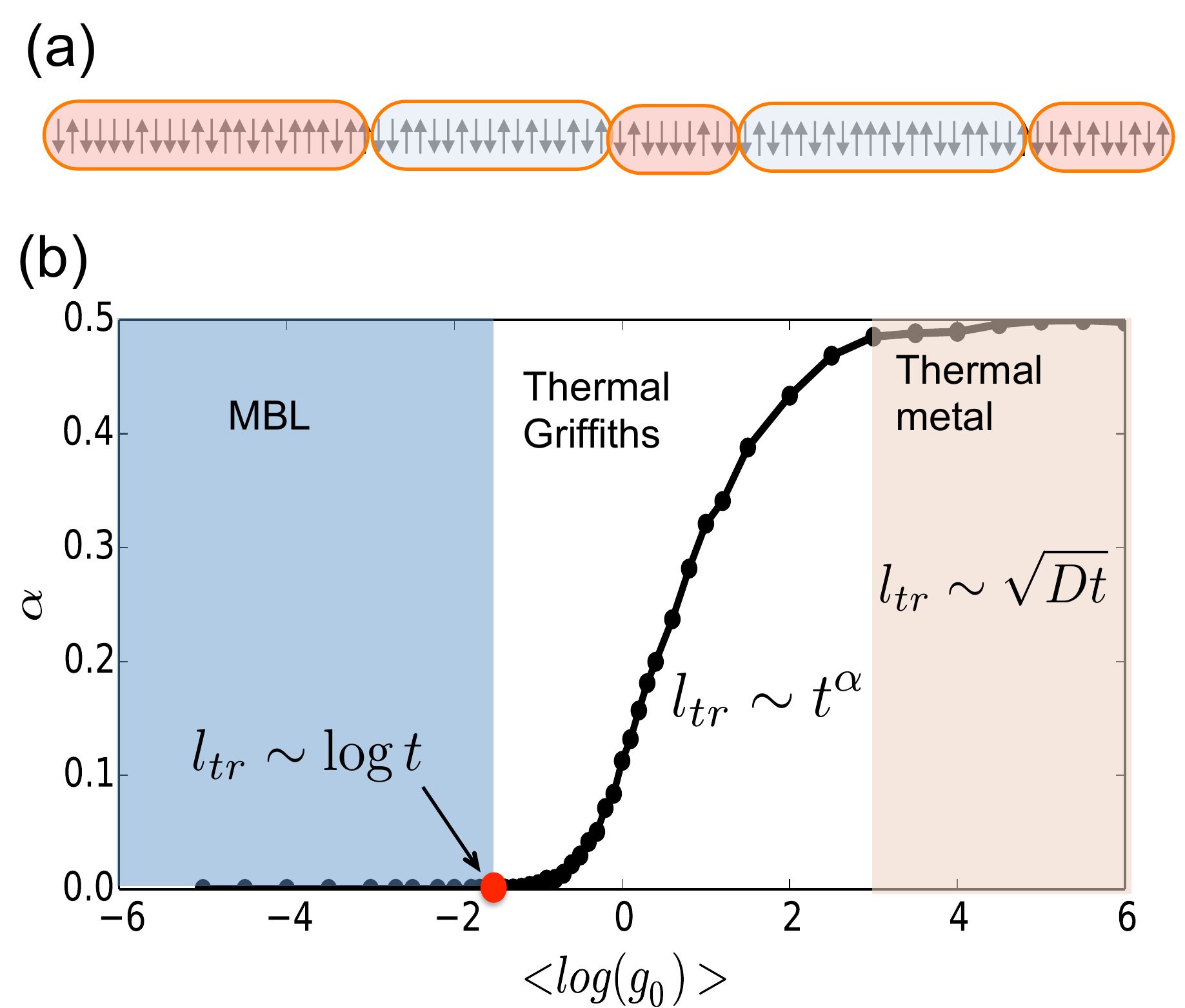}
 \caption{(a) Schematic picture of the system near the many-body localization transition. According to this picture the system is fragmented into incipient insulating and metallic puddles. The ultimate fate of the sample is decided on large scales depending on whether the metallic puddles are sufficiently large and dense to thermalize the intervening insulators. This competition and the resulting scaling near the critical point is described by a renormalization group analysis.  
 (b) One outcome of the RG is the scaling relation between length and time $l_tr\sim t^\alpha$, which describes the length over-which an energy fluctuation can be transported over a time $t$. Here $\a$ is plotted as a function of a microscopic tuning parameter (roughly disorder strength) for a one dimensional system. The continuous vanishing of  $\a$ at the critical point implies the existence of a sub-diffusive delocalized phase between the insulator and normal thermal fluid. \label{fig:Griffiths}}
\end{figure}

\subsection{The many body localization transition}

We have argued above that the many-body localized regime should be viewed as a distinct quantum-dynamical phase of matter, which provides a generic alternative to thermalization dynamics. These two phases must be separated by a dynamical phase transition, which represents a sharp boundary between ergodic dynamics that ultimately becomes classical and dynamics, which is inherently quantum even at long times. Theories of the many-body localization transition are only starting to appear\cite{Vosk2015,Potter2015} and the picture is far from being complete. Yet there is agreement on some interesting features of the transition, which could be highly relevant to experiments with ultra-cold atoms. In this section I briefly review recent progress in understanding some features of the many-body localization transition in one dimensional systems.

A plausible coarse grained picture of the system close to the MBL transition is shown in Fig. \ref{fig:Griffiths}(a).
Because of fluctuations in the disorder strength some regions (sub-chains containing many local degrees of freedom) may behave locally as insulators whereas others are incipient conductors. Because these regions are finite, they behave as insulators or conductors according to their local properties only if they are observed at some finite time scale. At longer scales, locally insulating regions may be thermalized by larger nearby conducting blocks, or conversely, isolated thermal regions may be absorbed into larger insulators around them. A semi-phenomenological renormalization group scheme that captures the  thermalization or absence of thermalization between the different regions at increasing scales has been presented by et. al. \cite{Vosk2015}. Whether a block is insulating or thermal is assumed to be determined by a single parameter $g=\G/\D$,  where $\G$ is the is the inverse time for entanglement propagation across the block and $\D$ is the many-body level spacing of the block.  
The scheme leads to a flow of the distribution of $g$ with two possible stable fixed points, insulating and thermalizing, in which $g$ respectively decreases or grows exponentially with the length scale.

One result of the renormalization group study is the relation between the length $l$ of a cluster (sub-chain) and the time $\tau$ it takes to transport energy across it $\tau\sim l^z$. In a conventional thermalizing system we expect diffusive transport, which implies $z=2$ (i.e. $l=\sqrt{D\tau}$). So, one might expect that the transition to the localized phase can be characterized by how the diffusion constant $D$ vanishes upon approaching the critical point.
But this turns out to be a misguided expectation,.
Instead, we find that the dynamical exponent $z$ diverges at the critical point, or as shown in Fig. \ref{fig:Griffiths}(b), $\a=z^{-1}$ vanishes continuously as the transition is approached. This result implies that the delocalized phase near the critical point exhibits sub-diffusive transport, as also supported by numerical studies of small systems using DMRG and exact diagonalization\cite{BarLev2014,Agarwal2014}.
At the critical point and in the insulating phase we have $\tau\sim \tau_0 e^l/l_0$, where $l_0$ and $\tau_0$ are microscopic time and length scales respectively (in a weakly interacting fermion system for example, $l_0$ could be the single particle localization length). 

It is easy to understand the origin of the sub-diffusive behavior even without delving into details of the renormalization group scheme. Lets assume that the many-body localization transition is controlled by a scale invariant critical point, as the renormalization group points to. As we approach this critical point from the delocalized side there must be a diverging correlation length $\xi$. At scales below $\xi$ the system has not yet "decided" if it is going to be insulating or localized on larger scales, therefore we are likely to find local insulating behavior on this scale. Insulating regions much larger than $\xi$ have a low probability per-unit length $p(l)\sim \xi^{-1}e^{-l/\xi}$, but they can occur with high probability in a sufficiently long sample. In a system of length $L$ Close to the critical point the transport will be dominated by the slowest (a.k.a longest) insulating cluster on the chain. We can estimate the length $l_m$ of this cluster by  requiring the probability  $L p(l_{m})$ to be approximately 1. Thus the longest bottleneck for transport is typically of size $l_m\approx \xi \ln(L/\xi)$ leading to a transport time
\be
\tau_{tr}\sim \tau_0e^{l_m/l_0}\approx \tau_0 (L/\xi)^{\xi/l_0}.
\ee
From this expression we can read off the dynamical exponent $z=\xi/l_0$, which indeed diverges at the critical point. Such a phase, dominated by rare slow clusters is called a Griffiths phase \cite{Griffiths1969}. 
%By a simple change of variables we can convert the distribution of insulating inclusion lengths in the Griffiths phase to a distribution of transport times $p(\tau_{tr})\sim \tau_0^{-1} (\tau_0/\tau)^{(z+1)/z}$. Hence at the critical point the distribution becomes maximally broad.

The Griffiths phase terminates at the critical point marking the transition to the many-body localized phase. Interestingly a good scaling variable to characterize this critical point is the entanglement entropy  of half the systems\cite{Vosk2014a}. In a finite system of size $L$, the results indicate a universal crossover from area-law entanglement entropy in the localized phase to fully thermal volume law entanglement in the delocalized phase. At the critical point itself the fluctuations of the half-system entanglement entropy diverge and scale with the volume $L$ (whereas they  approach constants at large $L$ on either side of the transition). A corollary of this result is that the Griffiths phase described above is ergodic; it is characterized by a fully thermal volume law entropy with vanishing fluctuations. 

\subsection{The experimental situation}

Because systems of ultra-cold atoms are  almost perfectly isolated from the environment they provide a promising platform to investigate the physics of many body localization. 
Until recently, experiments have focused on Anderson localization of non interacting particles\cite{Billy2008,Roati2008, Kondov2011, Jendrzejewski2012}. A typical experiment used a quench scheme to study the expansion of particles in a random potential background\cite{Billy2008}. The particles are prepared in a highly confined state induced by a dipole trap, then the cloud is released and expands in a potential created by a random speckle pattern. The interaction between the expanding particles is negligible compared to the large kinetic energy acquired from the initial confined state. A direct signature of localization is obtained by observing the density profile of the expanding cloud at long times, which is a proxy of the single particle wave function $|\psi(x,t)|^2$. Another set of experiments studied the zero temperature quantum phase transition of interacting bosons from a superfluid to a disordered Bose insulator using linear response transport-like measurements\cite{Pasienski2010,Gadway2011,DErrico2014} and using a quantum quench across the transition\cite{Meldgin2015}. 

The above experiments do not address the generic many-body localization problem. The goal in this case would be to demonstrate a robust non-ergodic state, which exists at finite energy and over a range of interaction strengths.
An early experiment to show evidence for MBL measured the global mass transport of interacting fermions on a disordered three dimensional lattice\cite{Kondov2013}. The relevant signature was apparent immobility of the atomic cloud, which persisted for a range of temperatures. One problem in measuring global transport, however, is that conserved quantities are slowest to equilibrate. Even in a thermalizing system, one may therefore need exceedingly long times to distinguish between slow transport and complete absence thereof. In addition global measurements are potentially sensitive to the inhomogeneity of the trap. For example a localized weakly interacting region at the wings of the cloud can block or slow down movement of a more mobile core. 

Some of these issues were overcome in a recent experiment designed to study many-body localization by observing the relaxation of local observables\cite{Schreiber2015}. The experiment was done with a system of interacting Fermions in a quasi-periodic one dimensional lattice, well described by the following model Hamiltonian:
\begin{equation}
\begin{split}
	\hat{H} = &-J \sum_{i,\sigma} \left(\hat{c}_{i,\sigma}^{\dagger} \hat{c}_{i+1,\sigma}+\text{h.c.}\right) \\ &+\Delta\sum_{i,\sigma} \cos (2\pi\beta i+\phi)\hat{c}_{i,\sigma}^{\dagger}\hat{c}_{i,\sigma} +U\sum_i \hat{n}_{i,\uparrow}\hat{n}_{i,\downarrow}.
	\label{AA_hamiltonian}
\end{split}	
\end{equation}
Here the magnitude $\D$ of the incommensurate lattice plays the role of effective disorder strength. Although this is not a true random potential the model without interactions exhibits Anderson localization  above a critical effective disorder $\D_c=2J$ \cite{Aubry1980}. The difference between this model and a one dimensional model with a true random potential is that in the latter all states are localized for arbitrarily weak disorder. Numerical results suggest that the interacting system with sufficiently large effective randomness also exhibits many-body localization with the same universal properties as the generic MBL phase. 

\begin{figure}[t]
 \includegraphics[width=1.0\textwidth]{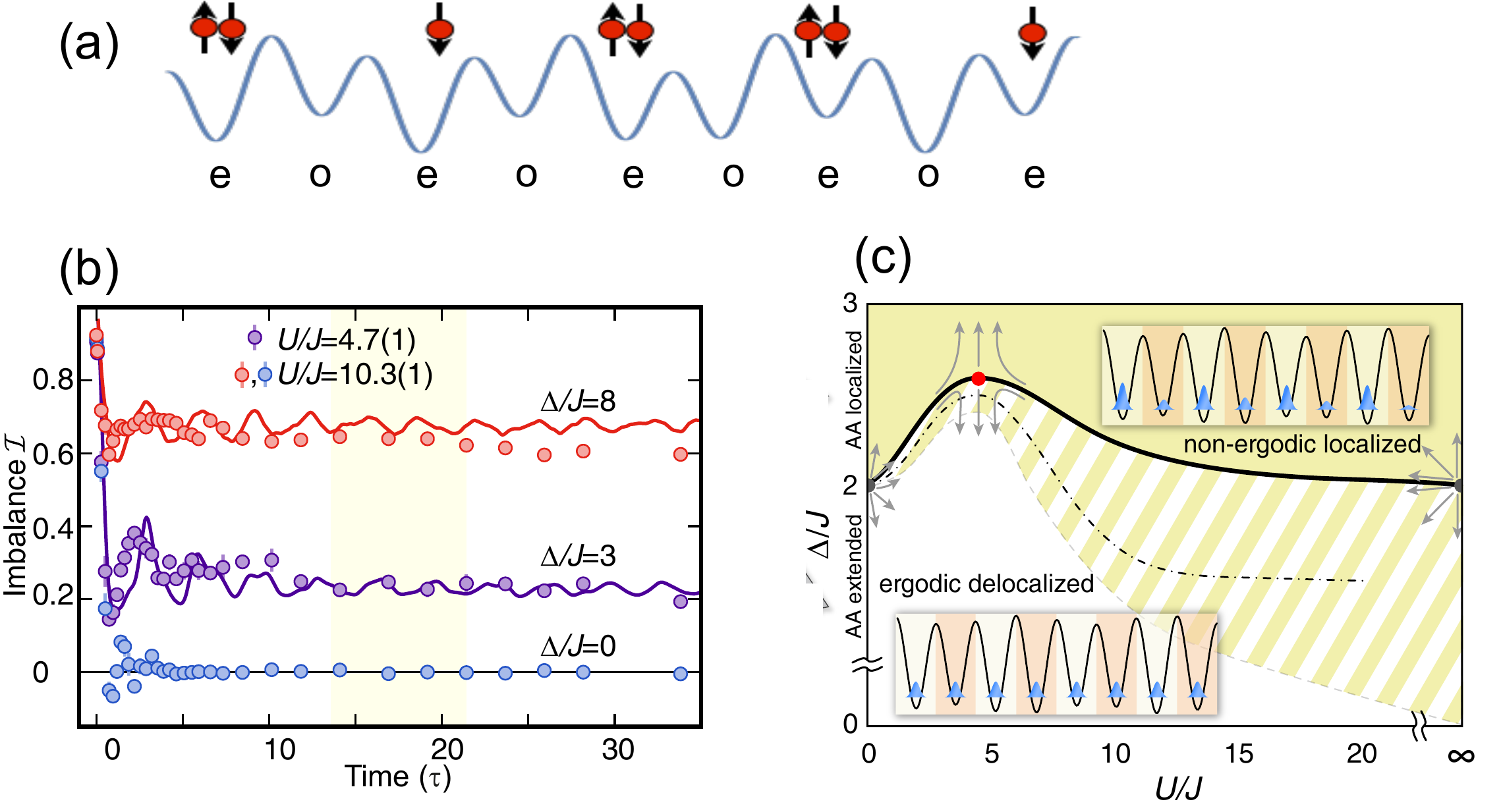}
 \caption{(a) Schematic illustration of the initial state for the quench dynamics studied experimentally in Ref. \cite{Schreiber2015}; the system is prepared in a density-wave configuration where every second site is empty.  
   (b) Time evolution of the relative density imbalance between even and odd sites $\mathcal{I}=(\av{N_e-N_o}/(N_e+N_o)$. Relaxation of the imbalance to a non-vanishing stationary value is a direct demonstration of ergodicity breaking from \cite{Schreiber2015}.  (c) Experimentally determined phase boundary between the ergodic fluid and the non-ergodic MBL phase. The bold line corresponds to initial state with no (or very few) doublons, while the lower line corresponds to an initial state with many doublons. \label{fig:Imbalance}. Panels (b) and (c) are reproduced from Ref. \cite{Schreiber2015}}\label{fig:exp}
\end{figure}

The many-body localized state is identified in the experiment and its properties are studied by implementing a quench scheme and observing the consequent relaxation dynamics of observables. Specifically, the system is prepared  in a far from equilibrium density wave state as depicted in Fig. \ref{fig:exp}(a). This initial state then begins to evolve subject to the full hamiltonian modeled by \ref{AA_hamiltonian} and we monitor the time dependence of the density wave order. The latter is captured by the normalized imbalance between the number of particles in even sites ($N_e$) and that in odd sites ($N_o$)
\be
\mathcal{I}={N_e-N_o\over N_e+N_o}.  
\ee
In an ergodic phase the system would relax to thermal equilibrium corresponding to a high temperature, in which the symmetry between even and odd sites must be restored so that $\mathcal{I}\to 0$. Instead the experimental time traces show, for a wide regime of interaction and disorder strengths) an initial rapid drop but then saturation of the imbalance to a stationary value different from zero at long times compared to the hopping rate. Examples of such time traces are shown in Fig. \ref{fig:exp}(b). This result is a direct signature of robust ergodicity breaking; it shows that there is a phase characterized by an infinite set of local conserved quantities, closely related to the local densities.  

The phase diagram shown in Fig. \ref{fig:exp}(c) is constructed after measuring time traces of the imbalance for a wide range of interaction and disorder strengths. 
The light gray flow lines depict the postulated RG flow, which controls the localization transitions on this phase diagram. The axis corresponding to $U=0$ exhibits the non generic critical point of the non-interacting Aubry-Andrey model\cite{Aubry1980}. On the other hand, for any finite interaction strength the transition is expected to be controlled by the generic fixed point addressed in Ref. \cite{Vosk2014a}.
The interactions also have a noticeable effect on the phase boundary. Moderate interactions have a delocalizing effect which leads to increase of the critical disorder strength. However, the critical disorder strength drops again when the interaction exceeds the single particle band-width.

The behavior at large $U$ has strong dependence on the nature of the initial state, which is controlled in the experiment by changing the strength of interaction $U_i$ in the preparation stage, when the atoms are loaded into the optical lattice. When $U_i$ is large and positive, there are almost no doubly occupied sites in the initial state, while if $U_i$ is large and negative there are many double occupancies . Now, if the interaction effective during time evolution is strong $U\to \pm \infty$, then due to energy conservation doublons present in the initial state cannot decay and new ones cannot form. The bold black phase boundary in Fig. \ref{fig:exp}(c) corresponds to the doubloon free situation corresponding to large positive $U_i$. In this case the dynamics of a state with no doublons can be mapped exactly to the dynamics of non-interacting (spineless) fermions\cite{Schreiber2015}, which explains the symmetry . Which explains the symmetry between the $U=0$ and $U\to\infty$ limits. On the other hand, if there are doublons in the initial state then they become extremely heavy particles (with effective hopping $J_D\sim J^2/U$ at large $U$) which are therefore easy to localize. The corresponding critical disorder is the lower boundary of the striped area in the phase diagram.

The above results have provided evidence for many-body localization but more experiments are needed to substantiate them  and to characterize the dynamics in the many-body localized phase and phase transition. One issue that came up in the recent experiment\cite{Schreiber2015} is that on time scales much larger than shown in Fig. \ref{fig:exp}(b), the imbalance slowly decays due to the slow destruction of MBL by scattering and loss of atoms interacting with the laser light. Recall that many-body localization crucially relies on having a closed system and is therefore very sensitive to the above loss processes that, in effect, open it. It may be interesting to learn about the MBL phase by investigating how susceptible it is to adding such dissipative couplings. This can be particularly useful in studying the many-body localization phase transition. Here an analogy with conventional quantum phase transitions (QPT) can be useful. A QPT is sharply defined only at zero temperature just as the MBL transition exists only for vanishing coupling to a bath. Nonetheless a QPT has a dramatic influence on the physics at finite temperatures, where sufficiently close to the critical point, the temperature provides the only infrared cutoff scale. Similarly we expect the MBL transition to broaden by the dissipative coupling into a universal crossover controlled by the putative localization critical point.  The dissipative coupling provides the only relevant infrared cutoff in this case. Hence by controlling it experiments should be able to access the critical properties, just as temperature dependence of observables reveals the critical properties of quantum phase transitions. 

Further experiments are also needed in order to characterize the dynamics in the insulating phase as well as the delocalized Griffiths phase. 
For example, as discussed in section \ref{sec:MBLeff}, the temporal fluctuations of the imbalance about its long time stationary value is expected to be suppressed as a power of the time in the MBL phase due to the correlated oscillations of a growing number of entangled particles\cite{Serbyn2014}. By contrast, in a non interacting localized system the fluctuations, due to independent oscillations of non entangled particles, do not decay in time. Anomalous power-laws are also expected to occur in the frequency dependence of dynamic-response functions due to Griffiths like effects in the insulating regime\cite{Pekker2014,Gopalakrishnan2015}.

\newpage

\section{Bibliography}
\bibliographystyle{phd}

\begin{thebibliography}{121}
\expandafter\ifx\csname natexlab\endcsname\relax\def\natexlab#1{#1}\fi
\expandafter\ifx\csname bibnamefont\endcsname\relax
  \def\bibnamefont#1{#1}\fi
\expandafter\ifx\csname bibfnamefont\endcsname\relax
  \def\bibfnamefont#1{#1}\fi
\expandafter\ifx\csname citenamefont\endcsname\relax
  \def\citenamefont#1{#1}\fi
\expandafter\ifx\csname url\endcsname\relax
  \def\url#1{\texttt{#1}}\fi
\expandafter\ifx\csname urlprefix\endcsname\relax\def\urlprefix{URL }\fi
\providecommand{\bibinfo}[2]{#2}
\providecommand{\eprint}[2][]{\url{#2}}

\bibitem[{\citenamefont{{Bakr} et~al.}(2009)\citenamefont{{Bakr}, {Gillen},
  {Peng}, {F{\"o}lling}, and {Greiner}}}]{Bakr2009}
\bibinfo{author}{\bibfnamefont{W.~S.} \bibnamefont{{Bakr}}},
  \bibinfo{author}{\bibfnamefont{J.~I.} \bibnamefont{{Gillen}}},
  \bibinfo{author}{\bibfnamefont{A.}~\bibnamefont{{Peng}}},
  \bibinfo{author}{\bibfnamefont{S.}~\bibnamefont{{F{\"o}lling}}},
  \bibnamefont{and}
  \bibinfo{author}{\bibfnamefont{M.}~\bibnamefont{{Greiner}}}, \emph{{A quantum
  gas microscope for detecting single atoms in a Hubbard-regime optical
  lattice}}, \bibinfo{journal}{\nat} \textbf{\bibinfo{volume}{462}},
  \bibinfo{pages}{74} (\bibinfo{year}{2009}), \eprint{0908.0174}.

\bibitem[{\citenamefont{{Sherson} et~al.}(2010)\citenamefont{{Sherson},
  {Weitenberg}, {Endres}, {Cheneau}, {Bloch}, and {Kuhr}}}]{Sherson2010}
\bibinfo{author}{\bibfnamefont{J.~F.} \bibnamefont{{Sherson}}},
  \bibinfo{author}{\bibfnamefont{C.}~\bibnamefont{{Weitenberg}}},
  \bibinfo{author}{\bibfnamefont{M.}~\bibnamefont{{Endres}}},
  \bibinfo{author}{\bibfnamefont{M.}~\bibnamefont{{Cheneau}}},
  \bibinfo{author}{\bibfnamefont{I.}~\bibnamefont{{Bloch}}}, \bibnamefont{and}
  \bibinfo{author}{\bibfnamefont{S.}~\bibnamefont{{Kuhr}}},
  \emph{{Single-atom-resolved fluorescence imaging of an atomic Mott
  insulator}}, \bibinfo{journal}{\nat} \textbf{\bibinfo{volume}{467}},
  \bibinfo{pages}{68} (\bibinfo{year}{2010}), \eprint{1006.3799}.

\bibitem[{\citenamefont{Polkovnikov et~al.}(2006)\citenamefont{Polkovnikov,
  Altman, and Demler}}]{Polkovnikov2006Fringe}
\bibinfo{author}{\bibfnamefont{A.}~\bibnamefont{Polkovnikov}},
  \bibinfo{author}{\bibfnamefont{E.}~\bibnamefont{Altman}}, \bibnamefont{and}
  \bibinfo{author}{\bibfnamefont{E.}~\bibnamefont{Demler}}, \emph{{Interference
  between independent fluctuating condensates.}}, \bibinfo{journal}{Proceedings
  of the National Academy of Sciences of the United States of America}
  \textbf{\bibinfo{volume}{103}}, \bibinfo{pages}{6125} (\bibinfo{year}{2006}),
  ISSN \bibinfo{issn}{0027-8424}.

\bibitem[{\citenamefont{Gritsev et~al.}(2006)\citenamefont{Gritsev, Altman,
  Demler, and Polkovnikov}}]{Gritsev2006}
\bibinfo{author}{\bibfnamefont{V.}~\bibnamefont{Gritsev}},
  \bibinfo{author}{\bibfnamefont{E.}~\bibnamefont{Altman}},
  \bibinfo{author}{\bibfnamefont{E.}~\bibnamefont{Demler}}, \bibnamefont{and}
  \bibinfo{author}{\bibfnamefont{A.}~\bibnamefont{Polkovnikov}}, \emph{{Full
  quantum distribution of contrast in interference experiments between
  interacting one-dimensional Bose liquids}}, \bibinfo{journal}{Nature Physics}
  \textbf{\bibinfo{volume}{2}}, \bibinfo{pages}{705} (\bibinfo{year}{2006}),
  ISSN \bibinfo{issn}{1745-2473}.

\bibitem[{\citenamefont{{Hadzibabic} et~al.}(2006)\citenamefont{{Hadzibabic},
  {Kr{\"u}ger}, {Cheneau}, {Battelier}, and {Dalibard}}}]{Hadzibabic2006}
\bibinfo{author}{\bibfnamefont{Z.}~\bibnamefont{{Hadzibabic}}},
  \bibinfo{author}{\bibfnamefont{P.}~\bibnamefont{{Kr{\"u}ger}}},
  \bibinfo{author}{\bibfnamefont{M.}~\bibnamefont{{Cheneau}}},
  \bibinfo{author}{\bibfnamefont{B.}~\bibnamefont{{Battelier}}},
  \bibnamefont{and}
  \bibinfo{author}{\bibfnamefont{J.}~\bibnamefont{{Dalibard}}},
  \emph{{Berezinskii-Kosterlitz-Thouless crossover in a trapped atomic gas}},
  \bibinfo{journal}{\nat} \textbf{\bibinfo{volume}{441}}, \bibinfo{pages}{1118}
  (\bibinfo{year}{2006}), \eprint{cond-mat/0605291}.

\bibitem[{\citenamefont{Gring et~al.}(2012)\citenamefont{Gring, Kuhnert,
  Langen, Kitagawa, Rauer, Schreitl, Mazets, Smith, Demler, and
  Schmiedmayer}}]{Gring2012}
\bibinfo{author}{\bibfnamefont{M.}~\bibnamefont{Gring}},
  \bibinfo{author}{\bibfnamefont{M.}~\bibnamefont{Kuhnert}},
  \bibinfo{author}{\bibfnamefont{T.}~\bibnamefont{Langen}},
  \bibinfo{author}{\bibfnamefont{T.}~\bibnamefont{Kitagawa}},
  \bibinfo{author}{\bibfnamefont{B.}~\bibnamefont{Rauer}},
  \bibinfo{author}{\bibfnamefont{M.}~\bibnamefont{Schreitl}},
  \bibinfo{author}{\bibfnamefont{I.}~\bibnamefont{Mazets}},
  \bibinfo{author}{\bibfnamefont{D.~A.} \bibnamefont{Smith}},
  \bibinfo{author}{\bibfnamefont{E.}~\bibnamefont{Demler}}, \bibnamefont{and}
  \bibinfo{author}{\bibfnamefont{J.}~\bibnamefont{Schmiedmayer}},
  \emph{Relaxation and Prethermalization in an Isolated Quantum System},
  \bibinfo{journal}{Science} \textbf{\bibinfo{volume}{337}},
  \bibinfo{pages}{1318} (\bibinfo{year}{2012}),
  \eprint{http://www.sciencemag.org/content/337/6100/1318.full.pdf}.

\bibitem[{\citenamefont{Altman et~al.}(2004)\citenamefont{Altman, Demler, and
  Lukin}}]{Altman2004a}
\bibinfo{author}{\bibfnamefont{E.}~\bibnamefont{Altman}},
  \bibinfo{author}{\bibfnamefont{E.}~\bibnamefont{Demler}}, \bibnamefont{and}
  \bibinfo{author}{\bibfnamefont{M.}~\bibnamefont{Lukin}}, \emph{{Probing
  many-body states of ultracold atoms via noise correlations}},
  \bibinfo{journal}{Physical Review A} \textbf{\bibinfo{volume}{70}},
  \bibinfo{pages}{013603} (\bibinfo{year}{2004}), ISSN
  \bibinfo{issn}{1050-2947}.

\bibitem[{\citenamefont{F\"{o}lling et~al.}(2005)\citenamefont{F\"{o}lling,
  Gerbier, Widera, Mandel, Gericke, and Bloch}}]{Folling2005}
\bibinfo{author}{\bibfnamefont{S.}~\bibnamefont{F\"{o}lling}},
  \bibinfo{author}{\bibfnamefont{F.}~\bibnamefont{Gerbier}},
  \bibinfo{author}{\bibfnamefont{A.}~\bibnamefont{Widera}},
  \bibinfo{author}{\bibfnamefont{O.}~\bibnamefont{Mandel}},
  \bibinfo{author}{\bibfnamefont{T.}~\bibnamefont{Gericke}}, \bibnamefont{and}
  \bibinfo{author}{\bibfnamefont{I.}~\bibnamefont{Bloch}}, \emph{{Spatial
  quantum noise interferometry in expanding ultracold atom clouds.}},
  \bibinfo{journal}{Nature} \textbf{\bibinfo{volume}{434}},
  \bibinfo{pages}{481} (\bibinfo{year}{2005}), ISSN \bibinfo{issn}{1476-4687}.

\bibitem[{\citenamefont{Kinoshita et~al.}(2006)\citenamefont{Kinoshita, Wenger,
  and Weiss}}]{Kinoshita2006}
\bibinfo{author}{\bibfnamefont{T.}~\bibnamefont{Kinoshita}},
  \bibinfo{author}{\bibfnamefont{T.}~\bibnamefont{Wenger}}, \bibnamefont{and}
  \bibinfo{author}{\bibfnamefont{D.~S.} \bibnamefont{Weiss}}, \emph{{A quantum
  Newton's cradle.}}, \bibinfo{journal}{Nature} \textbf{\bibinfo{volume}{440}},
  \bibinfo{pages}{900} (\bibinfo{year}{2006}), ISSN \bibinfo{issn}{1476-4687}.

\bibitem[{\citenamefont{{Rigol} et~al.}(2008)\citenamefont{{Rigol}, {Dunjko},
  and {Olshanii}}}]{Rigol2008}
\bibinfo{author}{\bibfnamefont{M.}~\bibnamefont{{Rigol}}},
  \bibinfo{author}{\bibfnamefont{V.}~\bibnamefont{{Dunjko}}}, \bibnamefont{and}
  \bibinfo{author}{\bibfnamefont{M.}~\bibnamefont{{Olshanii}}},
  \emph{{Thermalization and its mechanism for generic isolated quantum
  systems}}, \bibinfo{journal}{\nat} \textbf{\bibinfo{volume}{452}},
  \bibinfo{pages}{854} (\bibinfo{year}{2008}), \eprint{0708.1324}.

\bibitem[{\citenamefont{Basko et~al.}(2006)\citenamefont{Basko, Aleiner, and
  Altshuler}}]{Basko2006}
\bibinfo{author}{\bibfnamefont{D.}~\bibnamefont{Basko}},
  \bibinfo{author}{\bibfnamefont{I.}~\bibnamefont{Aleiner}}, \bibnamefont{and}
  \bibinfo{author}{\bibfnamefont{B.}~\bibnamefont{Altshuler}},
  \emph{{Metal–insulator transition in a weakly interacting many-electron
  system with localized single-particle states}}, \bibinfo{journal}{Annals of
  Physics} \textbf{\bibinfo{volume}{321}}, \bibinfo{pages}{1126}
  (\bibinfo{year}{2006}), ISSN \bibinfo{issn}{00034916}.

\bibitem[{\citenamefont{Gornyi et~al.}(2005)\citenamefont{Gornyi, Mirlin, and
  Polyakov}}]{Gornyi2005}
\bibinfo{author}{\bibfnamefont{I.~V.} \bibnamefont{Gornyi}},
  \bibinfo{author}{\bibfnamefont{A.~D.} \bibnamefont{Mirlin}},
  \bibnamefont{and} \bibinfo{author}{\bibfnamefont{D.~G.}
  \bibnamefont{Polyakov}}, \emph{Interacting Electrons in Disordered Wires:
  Anderson Localization and Low-$T$ Transport}, \bibinfo{journal}{Phys. Rev.
  Lett.} \textbf{\bibinfo{volume}{95}}, \bibinfo{pages}{206603}
  (\bibinfo{year}{2005}).

\bibitem[{\citenamefont{Anderson}(1958)}]{Anderson1958}
\bibinfo{author}{\bibfnamefont{P.~W.} \bibnamefont{Anderson}}, \emph{Absence of
  Diffusion in Certain Random Lattices}, \bibinfo{journal}{Phys. Rev.}
  \textbf{\bibinfo{volume}{109}}, \bibinfo{pages}{1492} (\bibinfo{year}{1958}).

\bibitem[{\citenamefont{{Kondov} et~al.}(2013)\citenamefont{{Kondov},
  {McGehee}, {Xu}, and {DeMarco}}}]{Kondov2013}
\bibinfo{author}{\bibfnamefont{S.~S.} \bibnamefont{{Kondov}}},
  \bibinfo{author}{\bibfnamefont{W.~R.} \bibnamefont{{McGehee}}},
  \bibinfo{author}{\bibfnamefont{W.}~\bibnamefont{{Xu}}}, \bibnamefont{and}
  \bibinfo{author}{\bibfnamefont{B.}~\bibnamefont{{DeMarco}}},
  \emph{{Disorder-induced Localization in a Strongly Correlated Atomic Hubbard
  Gas}}, \bibinfo{journal}{ArXiv e-prints}  (\bibinfo{year}{2013}),
  \eprint{1305.6072}.

\bibitem[{\citenamefont{Schreiber et~al.}(2015)\citenamefont{Schreiber,
  Hodgman, Bordia, Lüschen, Fischer, Vosk, Altman, Schneider, and
  Bloch}}]{Schreiber2015}
\bibinfo{author}{\bibfnamefont{M.}~\bibnamefont{Schreiber}},
  \bibinfo{author}{\bibfnamefont{S.~S.} \bibnamefont{Hodgman}},
  \bibinfo{author}{\bibfnamefont{P.}~\bibnamefont{Bordia}},
  \bibinfo{author}{\bibfnamefont{H.~P.} \bibnamefont{Lüschen}},
  \bibinfo{author}{\bibfnamefont{M.~H.} \bibnamefont{Fischer}},
  \bibinfo{author}{\bibfnamefont{R.}~\bibnamefont{Vosk}},
  \bibinfo{author}{\bibfnamefont{E.}~\bibnamefont{Altman}},
  \bibinfo{author}{\bibfnamefont{U.}~\bibnamefont{Schneider}},
  \bibnamefont{and} \bibinfo{author}{\bibfnamefont{I.}~\bibnamefont{Bloch}},
  \emph{Observation of many-body localization of interacting fermions in a
  quasirandom optical lattice}, \bibinfo{journal}{Science}
  \textbf{\bibinfo{volume}{349}}, \bibinfo{pages}{842} (\bibinfo{year}{2015}),
  \eprint{http://www.sciencemag.org/content/349/6250/842.full.pdf}.

\bibitem[{\citenamefont{{Schumm} et~al.}(2005)\citenamefont{{Schumm},
  {Hofferberth}, {Andersson}, {Wildermuth}, {Groth}, {Bar-Joseph},
  {Schmiedmayer}, and {Kr{\"u}ger}}}]{Schumm2005}
\bibinfo{author}{\bibfnamefont{T.}~\bibnamefont{{Schumm}}},
  \bibinfo{author}{\bibfnamefont{S.}~\bibnamefont{{Hofferberth}}},
  \bibinfo{author}{\bibfnamefont{L.~M.} \bibnamefont{{Andersson}}},
  \bibinfo{author}{\bibfnamefont{S.}~\bibnamefont{{Wildermuth}}},
  \bibinfo{author}{\bibfnamefont{S.}~\bibnamefont{{Groth}}},
  \bibinfo{author}{\bibfnamefont{I.}~\bibnamefont{{Bar-Joseph}}},
  \bibinfo{author}{\bibfnamefont{J.}~\bibnamefont{{Schmiedmayer}}},
  \bibnamefont{and}
  \bibinfo{author}{\bibfnamefont{P.}~\bibnamefont{{Kr{\"u}ger}}},
  \emph{{Matter-wave interferometry in a double well on an atom chip}},
  \bibinfo{journal}{Nature Physics} \textbf{\bibinfo{volume}{1}},
  \bibinfo{pages}{57} (\bibinfo{year}{2005}), \eprint{quant-ph/0507047}.

\bibitem[{\citenamefont{Widera et~al.}(2008)\citenamefont{Widera, Trotzky,
  Cheinet, F\"olling, Gerbier, Bloch, Gritsev, Lukin, and Demler}}]{Widera2008}
\bibinfo{author}{\bibfnamefont{A.}~\bibnamefont{Widera}},
  \bibinfo{author}{\bibfnamefont{S.}~\bibnamefont{Trotzky}},
  \bibinfo{author}{\bibfnamefont{P.}~\bibnamefont{Cheinet}},
  \bibinfo{author}{\bibfnamefont{S.}~\bibnamefont{F\"olling}},
  \bibinfo{author}{\bibfnamefont{F.}~\bibnamefont{Gerbier}},
  \bibinfo{author}{\bibfnamefont{I.}~\bibnamefont{Bloch}},
  \bibinfo{author}{\bibfnamefont{V.}~\bibnamefont{Gritsev}},
  \bibinfo{author}{\bibfnamefont{M.~D.} \bibnamefont{Lukin}}, \bibnamefont{and}
  \bibinfo{author}{\bibfnamefont{E.}~\bibnamefont{Demler}}, \emph{Quantum Spin
  Dynamics of Mode-Squeezed Luttinger Liquids in Two-Component Atomic Gases},
  \bibinfo{journal}{Phys. Rev. Lett.} \textbf{\bibinfo{volume}{100}},
  \bibinfo{pages}{140401} (\bibinfo{year}{2008}).

\bibitem[{\citenamefont{Hofferberth et~al.}(2007)\citenamefont{Hofferberth,
  Lesanovsky, Fischer, Schumm, and Schmiedmayer}}]{Hofferberth2007}
\bibinfo{author}{\bibfnamefont{S.}~\bibnamefont{Hofferberth}},
  \bibinfo{author}{\bibfnamefont{I.}~\bibnamefont{Lesanovsky}},
  \bibinfo{author}{\bibfnamefont{B.}~\bibnamefont{Fischer}},
  \bibinfo{author}{\bibfnamefont{T.}~\bibnamefont{Schumm}}, \bibnamefont{and}
  \bibinfo{author}{\bibfnamefont{J.}~\bibnamefont{Schmiedmayer}},
  \emph{{Non-equilibrium coherence dynamics in one-dimensional Bose gases.}},
  \bibinfo{journal}{Nature} \textbf{\bibinfo{volume}{449}},
  \bibinfo{pages}{324} (\bibinfo{year}{2007}), ISSN \bibinfo{issn}{1476-4687}.

\bibitem[{\citenamefont{Stern et~al.}(1990)\citenamefont{Stern, Aharonov, and
  Imry}}]{Stern1990}
\bibinfo{author}{\bibfnamefont{A.}~\bibnamefont{Stern}},
  \bibinfo{author}{\bibfnamefont{Y.}~\bibnamefont{Aharonov}}, \bibnamefont{and}
  \bibinfo{author}{\bibfnamefont{Y.}~\bibnamefont{Imry}}, \emph{Phase
  uncertainty and loss of interference: A general picture},
  \bibinfo{journal}{Physical Review A} \textbf{\bibinfo{volume}{41}},
  \bibinfo{pages}{3436} (\bibinfo{year}{1990}).

\bibitem[{\citenamefont{Andrews et~al.}(1997)\citenamefont{Andrews, Townsend,
  Miesner, Durfee, Kurn, and Ketterle}}]{Andrews1997}
\bibinfo{author}{\bibfnamefont{M.~R.} \bibnamefont{Andrews}},
  \bibinfo{author}{\bibfnamefont{C.~G.} \bibnamefont{Townsend}},
  \bibinfo{author}{\bibfnamefont{H.-J.} \bibnamefont{Miesner}},
  \bibinfo{author}{\bibfnamefont{D.~S.} \bibnamefont{Durfee}},
  \bibinfo{author}{\bibfnamefont{D.~M.} \bibnamefont{Kurn}}, \bibnamefont{and}
  \bibinfo{author}{\bibfnamefont{W.}~\bibnamefont{Ketterle}}, \emph{Observation
  of Interference Between Two Bose Condensates}, \bibinfo{journal}{Science}
  \textbf{\bibinfo{volume}{275}}, \bibinfo{pages}{637} (\bibinfo{year}{1997}),
  \eprint{http://www.sciencemag.org/content/275/5300/637.full.pdf}.

\bibitem[{\citenamefont{Castin and Dalibard}(1997)}]{Castin1997}
\bibinfo{author}{\bibfnamefont{Y.}~\bibnamefont{Castin}} \bibnamefont{and}
  \bibinfo{author}{\bibfnamefont{J.}~\bibnamefont{Dalibard}}, \emph{Relative
  phase of two Bose-Einstein condensates}, \bibinfo{journal}{Phys. Rev. A}
  \textbf{\bibinfo{volume}{55}}, \bibinfo{pages}{4330} (\bibinfo{year}{1997}).

\bibitem[{\citenamefont{Leggett and Sols}(1991)}]{Leggett1991}
\bibinfo{author}{\bibfnamefont{A.~J.} \bibnamefont{Leggett}} \bibnamefont{and}
  \bibinfo{author}{\bibfnamefont{F.}~\bibnamefont{Sols}}, \emph{{On the concept
  of spontaneously broken gauge symmetry in condensed matter physics}},
  \bibinfo{journal}{Foundations of Physics} \textbf{\bibinfo{volume}{21}},
  \bibinfo{pages}{353} (\bibinfo{year}{1991}), ISSN \bibinfo{issn}{0015-9018}.

\bibitem[{\citenamefont{Anderson}(1952)}]{Anderson1952}
\bibinfo{author}{\bibfnamefont{P.}~\bibnamefont{Anderson}}, \emph{{An
  Approximate Quantum Theory of the Antiferromagnetic Ground State}},
  \bibinfo{journal}{Physical Review} \textbf{\bibinfo{volume}{86}}
  (\bibinfo{year}{1952}).

\bibitem[{\citenamefont{Jo et~al.}(2007)\citenamefont{Jo, Shin, Will, Pasquini,
  Saba, Ketterle, Pritchard, Vengalattore, and Prentiss}}]{Jo2007}
\bibinfo{author}{\bibfnamefont{G.-B.} \bibnamefont{Jo}},
  \bibinfo{author}{\bibfnamefont{Y.}~\bibnamefont{Shin}},
  \bibinfo{author}{\bibfnamefont{S.}~\bibnamefont{Will}},
  \bibinfo{author}{\bibfnamefont{T.~A.} \bibnamefont{Pasquini}},
  \bibinfo{author}{\bibfnamefont{M.}~\bibnamefont{Saba}},
  \bibinfo{author}{\bibfnamefont{W.}~\bibnamefont{Ketterle}},
  \bibinfo{author}{\bibfnamefont{D.~E.} \bibnamefont{Pritchard}},
  \bibinfo{author}{\bibfnamefont{M.}~\bibnamefont{Vengalattore}},
  \bibnamefont{and} \bibinfo{author}{\bibfnamefont{M.}~\bibnamefont{Prentiss}},
  \emph{Long Phase Coherence Time and Number Squeezing of Two Bose-Einstein
  Condensates on an Atom Chip}, \bibinfo{journal}{Phys. Rev. Lett.}
  \textbf{\bibinfo{volume}{98}}, \bibinfo{pages}{030407}
  (\bibinfo{year}{2007}).

\bibitem[{\citenamefont{Bistritzer and Altman}(2007)}]{Bistritzer2007}
\bibinfo{author}{\bibfnamefont{R.}~\bibnamefont{Bistritzer}} \bibnamefont{and}
  \bibinfo{author}{\bibfnamefont{E.}~\bibnamefont{Altman}}, \emph{{Intrinsic
  dephasing in one-dimensional ultracold atom interferometers.}},
  \bibinfo{journal}{Proceedings of the National Academy of Sciences of the
  United States of America} \textbf{\bibinfo{volume}{104}},
  \bibinfo{pages}{9955} (\bibinfo{year}{2007}), ISSN \bibinfo{issn}{0027-8424}.

\bibitem[{\citenamefont{Popov}(1972)}]{Popov1972}
\bibinfo{author}{\bibfnamefont{V.~N.} \bibnamefont{Popov}}, \emph{On the theory
  of the superfluidity of two-and one-dimensional Bose systems},
  \bibinfo{journal}{Theoretical and Mathematical Physics}
  \textbf{\bibinfo{volume}{11}}, \bibinfo{pages}{565} (\bibinfo{year}{1972}).

\bibitem[{\citenamefont{Smith et~al.}(2013)\citenamefont{Smith, Gring, Langen,
  Kuhnert, Rauer, Geiger, Kitagawa, Mazets, Demler, and
  Schmiedmayer}}]{Smith2013}
\bibinfo{author}{\bibfnamefont{D.~A.} \bibnamefont{Smith}},
  \bibinfo{author}{\bibfnamefont{M.}~\bibnamefont{Gring}},
  \bibinfo{author}{\bibfnamefont{T.}~\bibnamefont{Langen}},
  \bibinfo{author}{\bibfnamefont{M.}~\bibnamefont{Kuhnert}},
  \bibinfo{author}{\bibfnamefont{B.}~\bibnamefont{Rauer}},
  \bibinfo{author}{\bibfnamefont{R.}~\bibnamefont{Geiger}},
  \bibinfo{author}{\bibfnamefont{T.}~\bibnamefont{Kitagawa}},
  \bibinfo{author}{\bibfnamefont{I.}~\bibnamefont{Mazets}},
  \bibinfo{author}{\bibfnamefont{E.}~\bibnamefont{Demler}}, \bibnamefont{and}
  \bibinfo{author}{\bibfnamefont{J.}~\bibnamefont{Schmiedmayer}},
  \emph{Prethermalization revealed by the relaxation dynamics of full
  distribution functions}, \bibinfo{journal}{New Journal of Physics}
  \textbf{\bibinfo{volume}{15}}, \bibinfo{pages}{075011}
  (\bibinfo{year}{2013}).

\bibitem[{\citenamefont{Kitagawa et~al.}(2011)\citenamefont{Kitagawa,
  Imambekov, Schmiedmayer, and Demler}}]{Kitagawa2011}
\bibinfo{author}{\bibfnamefont{T.}~\bibnamefont{Kitagawa}},
  \bibinfo{author}{\bibfnamefont{A.}~\bibnamefont{Imambekov}},
  \bibinfo{author}{\bibfnamefont{J.}~\bibnamefont{Schmiedmayer}},
  \bibnamefont{and} \bibinfo{author}{\bibfnamefont{E.}~\bibnamefont{Demler}},
  \emph{The dynamics and prethermalization of one-dimensional quantum systems
  probed through the full distributions of quantum noise},
  \bibinfo{journal}{New Journal of Physics} \textbf{\bibinfo{volume}{13}},
  \bibinfo{pages}{073018} (\bibinfo{year}{2011}).

\bibitem[{\citenamefont{Bistritzer}(2007)}]{Rafi-Thesis}
\bibinfo{author}{\bibfnamefont{R.}~\bibnamefont{Bistritzer}}, Ph.D. thesis,
  \bibinfo{school}{Weizmann Institute of Science} (\bibinfo{year}{2007}).

\bibitem[{\citenamefont{Haldane}(1981)}]{Haldane1981}
\bibinfo{author}{\bibfnamefont{F.~D.~M.} \bibnamefont{Haldane}},
  \emph{{Effective Harmonic-Fluid Approach to Low-Energy Properties of
  One-Dimensional Quantum Fluids}}, \bibinfo{journal}{Physical Review Letters}
  \textbf{\bibinfo{volume}{47}}, \bibinfo{pages}{1840} (\bibinfo{year}{1981}),
  ISSN \bibinfo{issn}{0031-9007}.

\bibitem[{\citenamefont{Burkov et~al.}(2007)\citenamefont{Burkov, Lukin, and
  Demler}}]{Burkov2007}
\bibinfo{author}{\bibfnamefont{A.}~\bibnamefont{Burkov}},
  \bibinfo{author}{\bibfnamefont{M.}~\bibnamefont{Lukin}}, \bibnamefont{and}
  \bibinfo{author}{\bibfnamefont{E.}~\bibnamefont{Demler}}, \emph{{Decoherence
  Dynamics in Low-Dimensional Cold Atom Interferometers}},
  \bibinfo{journal}{Physical Review Letters} \textbf{\bibinfo{volume}{98}},
  \bibinfo{pages}{200404} (\bibinfo{year}{2007}), ISSN
  \bibinfo{issn}{0031-9007}.

\bibitem[{\citenamefont{Arzamasovs et~al.}(2013)\citenamefont{Arzamasovs, Bovo,
  and Gangardt}}]{Arzamasovs2013}
\bibinfo{author}{\bibfnamefont{M.}~\bibnamefont{Arzamasovs}},
  \bibinfo{author}{\bibfnamefont{F.}~\bibnamefont{Bovo}}, \bibnamefont{and}
  \bibinfo{author}{\bibfnamefont{D.~M.} \bibnamefont{Gangardt}},
  \emph{{Kinetics of mobile impurities and correlation functions in
  one-dimensional superfluids at finite temperature}}, p.~\bibinfo{pages}{5}
  (\bibinfo{year}{2013}), \eprint{1309.2647}.

\bibitem[{\citenamefont{Andreev}(1980)}]{Andreev1980}
\bibinfo{author}{\bibfnamefont{D.}~\bibnamefont{Andreev}}, \emph{{Quantum
  dynamics of impurities in a 1D Bose gas}}, \bibinfo{journal}{Zh. Eksp. Teor.
  Fiz.} \textbf{\bibinfo{volume}{78}}, \bibinfo{pages}{2064}
  (\bibinfo{year}{1980}).

\bibitem[{\citenamefont{Kulkarni and Lamacraft}(2013)}]{Kulkarni2013}
\bibinfo{author}{\bibfnamefont{M.}~\bibnamefont{Kulkarni}} \bibnamefont{and}
  \bibinfo{author}{\bibfnamefont{A.}~\bibnamefont{Lamacraft}},
  \emph{{Finite-temperature dynamical structure factor of the one-dimensional
  Bose gas: From the Gross-Pitaevskii equation to the Kardar-Parisi-Zhang
  universality class of dynamical critical phenomena}},
  \bibinfo{journal}{Physical Review A} \textbf{\bibinfo{volume}{88}},
  \bibinfo{pages}{021603} (\bibinfo{year}{2013}), ISSN
  \bibinfo{issn}{1050-2947}.

\bibitem[{\citenamefont{Mazets et~al.}(2008)\citenamefont{Mazets, Schumm, and
  Schmiedmayer}}]{Mazets2008}
\bibinfo{author}{\bibfnamefont{I.}~\bibnamefont{Mazets}},
  \bibinfo{author}{\bibfnamefont{T.}~\bibnamefont{Schumm}}, \bibnamefont{and}
  \bibinfo{author}{\bibfnamefont{J.}~\bibnamefont{Schmiedmayer}},
  \emph{{Breakdown of Integrability in a Quasi-1D Ultracold Bosonic Gas}},
  \bibinfo{journal}{Physical Review Letters} \textbf{\bibinfo{volume}{100}},
  \bibinfo{pages}{2} (\bibinfo{year}{2008}), ISSN \bibinfo{issn}{0031-9007}.

\bibitem[{\citenamefont{Greiner
  et~al.}(2002{\natexlab{a}})\citenamefont{Greiner, Mandel, Esslinger,
  H\"{a}nsch, and Bloch}}]{Greiner2002a}
\bibinfo{author}{\bibfnamefont{M.}~\bibnamefont{Greiner}},
  \bibinfo{author}{\bibfnamefont{O.}~\bibnamefont{Mandel}},
  \bibinfo{author}{\bibfnamefont{T.}~\bibnamefont{Esslinger}},
  \bibinfo{author}{\bibfnamefont{T.~W.} \bibnamefont{H\"{a}nsch}},
  \bibnamefont{and} \bibinfo{author}{\bibfnamefont{I.}~\bibnamefont{Bloch}},
  \emph{{Quantum phase transition from a superfluid to a Mott insulator in a
  gas of ultracold atoms.}}, \bibinfo{journal}{Nature}
  \textbf{\bibinfo{volume}{415}}, \bibinfo{pages}{39}
  (\bibinfo{year}{2002}{\natexlab{a}}), ISSN \bibinfo{issn}{0028-0836}.

\bibitem[{\citenamefont{Giamarchi and Schulz}(1988)}]{Giamarchi1988}
\bibinfo{author}{\bibfnamefont{T.}~\bibnamefont{Giamarchi}} \bibnamefont{and}
  \bibinfo{author}{\bibfnamefont{H.~J.} \bibnamefont{Schulz}}, \emph{Anderson
  localization and interactions in one-dimensional metals},
  \bibinfo{journal}{Phys. Rev. B} \textbf{\bibinfo{volume}{37}},
  \bibinfo{pages}{325} (\bibinfo{year}{1988}).

\bibitem[{\citenamefont{Fisher}(1992)}]{Fisher1992}
\bibinfo{author}{\bibfnamefont{D.~S.} \bibnamefont{Fisher}}, \emph{Random
  transverse field Ising spin chains}, \bibinfo{journal}{Phys. Rev. Lett.}
  \textbf{\bibinfo{volume}{69}}, \bibinfo{pages}{534} (\bibinfo{year}{1992}).

\bibitem[{\citenamefont{Jaksch et~al.}(1998)\citenamefont{Jaksch, Bruder,
  Cirac, Gardiner, and Zoller}}]{Jaksch1998}
\bibinfo{author}{\bibfnamefont{D.}~\bibnamefont{Jaksch}},
  \bibinfo{author}{\bibfnamefont{C.}~\bibnamefont{Bruder}},
  \bibinfo{author}{\bibfnamefont{J.~I.} \bibnamefont{Cirac}},
  \bibinfo{author}{\bibfnamefont{C.~W.} \bibnamefont{Gardiner}},
  \bibnamefont{and} \bibinfo{author}{\bibfnamefont{P.}~\bibnamefont{Zoller}},
  \emph{Cold Bosonic Atoms in Optical Lattices}, \bibinfo{journal}{Phys. Rev.
  Lett.} \textbf{\bibinfo{volume}{81}}, \bibinfo{pages}{3108}
  (\bibinfo{year}{1998}).

\bibitem[{\citenamefont{Bloch et~al.}(2008)\citenamefont{Bloch, Dalibard, and
  Zwerger}}]{Bloch2008}
\bibinfo{author}{\bibfnamefont{I.}~\bibnamefont{Bloch}},
  \bibinfo{author}{\bibfnamefont{J.}~\bibnamefont{Dalibard}}, \bibnamefont{and}
  \bibinfo{author}{\bibfnamefont{W.}~\bibnamefont{Zwerger}}, \emph{{Many-body
  physics with ultracold gases}}, \bibinfo{journal}{Reviews of Modern Physics}
  \textbf{\bibinfo{volume}{80}}, \bibinfo{pages}{885} (\bibinfo{year}{2008}),
  ISSN \bibinfo{issn}{0034-6861}.

\bibitem[{\citenamefont{Sachdev}(2001)}]{Sachdev-Book}
\bibinfo{author}{\bibfnamefont{S.}~\bibnamefont{Sachdev}},
  \emph{\bibinfo{title}{Quantum Phase Transitions}}
  (\bibinfo{publisher}{Cambridge University Press}, \bibinfo{year}{2001}), ISBN
  \bibinfo{isbn}{9780521004541}.

\bibitem[{\citenamefont{Podolsky et~al.}(2011)\citenamefont{Podolsky, Auerbach,
  and Arovas}}]{Podolsky2011}
\bibinfo{author}{\bibfnamefont{D.}~\bibnamefont{Podolsky}},
  \bibinfo{author}{\bibfnamefont{A.}~\bibnamefont{Auerbach}}, \bibnamefont{and}
  \bibinfo{author}{\bibfnamefont{D.~P.} \bibnamefont{Arovas}}, \emph{Visibility
  of the amplitude (Higgs) mode in condensed matter}, \bibinfo{journal}{Phys.
  Rev. B} \textbf{\bibinfo{volume}{84}}, \bibinfo{pages}{174522}
  (\bibinfo{year}{2011}).

\bibitem[{\citenamefont{Podolsky and Sachdev}(2012)}]{Podolsky2012}
\bibinfo{author}{\bibfnamefont{D.}~\bibnamefont{Podolsky}} \bibnamefont{and}
  \bibinfo{author}{\bibfnamefont{S.}~\bibnamefont{Sachdev}}, \emph{Spectral
  functions of the Higgs mode near two-dimensional quantum critical points},
  \bibinfo{journal}{Phys. Rev. B} \textbf{\bibinfo{volume}{86}},
  \bibinfo{pages}{054508} (\bibinfo{year}{2012}).

\bibitem[{\citenamefont{Gazit et~al.}(2013)\citenamefont{Gazit, Podolsky, and
  Auerbach}}]{Gazit2013}
\bibinfo{author}{\bibfnamefont{S.}~\bibnamefont{Gazit}},
  \bibinfo{author}{\bibfnamefont{D.}~\bibnamefont{Podolsky}}, \bibnamefont{and}
  \bibinfo{author}{\bibfnamefont{A.}~\bibnamefont{Auerbach}}, \emph{Fate of the
  Higgs Mode Near Quantum Criticality}, \bibinfo{journal}{Phys. Rev. Lett.}
  \textbf{\bibinfo{volume}{110}}, \bibinfo{pages}{140401}
  (\bibinfo{year}{2013}).

\bibitem[{\citenamefont{Endres et~al.}(2012)\citenamefont{Endres, Fukuhara,
  Pekker, Cheneau, Schauss, Gross, Demler, Kuhr, and Bloch}}]{Endres2012}
\bibinfo{author}{\bibfnamefont{M.}~\bibnamefont{Endres}},
  \bibinfo{author}{\bibfnamefont{T.}~\bibnamefont{Fukuhara}},
  \bibinfo{author}{\bibfnamefont{D.}~\bibnamefont{Pekker}},
  \bibinfo{author}{\bibfnamefont{M.}~\bibnamefont{Cheneau}},
  \bibinfo{author}{\bibfnamefont{P.}~\bibnamefont{Schauss}},
  \bibinfo{author}{\bibfnamefont{C.}~\bibnamefont{Gross}},
  \bibinfo{author}{\bibfnamefont{E.}~\bibnamefont{Demler}},
  \bibinfo{author}{\bibfnamefont{S.}~\bibnamefont{Kuhr}}, \bibnamefont{and}
  \bibinfo{author}{\bibfnamefont{I.}~\bibnamefont{Bloch}}, \emph{{The 'Higgs'
  amplitude mode at the two-dimensional superfluid/Mott insulator
  transition.}}, \bibinfo{journal}{Nature} \textbf{\bibinfo{volume}{487}},
  \bibinfo{pages}{454} (\bibinfo{year}{2012}), ISSN \bibinfo{issn}{1476-4687}.

\bibitem[{\citenamefont{Schori et~al.}(2004)\citenamefont{Schori, St\"oferle,
  Moritz, K\"ohl, and Esslinger}}]{Schori2004}
\bibinfo{author}{\bibfnamefont{C.}~\bibnamefont{Schori}},
  \bibinfo{author}{\bibfnamefont{T.}~\bibnamefont{St\"oferle}},
  \bibinfo{author}{\bibfnamefont{H.}~\bibnamefont{Moritz}},
  \bibinfo{author}{\bibfnamefont{M.}~\bibnamefont{K\"ohl}}, \bibnamefont{and}
  \bibinfo{author}{\bibfnamefont{T.}~\bibnamefont{Esslinger}},
  \emph{Excitations of a Superfluid in a Three-Dimensional Optical Lattice},
  \bibinfo{journal}{Phys. Rev. Lett.} \textbf{\bibinfo{volume}{93}},
  \bibinfo{pages}{240402} (\bibinfo{year}{2004}).

\bibitem[{\citenamefont{Pollet and Prokof'ev}(2012)}]{Pollet2012}
\bibinfo{author}{\bibfnamefont{L.}~\bibnamefont{Pollet}} \bibnamefont{and}
  \bibinfo{author}{\bibfnamefont{N.}~\bibnamefont{Prokof'ev}}, \emph{Higgs Mode
  in a Two-Dimensional Superfluid}, \bibinfo{journal}{Phys. Rev. Lett.}
  \textbf{\bibinfo{volume}{109}}, \bibinfo{pages}{010401}
  (\bibinfo{year}{2012}).

\bibitem[{\citenamefont{Ran\ifmmode~\mbox{\c{c}}\else \c{c}\fi{}on and
  Dupuis}(2014)}]{Rancon2014}
\bibinfo{author}{\bibfnamefont{A.}~\bibnamefont{Ran\ifmmode~\mbox{\c{c}}\else
  \c{c}\fi{}on}} \bibnamefont{and}
  \bibinfo{author}{\bibfnamefont{N.}~\bibnamefont{Dupuis}}, \emph{Higgs
  amplitude mode in the vicinity of a $(2+1)$-dimensional quantum critical
  point}, \bibinfo{journal}{Phys. Rev. B} \textbf{\bibinfo{volume}{89}},
  \bibinfo{pages}{180501} (\bibinfo{year}{2014}).

\bibitem[{\citenamefont{Katan and Podolsky}(2015)}]{Katan2015}
\bibinfo{author}{\bibfnamefont{Y.~T.} \bibnamefont{Katan}} \bibnamefont{and}
  \bibinfo{author}{\bibfnamefont{D.}~\bibnamefont{Podolsky}}, \emph{Spectral
  function of the Higgs mode in 4-epsilon dimensions}, \bibinfo{journal}{Phys.
  Rev. B} \textbf{\bibinfo{volume}{91}}, \bibinfo{pages}{075132}
  (\bibinfo{year}{2015}).

\bibitem[{\citenamefont{Wu and Niu}(2001)}]{Wu2001}
\bibinfo{author}{\bibfnamefont{B.}~\bibnamefont{Wu}} \bibnamefont{and}
  \bibinfo{author}{\bibfnamefont{Q.}~\bibnamefont{Niu}}, \emph{Landau and
  dynamical instabilities of the superflow of Bose-Einstein condensates in
  optical lattices}, \bibinfo{journal}{Phys. Rev. A}
  \textbf{\bibinfo{volume}{64}}, \bibinfo{pages}{061603}
  (\bibinfo{year}{2001}).

\bibitem[{\citenamefont{Fallani et~al.}(2004)\citenamefont{Fallani, De~Sarlo,
  Lye, Modugno, Saers, Fort, and Inguscio}}]{Fallani2004}
\bibinfo{author}{\bibfnamefont{L.}~\bibnamefont{Fallani}},
  \bibinfo{author}{\bibfnamefont{L.}~\bibnamefont{De~Sarlo}},
  \bibinfo{author}{\bibfnamefont{J.~E.} \bibnamefont{Lye}},
  \bibinfo{author}{\bibfnamefont{M.}~\bibnamefont{Modugno}},
  \bibinfo{author}{\bibfnamefont{R.}~\bibnamefont{Saers}},
  \bibinfo{author}{\bibfnamefont{C.}~\bibnamefont{Fort}}, \bibnamefont{and}
  \bibinfo{author}{\bibfnamefont{M.}~\bibnamefont{Inguscio}}, \emph{Observation
  of Dynamical Instability for a Bose-Einstein Condensate in a Moving 1D
  Optical Lattice}, \bibinfo{journal}{Phys. Rev. Lett.}
  \textbf{\bibinfo{volume}{93}}, \bibinfo{pages}{140406}
  (\bibinfo{year}{2004}).

\bibitem[{\citenamefont{Altman et~al.}(2005)\citenamefont{Altman, Polkovnikov,
  Demler, Halperin, and Lukin}}]{Altman2005a}
\bibinfo{author}{\bibfnamefont{E.}~\bibnamefont{Altman}},
  \bibinfo{author}{\bibfnamefont{A.}~\bibnamefont{Polkovnikov}},
  \bibinfo{author}{\bibfnamefont{E.}~\bibnamefont{Demler}},
  \bibinfo{author}{\bibfnamefont{B.~I.} \bibnamefont{Halperin}},
  \bibnamefont{and} \bibinfo{author}{\bibfnamefont{M.~D.} \bibnamefont{Lukin}},
  \emph{Superfluid-Insulator Transition in a Moving System of Interacting
  Bosons}, \bibinfo{journal}{Phys. Rev. Lett.} \textbf{\bibinfo{volume}{95}},
  \bibinfo{pages}{020402} (\bibinfo{year}{2005}).

\bibitem[{\citenamefont{Mun et~al.}(2007)\citenamefont{Mun, Medley, Campbell,
  Marcassa, Pritchard, and Ketterle}}]{Mun2007}
\bibinfo{author}{\bibfnamefont{J.}~\bibnamefont{Mun}},
  \bibinfo{author}{\bibfnamefont{P.}~\bibnamefont{Medley}},
  \bibinfo{author}{\bibfnamefont{G.~K.} \bibnamefont{Campbell}},
  \bibinfo{author}{\bibfnamefont{L.~G.} \bibnamefont{Marcassa}},
  \bibinfo{author}{\bibfnamefont{D.~E.} \bibnamefont{Pritchard}},
  \bibnamefont{and} \bibinfo{author}{\bibfnamefont{W.}~\bibnamefont{Ketterle}},
  \emph{Phase Diagram for a Bose-Einstein Condensate Moving in an Optical
  Lattice}, \bibinfo{journal}{Phys. Rev. Lett.} \textbf{\bibinfo{volume}{99}},
  \bibinfo{pages}{150604} (\bibinfo{year}{2007}).

\bibitem[{\citenamefont{Coleman}(1977{\natexlab{a}})}]{Coleman1977}
\bibinfo{author}{\bibfnamefont{S.}~\bibnamefont{Coleman}}, \emph{Fate of the
  false vacuum: Semiclassical theory}, \bibinfo{journal}{Physical Review D}
  \textbf{\bibinfo{volume}{15}}, \bibinfo{pages}{2929}
  (\bibinfo{year}{1977}{\natexlab{a}}).

\bibitem[{\citenamefont{Coleman}(1977{\natexlab{b}})}]{Coleman1977erratum}
\bibinfo{author}{\bibfnamefont{S.}~\bibnamefont{Coleman}}, \emph{Erratum: Fate
  of the false vacuum: semiclassical theory}, \bibinfo{journal}{Physical Review
  D} \textbf{\bibinfo{volume}{16}}, \bibinfo{pages}{1248}
  (\bibinfo{year}{1977}{\natexlab{b}}).

\bibitem[{\citenamefont{Callan~Jr and Coleman}(1977)}]{Callan1977}
\bibinfo{author}{\bibfnamefont{C.~G.} \bibnamefont{Callan~Jr}}
  \bibnamefont{and} \bibinfo{author}{\bibfnamefont{S.}~\bibnamefont{Coleman}},
  \emph{Fate of the false vacuum. II. First quantum corrections},
  \bibinfo{journal}{Physical Review D} \textbf{\bibinfo{volume}{16}},
  \bibinfo{pages}{1762} (\bibinfo{year}{1977}).

\bibitem[{\citenamefont{Langer and Ambegaokar}(1967)}]{Langer1967}
\bibinfo{author}{\bibfnamefont{J.~S.} \bibnamefont{Langer}} \bibnamefont{and}
  \bibinfo{author}{\bibfnamefont{V.}~\bibnamefont{Ambegaokar}}, \emph{Intrinsic
  Resistive Transition in Narrow Superconducting Channels},
  \bibinfo{journal}{Phys. Rev.} \textbf{\bibinfo{volume}{164}},
  \bibinfo{pages}{498} (\bibinfo{year}{1967}).

\bibitem[{\citenamefont{McCumber and Halperin}(1970)}]{McCumber1970}
\bibinfo{author}{\bibfnamefont{D.~E.} \bibnamefont{McCumber}} \bibnamefont{and}
  \bibinfo{author}{\bibfnamefont{B.~I.} \bibnamefont{Halperin}}, \emph{Time
  Scale of Intrinsic Resistive Fluctuations in Thin Superconducting Wires},
  \bibinfo{journal}{Phys. Rev. B} \textbf{\bibinfo{volume}{1}},
  \bibinfo{pages}{1054} (\bibinfo{year}{1970}).

\bibitem[{\citenamefont{Polkovnikov et~al.}(2005)\citenamefont{Polkovnikov,
  Altman, Demler, Halperin, and Lukin}}]{Polkovnikov2005}
\bibinfo{author}{\bibfnamefont{A.}~\bibnamefont{Polkovnikov}},
  \bibinfo{author}{\bibfnamefont{E.}~\bibnamefont{Altman}},
  \bibinfo{author}{\bibfnamefont{E.}~\bibnamefont{Demler}},
  \bibinfo{author}{\bibfnamefont{B.}~\bibnamefont{Halperin}}, \bibnamefont{and}
  \bibinfo{author}{\bibfnamefont{M.~D.} \bibnamefont{Lukin}}, \emph{Decay of
  superfluid currents in a moving system of strongly interacting bosons},
  \bibinfo{journal}{Phys. Rev. A} \textbf{\bibinfo{volume}{71}},
  \bibinfo{pages}{063613} (\bibinfo{year}{2005}).

\bibitem[{\citenamefont{Alexander and Simons}(2010)}]{AltlandBook}
\bibinfo{author}{\bibfnamefont{A.}~\bibnamefont{Alexander}} \bibnamefont{and}
  \bibinfo{author}{\bibfnamefont{B.}~\bibnamefont{Simons}}, \emph{Condensed
  Matter Field Theory}, \bibinfo{journal}{Cambridge: Cambridge University
  Press} \textbf{\bibinfo{volume}{227}}, \bibinfo{pages}{310}
  (\bibinfo{year}{2010}).

\bibitem[{\citenamefont{Cazalilla et~al.}(2011)\citenamefont{Cazalilla, Citro,
  Giamarchi, Orignac, and Rigol}}]{Cazalilla2011}
\bibinfo{author}{\bibfnamefont{M.~A.} \bibnamefont{Cazalilla}},
  \bibinfo{author}{\bibfnamefont{R.}~\bibnamefont{Citro}},
  \bibinfo{author}{\bibfnamefont{T.}~\bibnamefont{Giamarchi}},
  \bibinfo{author}{\bibfnamefont{E.}~\bibnamefont{Orignac}}, \bibnamefont{and}
  \bibinfo{author}{\bibfnamefont{M.}~\bibnamefont{Rigol}}, \emph{One
  dimensional bosons: From condensed matter systems to ultracold gases},
  \bibinfo{journal}{Rev. Mod. Phys.} \textbf{\bibinfo{volume}{83}},
  \bibinfo{pages}{1405} (\bibinfo{year}{2011}).

\bibitem[{\citenamefont{Greiner
  et~al.}(2002{\natexlab{b}})\citenamefont{Greiner, Mandel, H\"{a}nsch, and
  Bloch}}]{Greiner2002b}
\bibinfo{author}{\bibfnamefont{M.}~\bibnamefont{Greiner}},
  \bibinfo{author}{\bibfnamefont{O.}~\bibnamefont{Mandel}},
  \bibinfo{author}{\bibfnamefont{T.~W.} \bibnamefont{H\"{a}nsch}},
  \bibnamefont{and} \bibinfo{author}{\bibfnamefont{I.}~\bibnamefont{Bloch}},
  \emph{{Collapse and revival of the matter wave field of a Bose-Einstein
  condensate}}, \bibinfo{journal}{$\backslash$nat}
  \textbf{\bibinfo{volume}{419}}, \bibinfo{pages}{51}
  (\bibinfo{year}{2002}{\natexlab{b}}).

\bibitem[{\citenamefont{Sciolla and Biroli}(2010)}]{Sciolla2010}
\bibinfo{author}{\bibfnamefont{B.}~\bibnamefont{Sciolla}} \bibnamefont{and}
  \bibinfo{author}{\bibfnamefont{G.}~\bibnamefont{Biroli}}, \emph{{Quantum
  Quenches and Off-Equilibrium Dynamical Transition in the Infinite-Dimensional
  Bose-Hubbard Model}}, \bibinfo{journal}{Physical Review Letters}
  \textbf{\bibinfo{volume}{105}}, \bibinfo{pages}{2} (\bibinfo{year}{2010}),
  ISSN \bibinfo{issn}{0031-9007}.

\bibitem[{\citenamefont{Polkovnikov}(2003)}]{Polkovnikov2003}
\bibinfo{author}{\bibfnamefont{A.}~\bibnamefont{Polkovnikov}}, \emph{Evolution
  of the macroscopically entangled states in optical lattices},
  \bibinfo{journal}{Phys. Rev. A} \textbf{\bibinfo{volume}{68}},
  \bibinfo{pages}{033609} (\bibinfo{year}{2003}).

\bibitem[{\citenamefont{{Lamacraft} and
  {Moore}}(2011)}]{Lamacraft-Moore-Review}
\bibinfo{author}{\bibfnamefont{A.}~\bibnamefont{{Lamacraft}}} \bibnamefont{and}
  \bibinfo{author}{\bibfnamefont{J.}~\bibnamefont{{Moore}}}, \emph{{Potential
  insights into non-equilibrium behavior from atomic physics}},
  \bibinfo{journal}{ArXiv e-prints}  (\bibinfo{year}{2011}),
  \eprint{1106.3567}.

\bibitem[{\citenamefont{Lamacraft}(2007)}]{Lamacraft2007}
\bibinfo{author}{\bibfnamefont{A.}~\bibnamefont{Lamacraft}}, \emph{{Quantum
  Quenches in a Spinor Condensate}}, \bibinfo{journal}{Physical Review Letters}
  \textbf{\bibinfo{volume}{98}}, \bibinfo{pages}{1} (\bibinfo{year}{2007}),
  ISSN \bibinfo{issn}{0031-9007}.

\bibitem[{\citenamefont{Sadler et~al.}(2006)\citenamefont{Sadler, Higbie,
  Leslie, Vengalattore, and Stamper-Kurn}}]{sadler2006}
\bibinfo{author}{\bibfnamefont{L.~E.} \bibnamefont{Sadler}},
  \bibinfo{author}{\bibfnamefont{J.~M.} \bibnamefont{Higbie}},
  \bibinfo{author}{\bibfnamefont{S.~R.} \bibnamefont{Leslie}},
  \bibinfo{author}{\bibfnamefont{M.}~\bibnamefont{Vengalattore}},
  \bibnamefont{and} \bibinfo{author}{\bibfnamefont{D.~M.}
  \bibnamefont{Stamper-Kurn}}, \emph{{Spontaneous symmetry breaking in a
  quenched ferromagnetic spinor Bose-Einstein condensate.}},
  \bibinfo{journal}{Nature} \textbf{\bibinfo{volume}{443}},
  \bibinfo{pages}{312} (\bibinfo{year}{2006}), ISSN \bibinfo{issn}{1476-4687}.

\bibitem[{\citenamefont{Liu and Mazenko}(1992)}]{Liu1992}
\bibinfo{author}{\bibfnamefont{F.}~\bibnamefont{Liu}} \bibnamefont{and}
  \bibinfo{author}{\bibfnamefont{G.~F.} \bibnamefont{Mazenko}},
  \emph{Defect-defect correlation in the dynamics of first-order phase
  transitions}, \bibinfo{journal}{Phys. Rev. B} \textbf{\bibinfo{volume}{46}},
  \bibinfo{pages}{5963} (\bibinfo{year}{1992}).

\bibitem[{\citenamefont{Hohenberg and Halperin}(1977)}]{Hohenberg1977}
\bibinfo{author}{\bibfnamefont{P.}~\bibnamefont{Hohenberg}} \bibnamefont{and}
  \bibinfo{author}{\bibfnamefont{B.}~\bibnamefont{Halperin}}, \emph{{Theory of
  dynamic critical phenomena}}, \bibinfo{journal}{Reviews of Modern Physics}
  \textbf{\bibinfo{volume}{49}}, \bibinfo{pages}{435} (\bibinfo{year}{1977}),
  ISSN \bibinfo{issn}{0034-6861}.

\bibitem[{\citenamefont{Polkovnikov}(2005)}]{Polkovnikov2005universal}
\bibinfo{author}{\bibfnamefont{A.}~\bibnamefont{Polkovnikov}}, \emph{Universal
  adiabatic dynamics in the vicinity of a quantum critical point},
  \bibinfo{journal}{Physical Review B} \textbf{\bibinfo{volume}{72}},
  \bibinfo{pages}{161201} (\bibinfo{year}{2005}).

\bibitem[{\citenamefont{Zurek}(1985)}]{Zurek1985}
\bibinfo{author}{\bibfnamefont{W.}~\bibnamefont{Zurek}}, \emph{Cosmological
  experiments in superfluid helium?}, \bibinfo{journal}{Nature}
  \textbf{\bibinfo{volume}{317}}, \bibinfo{pages}{505} (\bibinfo{year}{1985}).

\bibitem[{\citenamefont{{Schollw{\"o}ck}}(2011)}]{Schol2011}
\bibinfo{author}{\bibfnamefont{U.}~\bibnamefont{{Schollw{\"o}ck}}}, \emph{{The
  density-matrix renormalization group in the age of matrix product states}},
  \bibinfo{journal}{Annals of Physics} \textbf{\bibinfo{volume}{326}},
  \bibinfo{pages}{96} (\bibinfo{year}{2011}), \eprint{1008.3477}.

\bibitem[{\citenamefont{Lieb and Robinson}(1972)}]{Lieb1972}
\bibinfo{author}{\bibfnamefont{E.}~\bibnamefont{Lieb}} \bibnamefont{and}
  \bibinfo{author}{\bibfnamefont{D.}~\bibnamefont{Robinson}}, \emph{The finite
  group velocity of quantum spin systems}, \bibinfo{journal}{Communications in
  Mathematical Physics} \textbf{\bibinfo{volume}{28}}, \bibinfo{pages}{251}
  (\bibinfo{year}{1972}), ISSN \bibinfo{issn}{0010-3616}.

\bibitem[{\citenamefont{Calabrese and Cardy}(2005)}]{calabrese2005evolution}
\bibinfo{author}{\bibfnamefont{P.}~\bibnamefont{Calabrese}} \bibnamefont{and}
  \bibinfo{author}{\bibfnamefont{J.}~\bibnamefont{Cardy}}, \emph{Evolution of
  entanglement entropy in one-dimensional systems}, \bibinfo{journal}{Journal
  of Statistical Mechanics: Theory and Experiment}
  \textbf{\bibinfo{volume}{2005}}, \bibinfo{pages}{P04010}
  (\bibinfo{year}{2005}).

\bibitem[{\citenamefont{Bravyi et~al.}(2006)\citenamefont{Bravyi, Hastings, and
  Verstraete}}]{Bravyi2006}
\bibinfo{author}{\bibfnamefont{S.}~\bibnamefont{Bravyi}},
  \bibinfo{author}{\bibfnamefont{M.}~\bibnamefont{Hastings}}, \bibnamefont{and}
  \bibinfo{author}{\bibfnamefont{F.}~\bibnamefont{Verstraete}},
  \emph{Lieb-Robinson bounds and the generation of correlations and topological
  quantum order}, \bibinfo{journal}{Physical review letters}
  \textbf{\bibinfo{volume}{97}}, \bibinfo{pages}{050401}
  (\bibinfo{year}{2006}).

\bibitem[{\citenamefont{Cheneau et~al.}(2012)\citenamefont{Cheneau, Barmettler,
  Poletti, Endres, Schau{\ss}, Fukuhara, Gross, Bloch, Kollath, and
  Kuhr}}]{Cheneau2012}
\bibinfo{author}{\bibfnamefont{M.}~\bibnamefont{Cheneau}},
  \bibinfo{author}{\bibfnamefont{P.}~\bibnamefont{Barmettler}},
  \bibinfo{author}{\bibfnamefont{D.}~\bibnamefont{Poletti}},
  \bibinfo{author}{\bibfnamefont{M.}~\bibnamefont{Endres}},
  \bibinfo{author}{\bibfnamefont{P.}~\bibnamefont{Schau{\ss}}},
  \bibinfo{author}{\bibfnamefont{T.}~\bibnamefont{Fukuhara}},
  \bibinfo{author}{\bibfnamefont{C.}~\bibnamefont{Gross}},
  \bibinfo{author}{\bibfnamefont{I.}~\bibnamefont{Bloch}},
  \bibinfo{author}{\bibfnamefont{C.}~\bibnamefont{Kollath}}, \bibnamefont{and}
  \bibinfo{author}{\bibfnamefont{S.}~\bibnamefont{Kuhr}}, \emph{Light-cone-like
  spreading of correlations in a quantum many-body system},
  \bibinfo{journal}{Nature} \textbf{\bibinfo{volume}{481}},
  \bibinfo{pages}{484} (\bibinfo{year}{2012}).

\bibitem[{\citenamefont{{Trotzky} et~al.}(2012)\citenamefont{{Trotzky}, {Chen},
  {Flesch}, {McCulloch}, {Schollw{\"o}ck}, {Eisert}, and
  {Bloch}}}]{Trotzky2012}
\bibinfo{author}{\bibfnamefont{S.}~\bibnamefont{{Trotzky}}},
  \bibinfo{author}{\bibfnamefont{Y.-A.} \bibnamefont{{Chen}}},
  \bibinfo{author}{\bibfnamefont{A.}~\bibnamefont{{Flesch}}},
  \bibinfo{author}{\bibfnamefont{I.~P.} \bibnamefont{{McCulloch}}},
  \bibinfo{author}{\bibfnamefont{U.}~\bibnamefont{{Schollw{\"o}ck}}},
  \bibinfo{author}{\bibfnamefont{J.}~\bibnamefont{{Eisert}}}, \bibnamefont{and}
  \bibinfo{author}{\bibfnamefont{I.}~\bibnamefont{{Bloch}}}, \emph{{Probing the
  relaxation towards equilibrium in an isolated strongly correlated
  one-dimensional Bose gas}}, \bibinfo{journal}{Nature Physics}
  \textbf{\bibinfo{volume}{8}}, \bibinfo{pages}{325} (\bibinfo{year}{2012}),
  \eprint{1101.2659}.

\bibitem[{\citenamefont{Calabrese and Cardy}(2006)}]{Calabrese2006}
\bibinfo{author}{\bibfnamefont{P.}~\bibnamefont{Calabrese}} \bibnamefont{and}
  \bibinfo{author}{\bibfnamefont{J.}~\bibnamefont{Cardy}}, \emph{Time
  dependence of correlation functions following a quantum quench},
  \bibinfo{journal}{Physical review letters} \textbf{\bibinfo{volume}{96}},
  \bibinfo{pages}{136801} (\bibinfo{year}{2006}).

\bibitem[{\citenamefont{Cazalilla}(2006)}]{Cazalilla2006}
\bibinfo{author}{\bibfnamefont{M.~A.} \bibnamefont{Cazalilla}}, \emph{Effect of
  Suddenly Turning on Interactions in the Luttinger Model},
  \bibinfo{journal}{Phys. Rev. Lett.} \textbf{\bibinfo{volume}{97}},
  \bibinfo{pages}{156403} (\bibinfo{year}{2006}).

\bibitem[{\citenamefont{Barmettler et~al.}(2009)\citenamefont{Barmettler, Punk,
  Gritsev, Demler, and Altman}}]{Barmettler2009}
\bibinfo{author}{\bibfnamefont{P.}~\bibnamefont{Barmettler}},
  \bibinfo{author}{\bibfnamefont{M.}~\bibnamefont{Punk}},
  \bibinfo{author}{\bibfnamefont{V.}~\bibnamefont{Gritsev}},
  \bibinfo{author}{\bibfnamefont{E.}~\bibnamefont{Demler}}, \bibnamefont{and}
  \bibinfo{author}{\bibfnamefont{E.}~\bibnamefont{Altman}}, \emph{Relaxation of
  antiferromagnetic order in spin-1/2 chains following a quantum quench},
  \bibinfo{journal}{Physical review letters} \textbf{\bibinfo{volume}{102}},
  \bibinfo{pages}{130603} (\bibinfo{year}{2009}).

\bibitem[{\citenamefont{Barmettler et~al.}(2010)\citenamefont{Barmettler, Punk,
  Gritsev, Demler, and Altman}}]{Barmettler2010}
\bibinfo{author}{\bibfnamefont{P.}~\bibnamefont{Barmettler}},
  \bibinfo{author}{\bibfnamefont{M.}~\bibnamefont{Punk}},
  \bibinfo{author}{\bibfnamefont{V.}~\bibnamefont{Gritsev}},
  \bibinfo{author}{\bibfnamefont{E.}~\bibnamefont{Demler}}, \bibnamefont{and}
  \bibinfo{author}{\bibfnamefont{E.}~\bibnamefont{Altman}}, \emph{Quantum
  quenches in the anisotropic spin-1/2 Heisenberg chain: different approaches
  to many-body dynamics far from equilibrium}, \bibinfo{journal}{New Journal of
  Physics} \textbf{\bibinfo{volume}{12}}, \bibinfo{pages}{055017}
  (\bibinfo{year}{2010}).

\bibitem[{\citenamefont{Lux et~al.}(2014)\citenamefont{Lux, M\"uller, Mitra,
  and Rosch}}]{Lux2014}
\bibinfo{author}{\bibfnamefont{J.}~\bibnamefont{Lux}},
  \bibinfo{author}{\bibfnamefont{J.}~\bibnamefont{M\"uller}},
  \bibinfo{author}{\bibfnamefont{A.}~\bibnamefont{Mitra}}, \bibnamefont{and}
  \bibinfo{author}{\bibfnamefont{A.}~\bibnamefont{Rosch}}, \emph{Hydrodynamic
  long-time tails after a quantum quench}, \bibinfo{journal}{Phys. Rev. A}
  \textbf{\bibinfo{volume}{89}}, \bibinfo{pages}{053608}
  (\bibinfo{year}{2014}).

\bibitem[{\citenamefont{{Imbrie}}(2014)}]{Imbrie2014}
\bibinfo{author}{\bibfnamefont{J.~Z.} \bibnamefont{{Imbrie}}}, \emph{{On
  Many-Body Localization for Quantum Spin Chains}}, \bibinfo{journal}{ArXiv
  e-prints}  (\bibinfo{year}{2014}), \eprint{1403.7837}.

\bibitem[{\citenamefont{\v{Z}nidari\v{c}
  et~al.}(2008)\citenamefont{\v{Z}nidari\v{c}, Prosen, and
  Prelov\v{s}ek}}]{Znidaric2008}
\bibinfo{author}{\bibfnamefont{M.}~\bibnamefont{\v{Z}nidari\v{c}}},
  \bibinfo{author}{\bibfnamefont{T.}~\bibnamefont{Prosen}}, \bibnamefont{and}
  \bibinfo{author}{\bibfnamefont{P.}~\bibnamefont{Prelov\v{s}ek}},
  \emph{{Many-body localization in the Heisenberg XXZ magnet in a random
  field}}, \bibinfo{journal}{Physical Review B} \textbf{\bibinfo{volume}{77}},
  \bibinfo{pages}{1} (\bibinfo{year}{2008}), ISSN \bibinfo{issn}{1098-0121}.

\bibitem[{\citenamefont{Pal and Huse}(2010)}]{Pal2010}
\bibinfo{author}{\bibfnamefont{A.}~\bibnamefont{Pal}} \bibnamefont{and}
  \bibinfo{author}{\bibfnamefont{D.~A.} \bibnamefont{Huse}}, \emph{Many-body
  localization phase transition}, \bibinfo{journal}{Phys. Rev. B}
  \textbf{\bibinfo{volume}{82}}, \bibinfo{pages}{174411}
  (\bibinfo{year}{2010}).

\bibitem[{\citenamefont{Bardarson et~al.}(2012)\citenamefont{Bardarson,
  Pollmann, and Moore}}]{Bardarson2012}
\bibinfo{author}{\bibfnamefont{J.~H.} \bibnamefont{Bardarson}},
  \bibinfo{author}{\bibfnamefont{F.}~\bibnamefont{Pollmann}}, \bibnamefont{and}
  \bibinfo{author}{\bibfnamefont{J.~E.} \bibnamefont{Moore}}, \emph{Unbounded
  Growth of Entanglement in Models of Many-Body Localization},
  \bibinfo{journal}{Phys. Rev. Lett.} \textbf{\bibinfo{volume}{109}},
  \bibinfo{pages}{017202} (\bibinfo{year}{2012}).

\bibitem[{\citenamefont{Bauer and Nayak}(2013)}]{Bauer2013}
\bibinfo{author}{\bibfnamefont{B.}~\bibnamefont{Bauer}} \bibnamefont{and}
  \bibinfo{author}{\bibfnamefont{C.}~\bibnamefont{Nayak}}, \emph{{Area laws in
  a many-body localized state and its implications for topological order}},
  p.~\bibinfo{pages}{18} (\bibinfo{year}{2013}), \eprint{1306.5753}.

\bibitem[{\citenamefont{Vosk and Altman}(2013)}]{Vosk2013}
\bibinfo{author}{\bibfnamefont{R.}~\bibnamefont{Vosk}} \bibnamefont{and}
  \bibinfo{author}{\bibfnamefont{E.}~\bibnamefont{Altman}}, \emph{Many-Body
  Localization in One Dimension as a Dynamical Renormalization Group Fixed
  Point}, \bibinfo{journal}{Phys. Rev. Lett.} \textbf{\bibinfo{volume}{110}},
  \bibinfo{pages}{067204} (\bibinfo{year}{2013}).

\bibitem[{\citenamefont{Serbyn et~al.}(2013{\natexlab{a}})\citenamefont{Serbyn,
  Papi\ifmmode~\acute{c}\else \'{c}\fi{}, and Abanin}}]{Serbyn2013}
\bibinfo{author}{\bibfnamefont{M.}~\bibnamefont{Serbyn}},
  \bibinfo{author}{\bibfnamefont{Z.}~\bibnamefont{Papi\ifmmode~\acute{c}\else
  \'{c}\fi{}}}, \bibnamefont{and} \bibinfo{author}{\bibfnamefont{D.~A.}
  \bibnamefont{Abanin}}, \emph{Universal Slow Growth of Entanglement in
  Interacting Strongly Disordered Systems}, \bibinfo{journal}{Phys. Rev. Lett.}
  \textbf{\bibinfo{volume}{110}}, \bibinfo{pages}{260601}
  (\bibinfo{year}{2013}{\natexlab{a}}).

\bibitem[{\citenamefont{Serbyn et~al.}(2013{\natexlab{b}})\citenamefont{Serbyn,
  Papi\ifmmode~\acute{c}\else \'{c}\fi{}, and Abanin}}]{Serbyn2013a}
\bibinfo{author}{\bibfnamefont{M.}~\bibnamefont{Serbyn}},
  \bibinfo{author}{\bibfnamefont{Z.}~\bibnamefont{Papi\ifmmode~\acute{c}\else
  \'{c}\fi{}}}, \bibnamefont{and} \bibinfo{author}{\bibfnamefont{D.~A.}
  \bibnamefont{Abanin}}, \emph{Local Conservation Laws and the Structure of the
  Many-Body Localized States}, \bibinfo{journal}{Phys. Rev. Lett.}
  \textbf{\bibinfo{volume}{111}}, \bibinfo{pages}{127201}
  (\bibinfo{year}{2013}{\natexlab{b}}).

\bibitem[{\citenamefont{{Nandkishore} and {Huse}}(2014)}]{Nandkishore2014}
\bibinfo{author}{\bibfnamefont{R.}~\bibnamefont{{Nandkishore}}}
  \bibnamefont{and} \bibinfo{author}{\bibfnamefont{D.~A.}
  \bibnamefont{{Huse}}}, \emph{{Many body localization and thermalization in
  quantum statistical mechanics}}, \bibinfo{journal}{ArXiv e-prints}
  (\bibinfo{year}{2014}), \eprint{1404.0686}.

\bibitem[{\citenamefont{{Andraschko} et~al.}(2014)\citenamefont{{Andraschko},
  {Enss}, and {Sirker}}}]{Andraschko2014}
\bibinfo{author}{\bibfnamefont{F.}~\bibnamefont{{Andraschko}}},
  \bibinfo{author}{\bibfnamefont{T.}~\bibnamefont{{Enss}}}, \bibnamefont{and}
  \bibinfo{author}{\bibfnamefont{J.}~\bibnamefont{{Sirker}}},
  \emph{{Purification and many-body localization in cold atomic gases}},
  \bibinfo{journal}{ArXiv e-prints}  (\bibinfo{year}{2014}),
  \eprint{1407.4251}.

\bibitem[{\citenamefont{{Vasseur} et~al.}(2014)\citenamefont{{Vasseur},
  {Parameswaran}, and {Moore}}}]{Vasseur2014}
\bibinfo{author}{\bibfnamefont{R.}~\bibnamefont{{Vasseur}}},
  \bibinfo{author}{\bibfnamefont{S.~A.} \bibnamefont{{Parameswaran}}},
  \bibnamefont{and} \bibinfo{author}{\bibfnamefont{J.~E.}
  \bibnamefont{{Moore}}}, \emph{{Quantum revivals and many-body localization}},
  \bibinfo{journal}{ArXiv e-prints}  (\bibinfo{year}{2014}),
  \eprint{1407.4476}.

\bibitem[{\citenamefont{{Serbyn} et~al.}(2014)\citenamefont{{Serbyn},
  {Papi{\'c}}, and {Abanin}}}]{Serbyn2014a}
\bibinfo{author}{\bibfnamefont{M.}~\bibnamefont{{Serbyn}}},
  \bibinfo{author}{\bibfnamefont{Z.}~\bibnamefont{{Papi{\'c}}}},
  \bibnamefont{and} \bibinfo{author}{\bibfnamefont{D.~A.}
  \bibnamefont{{Abanin}}}, \emph{{Quantum quenches in the many-body localized
  phase}}, \bibinfo{journal}{ArXiv e-prints}  (\bibinfo{year}{2014}),
  \eprint{1408.4105}.

\bibitem[{\citenamefont{Bar~Lev and Reichman}(2014)}]{BarLev2014}
\bibinfo{author}{\bibfnamefont{Y.}~\bibnamefont{Bar~Lev}} \bibnamefont{and}
  \bibinfo{author}{\bibfnamefont{D.~R.} \bibnamefont{Reichman}}, \emph{Dynamics
  of many-body localization}, \bibinfo{journal}{Phys. Rev. B}
  \textbf{\bibinfo{volume}{89}}, \bibinfo{pages}{220201}
  (\bibinfo{year}{2014}).

\bibitem[{\citenamefont{{Agarwal} et~al.}(2014)\citenamefont{{Agarwal},
  {Gopalakrishnan}, {Knap}, {Mueller}, and {Demler}}}]{Agarwal2014}
\bibinfo{author}{\bibfnamefont{K.}~\bibnamefont{{Agarwal}}},
  \bibinfo{author}{\bibfnamefont{S.}~\bibnamefont{{Gopalakrishnan}}},
  \bibinfo{author}{\bibfnamefont{M.}~\bibnamefont{{Knap}}},
  \bibinfo{author}{\bibfnamefont{M.}~\bibnamefont{{Mueller}}},
  \bibnamefont{and} \bibinfo{author}{\bibfnamefont{E.}~\bibnamefont{{Demler}}},
  \emph{{Anomalous diffusion and Griffiths effects near the many-body
  localization transition}}, \bibinfo{journal}{ArXiv e-prints}
  (\bibinfo{year}{2014}), \eprint{1408.3413}.

\bibitem[{\citenamefont{{Grover}}(2014)}]{Grover2014}
\bibinfo{author}{\bibfnamefont{T.}~\bibnamefont{{Grover}}}, \emph{{Certain
  General Constraints on the Many-Body Localization Transition}},
  \bibinfo{journal}{ArXiv e-prints}  (\bibinfo{year}{2014}),
  \eprint{1405.1471}.

\bibitem[{\citenamefont{Vosk and Altman}(2014)}]{Vosk2014}
\bibinfo{author}{\bibfnamefont{R.}~\bibnamefont{Vosk}} \bibnamefont{and}
  \bibinfo{author}{\bibfnamefont{E.}~\bibnamefont{Altman}}, \emph{Dynamical
  Quantum Phase Transitions in Random Spin Chains}, \bibinfo{journal}{Phys.
  Rev. Lett.} \textbf{\bibinfo{volume}{112}}, \bibinfo{pages}{217204}
  (\bibinfo{year}{2014}).

\bibitem[{\citenamefont{{Ovadia} et~al.}(2014)\citenamefont{{Ovadia}, {Kalok},
  {Tamir}, {Mitra}, {Sacepe}, and {Shahar}}}]{Ovadia2014}
\bibinfo{author}{\bibfnamefont{M.}~\bibnamefont{{Ovadia}}},
  \bibinfo{author}{\bibfnamefont{D.}~\bibnamefont{{Kalok}}},
  \bibinfo{author}{\bibfnamefont{I.}~\bibnamefont{{Tamir}}},
  \bibinfo{author}{\bibfnamefont{S.}~\bibnamefont{{Mitra}}},
  \bibinfo{author}{\bibfnamefont{B.}~\bibnamefont{{Sacepe}}}, \bibnamefont{and}
  \bibinfo{author}{\bibfnamefont{D.}~\bibnamefont{{Shahar}}}, \emph{{Evidence
  for a Finite Temperature Insulator}}, \bibinfo{journal}{ArXiv e-prints}
  (\bibinfo{year}{2014}), \eprint{1406.7510}.

\bibitem[{\citenamefont{{Altman} and {Vosk}}(2014)}]{Altman2014}
\bibinfo{author}{\bibfnamefont{E.}~\bibnamefont{{Altman}}} \bibnamefont{and}
  \bibinfo{author}{\bibfnamefont{R.}~\bibnamefont{{Vosk}}}, \emph{{Universal
  dynamics and renormalization in many body localized systems}},
  \bibinfo{journal}{ArXiv e-prints}  (\bibinfo{year}{2014}),
  \eprint{1408.2834}.

\bibitem[{\citenamefont{Oganesyan and Huse}(2007)}]{Oganesyan2007}
\bibinfo{author}{\bibfnamefont{V.}~\bibnamefont{Oganesyan}} \bibnamefont{and}
  \bibinfo{author}{\bibfnamefont{D.}~\bibnamefont{Huse}}, \emph{{Localization
  of interacting fermions at high temperature}}, \bibinfo{journal}{Physical
  Review B} \textbf{\bibinfo{volume}{75}}, \bibinfo{pages}{1}
  (\bibinfo{year}{2007}), ISSN \bibinfo{issn}{1098-0121}.

\bibitem[{\citenamefont{{Shankar}}(1994)}]{Shankar1994}
\bibinfo{author}{\bibfnamefont{R.}~\bibnamefont{{Shankar}}},
  \emph{{Renormalization-group approach to interacting fermions}},
  \bibinfo{journal}{Reviews of Modern Physics} \textbf{\bibinfo{volume}{66}},
  \bibinfo{pages}{129} (\bibinfo{year}{1994}), \eprint{cond-mat/9307009}.

\bibitem[{\citenamefont{Pekker et~al.}(2014)\citenamefont{Pekker, Refael,
  Altman, Demler, and Oganesyan}}]{Pekker2014}
\bibinfo{author}{\bibfnamefont{D.}~\bibnamefont{Pekker}},
  \bibinfo{author}{\bibfnamefont{G.}~\bibnamefont{Refael}},
  \bibinfo{author}{\bibfnamefont{E.}~\bibnamefont{Altman}},
  \bibinfo{author}{\bibfnamefont{E.}~\bibnamefont{Demler}}, \bibnamefont{and}
  \bibinfo{author}{\bibfnamefont{V.}~\bibnamefont{Oganesyan}},
  \emph{Hilbert-Glass Transition: New Universality of Temperature-Tuned
  Many-Body Dynamical Quantum Criticality}, \bibinfo{journal}{Phys. Rev. X}
  \textbf{\bibinfo{volume}{4}}, \bibinfo{pages}{011052} (\bibinfo{year}{2014}).

\bibitem[{\citenamefont{Mathey and Polkovnikov}(2010)}]{Mathey2010}
\bibinfo{author}{\bibfnamefont{L.}~\bibnamefont{Mathey}} \bibnamefont{and}
  \bibinfo{author}{\bibfnamefont{A.}~\bibnamefont{Polkovnikov}}, \emph{Light
  cone dynamics and reverse Kibble-Zurek mechanism in two-dimensional
  superfluids following a quantum quench}, \bibinfo{journal}{Phys. Rev. A}
  \textbf{\bibinfo{volume}{81}}, \bibinfo{pages}{033605}
  (\bibinfo{year}{2010}).

\bibitem[{\citenamefont{Huse et~al.}(2014)\citenamefont{Huse, Nandkishore, and
  Oganesyan}}]{Huse2014}
\bibinfo{author}{\bibfnamefont{D.~A.} \bibnamefont{Huse}},
  \bibinfo{author}{\bibfnamefont{R.}~\bibnamefont{Nandkishore}},
  \bibnamefont{and}
  \bibinfo{author}{\bibfnamefont{V.}~\bibnamefont{Oganesyan}},
  \emph{Phenomenology of fully many-body-localized systems},
  \bibinfo{journal}{Phys. Rev. B} \textbf{\bibinfo{volume}{90}},
  \bibinfo{pages}{174202} (\bibinfo{year}{2014}).

\bibitem[{\citenamefont{Dasgupta and Ma}(1980)}]{Dasgupta1980}
\bibinfo{author}{\bibfnamefont{C.}~\bibnamefont{Dasgupta}} \bibnamefont{and}
  \bibinfo{author}{\bibfnamefont{S.-k.} \bibnamefont{Ma}},
  \emph{{Low-temperature properties of the random Heisenberg antiferromagnetic
  chain}}, \bibinfo{journal}{Physical Review B} \textbf{\bibinfo{volume}{22}},
  \bibinfo{pages}{1305} (\bibinfo{year}{1980}), ISSN \bibinfo{issn}{0163-1829}.

\bibitem[{\citenamefont{Serbyn et~al.}(2014)\citenamefont{Serbyn, Knap,
  Gopalakrishnan, Papi\ifmmode~\acute{c}\else \'{c}\fi{}, Yao, Laumann, Abanin,
  Lukin, and Demler}}]{Serbyn2014}
\bibinfo{author}{\bibfnamefont{M.}~\bibnamefont{Serbyn}},
  \bibinfo{author}{\bibfnamefont{M.}~\bibnamefont{Knap}},
  \bibinfo{author}{\bibfnamefont{S.}~\bibnamefont{Gopalakrishnan}},
  \bibinfo{author}{\bibfnamefont{Z.}~\bibnamefont{Papi\ifmmode~\acute{c}\else
  \'{c}\fi{}}}, \bibinfo{author}{\bibfnamefont{N.~Y.} \bibnamefont{Yao}},
  \bibinfo{author}{\bibfnamefont{C.~R.} \bibnamefont{Laumann}},
  \bibinfo{author}{\bibfnamefont{D.~A.} \bibnamefont{Abanin}},
  \bibinfo{author}{\bibfnamefont{M.~D.} \bibnamefont{Lukin}}, \bibnamefont{and}
  \bibinfo{author}{\bibfnamefont{E.~A.} \bibnamefont{Demler}},
  \emph{Interferometric Probes of Many-Body Localization},
  \bibinfo{journal}{Phys. Rev. Lett.} \textbf{\bibinfo{volume}{113}},
  \bibinfo{pages}{147204} (\bibinfo{year}{2014}).

\bibitem[{\citenamefont{{Bahri} et~al.}(2013)\citenamefont{{Bahri}, {Vosk},
  {Altman}, and {Vishwanath}}}]{Bahri2013}
\bibinfo{author}{\bibfnamefont{Y.}~\bibnamefont{{Bahri}}},
  \bibinfo{author}{\bibfnamefont{R.}~\bibnamefont{{Vosk}}},
  \bibinfo{author}{\bibfnamefont{E.}~\bibnamefont{{Altman}}}, \bibnamefont{and}
  \bibinfo{author}{\bibfnamefont{A.}~\bibnamefont{{Vishwanath}}},
  \emph{{Localization and topology protected quantum coherence at the edge of
  'hot' matter}}, \bibinfo{journal}{ArXiv e-prints}  (\bibinfo{year}{2013}),
  \eprint{1307.4092}.

\bibitem[{\citenamefont{Huse et~al.}(2013)\citenamefont{Huse, Nandkishore,
  Oganesyan, Pal, and Sondhi}}]{Huse2013}
\bibinfo{author}{\bibfnamefont{D.~A.} \bibnamefont{Huse}},
  \bibinfo{author}{\bibfnamefont{R.}~\bibnamefont{Nandkishore}},
  \bibinfo{author}{\bibfnamefont{V.}~\bibnamefont{Oganesyan}},
  \bibinfo{author}{\bibfnamefont{A.}~\bibnamefont{Pal}}, \bibnamefont{and}
  \bibinfo{author}{\bibfnamefont{S.~L.} \bibnamefont{Sondhi}},
  \emph{Localization-protected quantum order}, \bibinfo{journal}{Phys. Rev. B}
  \textbf{\bibinfo{volume}{88}}, \bibinfo{pages}{014206}
  (\bibinfo{year}{2013}).

\bibitem[{\citenamefont{Vosk et~al.}(2015)\citenamefont{Vosk, Huse, and
  Altman}}]{Vosk2015}
\bibinfo{author}{\bibfnamefont{R.}~\bibnamefont{Vosk}},
  \bibinfo{author}{\bibfnamefont{D.~A.} \bibnamefont{Huse}}, \bibnamefont{and}
  \bibinfo{author}{\bibfnamefont{E.}~\bibnamefont{Altman}}, \emph{Theory of the
  Many-Body Localization Transition in One-Dimensional Systems},
  \bibinfo{journal}{Phys. Rev. X} \textbf{\bibinfo{volume}{5}},
  \bibinfo{pages}{031032} (\bibinfo{year}{2015}).

\bibitem[{\citenamefont{Griffiths}(1969)}]{Griffiths1969}
\bibinfo{author}{\bibfnamefont{R.~B.} \bibnamefont{Griffiths}},
  \emph{Nonanalytic Behavior Above the Critical Point in a Random Ising
  Ferromagnet}, \bibinfo{journal}{Phys. Rev. Lett.}
  \textbf{\bibinfo{volume}{23}}, \bibinfo{pages}{17} (\bibinfo{year}{1969}).

\bibitem[{\citenamefont{{Billy} et~al.}(2008)\citenamefont{{Billy}, {Josse},
  {Zuo}, {Bernard}, {Hambrecht}, {Lugan}, {Cl{\'e}ment}, {Sanchez-Palencia},
  {Bouyer}, and {Aspect}}}]{Billy2008}
\bibinfo{author}{\bibfnamefont{J.}~\bibnamefont{{Billy}}},
  \bibinfo{author}{\bibfnamefont{V.}~\bibnamefont{{Josse}}},
  \bibinfo{author}{\bibfnamefont{Z.}~\bibnamefont{{Zuo}}},
  \bibinfo{author}{\bibfnamefont{A.}~\bibnamefont{{Bernard}}},
  \bibinfo{author}{\bibfnamefont{B.}~\bibnamefont{{Hambrecht}}},
  \bibinfo{author}{\bibfnamefont{P.}~\bibnamefont{{Lugan}}},
  \bibinfo{author}{\bibfnamefont{D.}~\bibnamefont{{Cl{\'e}ment}}},
  \bibinfo{author}{\bibfnamefont{L.}~\bibnamefont{{Sanchez-Palencia}}},
  \bibinfo{author}{\bibfnamefont{P.}~\bibnamefont{{Bouyer}}}, \bibnamefont{and}
  \bibinfo{author}{\bibfnamefont{A.}~\bibnamefont{{Aspect}}}, \emph{{Direct
  observation of Anderson localization of matter waves in a controlled
  disorder}}, \bibinfo{journal}{\nat} \textbf{\bibinfo{volume}{453}},
  \bibinfo{pages}{891} (\bibinfo{year}{2008}), \eprint{0804.1621}.

\bibitem[{\citenamefont{{Roati} et~al.}(2008)\citenamefont{{Roati}, {D'Errico},
  {Fallani}, {Fattori}, {Fort}, {Zaccanti}, {Modugno}, {Modugno}, and
  {Inguscio}}}]{Roati2008}
\bibinfo{author}{\bibfnamefont{G.}~\bibnamefont{{Roati}}},
  \bibinfo{author}{\bibfnamefont{C.}~\bibnamefont{{D'Errico}}},
  \bibinfo{author}{\bibfnamefont{L.}~\bibnamefont{{Fallani}}},
  \bibinfo{author}{\bibfnamefont{M.}~\bibnamefont{{Fattori}}},
  \bibinfo{author}{\bibfnamefont{C.}~\bibnamefont{{Fort}}},
  \bibinfo{author}{\bibfnamefont{M.}~\bibnamefont{{Zaccanti}}},
  \bibinfo{author}{\bibfnamefont{G.}~\bibnamefont{{Modugno}}},
  \bibinfo{author}{\bibfnamefont{M.}~\bibnamefont{{Modugno}}},
  \bibnamefont{and}
  \bibinfo{author}{\bibfnamefont{M.}~\bibnamefont{{Inguscio}}}, \emph{{Anderson
  localization of a non-interacting Bose-Einstein condensate}},
  \bibinfo{journal}{\nat} \textbf{\bibinfo{volume}{453}}, \bibinfo{pages}{895}
  (\bibinfo{year}{2008}), \eprint{0804.2609}.

\bibitem[{\citenamefont{Kondov et~al.}(2011)\citenamefont{Kondov, McGehee,
  Zirbel, and DeMarco}}]{Kondov2011}
\bibinfo{author}{\bibfnamefont{S.~S.} \bibnamefont{Kondov}},
  \bibinfo{author}{\bibfnamefont{W.~R.} \bibnamefont{McGehee}},
  \bibinfo{author}{\bibfnamefont{J.~J.} \bibnamefont{Zirbel}},
  \bibnamefont{and} \bibinfo{author}{\bibfnamefont{B.}~\bibnamefont{DeMarco}},
  \emph{Three-Dimensional Anderson Localization of Ultracold Matter},
  \bibinfo{journal}{Science} \textbf{\bibinfo{volume}{334}},
  \bibinfo{pages}{66} (\bibinfo{year}{2011}),
  \eprint{http://www.sciencemag.org/content/334/6052/66.full.pdf}.

\bibitem[{\citenamefont{{Jendrzejewski}
  et~al.}(2012)\citenamefont{{Jendrzejewski}, {Bernard}, {M{\"u}ller},
  {Cheinet}, {Josse}, {Piraud}, {Pezz{\'e}}, {Sanchez-Palencia}, {Aspect}, and
  {Bouyer}}}]{Jendrzejewski2012}
\bibinfo{author}{\bibfnamefont{F.}~\bibnamefont{{Jendrzejewski}}},
  \bibinfo{author}{\bibfnamefont{A.}~\bibnamefont{{Bernard}}},
  \bibinfo{author}{\bibfnamefont{K.}~\bibnamefont{{M{\"u}ller}}},
  \bibinfo{author}{\bibfnamefont{P.}~\bibnamefont{{Cheinet}}},
  \bibinfo{author}{\bibfnamefont{V.}~\bibnamefont{{Josse}}},
  \bibinfo{author}{\bibfnamefont{M.}~\bibnamefont{{Piraud}}},
  \bibinfo{author}{\bibfnamefont{L.}~\bibnamefont{{Pezz{\'e}}}},
  \bibinfo{author}{\bibfnamefont{L.}~\bibnamefont{{Sanchez-Palencia}}},
  \bibinfo{author}{\bibfnamefont{A.}~\bibnamefont{{Aspect}}}, \bibnamefont{and}
  \bibinfo{author}{\bibfnamefont{P.}~\bibnamefont{{Bouyer}}},
  \emph{{Three-dimensional localization of ultracold atoms in an optical
  disordered potential}}, \bibinfo{journal}{Nature Physics}
  \textbf{\bibinfo{volume}{8}}, \bibinfo{pages}{398} (\bibinfo{year}{2012}),
  \eprint{1108.0137}.

\bibitem[{\citenamefont{{Pasienski} et~al.}(2010)\citenamefont{{Pasienski},
  {McKay}, {White}, and {Demarco}}}]{Pasienski2010}
\bibinfo{author}{\bibfnamefont{M.}~\bibnamefont{{Pasienski}}},
  \bibinfo{author}{\bibfnamefont{D.}~\bibnamefont{{McKay}}},
  \bibinfo{author}{\bibfnamefont{M.}~\bibnamefont{{White}}}, \bibnamefont{and}
  \bibinfo{author}{\bibfnamefont{B.}~\bibnamefont{{Demarco}}}, \emph{{A
  disordered insulator in an optical lattice}}, \bibinfo{journal}{Nature
  Physics} \textbf{\bibinfo{volume}{6}}, \bibinfo{pages}{677}
  (\bibinfo{year}{2010}), \eprint{0908.1182}.

\bibitem[{\citenamefont{Gadway et~al.}(2011)\citenamefont{Gadway, Pertot,
  Reeves, Vogt, and Schneble}}]{Gadway2011}
\bibinfo{author}{\bibfnamefont{B.}~\bibnamefont{Gadway}},
  \bibinfo{author}{\bibfnamefont{D.}~\bibnamefont{Pertot}},
  \bibinfo{author}{\bibfnamefont{J.}~\bibnamefont{Reeves}},
  \bibinfo{author}{\bibfnamefont{M.}~\bibnamefont{Vogt}}, \bibnamefont{and}
  \bibinfo{author}{\bibfnamefont{D.}~\bibnamefont{Schneble}}, \emph{Glassy
  Behavior in a Binary Atomic Mixture}, \bibinfo{journal}{Phys. Rev. Lett.}
  \textbf{\bibinfo{volume}{107}}, \bibinfo{pages}{145306}
  (\bibinfo{year}{2011}).

\bibitem[{\citenamefont{D'Errico et~al.}(2014)\citenamefont{D'Errico, Lucioni,
  Tanzi, Gori, Roux, McCulloch, Giamarchi, Inguscio, and
  Modugno}}]{DErrico2014}
\bibinfo{author}{\bibfnamefont{C.}~\bibnamefont{D'Errico}},
  \bibinfo{author}{\bibfnamefont{E.}~\bibnamefont{Lucioni}},
  \bibinfo{author}{\bibfnamefont{L.}~\bibnamefont{Tanzi}},
  \bibinfo{author}{\bibfnamefont{L.}~\bibnamefont{Gori}},
  \bibinfo{author}{\bibfnamefont{G.}~\bibnamefont{Roux}},
  \bibinfo{author}{\bibfnamefont{I.~P.} \bibnamefont{McCulloch}},
  \bibinfo{author}{\bibfnamefont{T.}~\bibnamefont{Giamarchi}},
  \bibinfo{author}{\bibfnamefont{M.}~\bibnamefont{Inguscio}}, \bibnamefont{and}
  \bibinfo{author}{\bibfnamefont{G.}~\bibnamefont{Modugno}}, \emph{Observation
  of a Disordered Bosonic Insulator from Weak to Strong Interactions},
  \bibinfo{journal}{Phys. Rev. Lett.} \textbf{\bibinfo{volume}{113}},
  \bibinfo{pages}{095301} (\bibinfo{year}{2014}).

\bibitem[{\citenamefont{{Meldgin} et~al.}(2015)\citenamefont{{Meldgin}, {Ray},
  {Russ}, {Ceperley}, and {DeMarco}}}]{Meldgin2015}
\bibinfo{author}{\bibfnamefont{C.}~\bibnamefont{{Meldgin}}},
  \bibinfo{author}{\bibfnamefont{U.}~\bibnamefont{{Ray}}},
  \bibinfo{author}{\bibfnamefont{P.}~\bibnamefont{{Russ}}},
  \bibinfo{author}{\bibfnamefont{D.}~\bibnamefont{{Ceperley}}},
  \bibnamefont{and}
  \bibinfo{author}{\bibfnamefont{B.}~\bibnamefont{{DeMarco}}}, \emph{{Probing
  the Bose-Glass--Superfluid Transition using Quantum Quenches of Disorder}},
  \bibinfo{journal}{ArXiv e-prints}  (\bibinfo{year}{2015}),
  \eprint{1502.02333}.

\bibitem[{\citenamefont{Aubry and Andr{\'e}}(1980)}]{Aubry1980}
\bibinfo{author}{\bibfnamefont{S.}~\bibnamefont{Aubry}} \bibnamefont{and}
  \bibinfo{author}{\bibfnamefont{G.}~\bibnamefont{Andr{\'e}}},
  \emph{Analyticity breaking and Anderson localization in incommensurate
  lattices}, \bibinfo{journal}{Ann. Israel Phys. Soc}
  \textbf{\bibinfo{volume}{3}}, \bibinfo{pages}{18} (\bibinfo{year}{1980}).

\bibitem[{\citenamefont{{Gopalakrishnan}
  et~al.}(2015)\citenamefont{{Gopalakrishnan}, {Mueller}, {Khemani}, {Knap},
  {Demler}, and {Huse}}}]{Gopalakrishnan2015}
\bibinfo{author}{\bibfnamefont{S.}~\bibnamefont{{Gopalakrishnan}}},
  \bibinfo{author}{\bibfnamefont{M.}~\bibnamefont{{Mueller}}},
  \bibinfo{author}{\bibfnamefont{V.}~\bibnamefont{{Khemani}}},
  \bibinfo{author}{\bibfnamefont{M.}~\bibnamefont{{Knap}}},
  \bibinfo{author}{\bibfnamefont{E.}~\bibnamefont{{Demler}}}, \bibnamefont{and}
  \bibinfo{author}{\bibfnamefont{D.~A.} \bibnamefont{{Huse}}},
  \emph{{Low-frequency conductivity in many-body localized systems}},
  \bibinfo{journal}{ArXiv e-prints}  (\bibinfo{year}{2015}),
  \eprint{1502.07712}.

\end{thebibliography}

\end{document}